## Title:

# Quantum diamond spectrometer for nanoscale NMR and ESR spectroscopy

## Authors:

Dominik B. Bucher[1,2], Diana P. L. Aude Craik[2], Mikael P. Backlund[1,2], Matthew J. Turner[2], Oren Ben-Dor[1,2], David R. Glenn[2], and Ronald L. Walsworth[1,2]

1. Harvard-Smithsonian Center for Astrophysics, Cambridge, MA
2. Department of Physics, Harvard University, Cambridge, MA

Correspondence should be sent to rwalsworth@cfa.harvard.edu or dominik.bucher@cfa.harvard.edu

## Abstract:

Nitrogen-vacancy (NV) quantum defects in diamond are sensitive detectors of magnetic fields. Due to their atomic size and optical readout capability, they have been used for magnetic resonance spectroscopy of nanoscale samples on diamond surfaces. Here we present a protocol for fabricating NV-diamond chips and for constructing and operating a simple, low-cost "quantum diamond spectrometer" for performing nuclear magnetic resonance (NMR) and electron spin resonance (ESR) spectroscopy in nanoscale volumes. The instrument is based on a commercially-available diamond chip, with an ion-implanted NV-ensemble at a depth of ~ 10 nm below the diamond surface. The spectrometer operates at low magnetic fields (~ 300 G) and requires standard optical and microwave components for NV spin preparation, manipulation and readout. We demonstrate the utility of this device for nanoscale proton and fluorine NMR spectroscopy, as well as for the detection of transition metals via ESR noise spectroscopy. We estimate that the full protocol requires 2-3 months to implement, depending on the availability of equipment, diamond substrates, and user experience.

## Introduction:

Magnetic resonance spectroscopy of electrons and nuclei comprises a family of ubiquitous and essential analytical tools in modern chemical and biological research[1]. Electron spin resonance (ESR), also known as electron paramagnetic resonance (EPR), spectroscopy is a useful means for probing molecules possessing unpaired electrons, such as transition metal complexes and radicals[2]. (Bio)molecules that lack an unpaired electronic spin can be probed via ESR-active spin labels. Nuclear magnetic resonance (NMR), on the other hand, is a more widely utilized technique, as NMR-active nuclei (e.g., $^{1}H$, $^{13}C$, $^{14}N$, and $^{31}P$) are commonly encountered in organic and biological chemistry. The narrow spectral lines of NMR afford unprecedented information about molecular structure and dynamics. NMR is less sensitive than ESR, however, owing to the lower gyromagnetic ratios of nuclei compared to that of the electron. In fact, both NMR and ESR are relatively insensitive when compared to the state of the art in other analytical techniques like mass spectrometry or fluorescence microscopy. The low sensitivity of magnetic resonance is particularly challenging for life science applications, where biomolecules of interest commonly occur in small absolute quantities or concentrations. Thus, there is great interest in new techniques to increase the sensitivity of magnetic resonance spectroscopy[3–5]. One promising approach employs a magnetic sensor based on fluorescent quantum defects in diamond, such as nitrogen-vacancy (NV) color centers, enabling interrogation of sample volumes down to the nanoscale[6,7], including single proteins[8,9], single protons[10] and 2D materials[11]. In this protocol, we describe the procedure for generating NV-diamond sensor chips and the construction of a "quantum diamond spectrometer" for NMR and ESR of nanoscale samples placed on the diamond chip.

## Physical background

**NV diamond quantum sensor.** NV color centers are atomic-scale quantum defects that provide high-resolution magnetic field sensing and imaging via optically detected magnetic resonance (ODMR), with broad applicability in both the physical and life sciences. NVs have been reviewed extensively elsewhere[12–

[16]; and so we give only a brief overview here to introduce the most important concepts for the present work. An NV center is created when two neighboring carbon atoms in the diamond lattice are replaced by a nitrogen atom and a vacancy (Figure 1b), resulting in local $C_{3v}$ symmetry. The point-like defect has electronic states that sit within the band gap of diamond, a fact that allows one to address the energy levels of the NV in a manner analogous to the manipulation of molecular or atomic levels. The NV center can exist in several charge states, the most widely studied of which is the negatively charged $NV^-$, with electronic spin $S = 1$ in its ground state. Throughout this manuscript we use the term "NV center" to be synonymous with $NV^-$. A zero-phonon splitting of 637 nm separates the electronic spin-triplet ground and excited states. Each of these states is further split by higher order interactions, some of which are described below. A broad phonon side band allows one to prepare and read out the NV spin state with absorption of blue-shifted light (e.g., a 532-nm laser) and then detection of the red-shifted photoluminescence (PL) (Figure 1a).

NV centers have drawn considerable interest in recent years as a tool for sensing, especially of magnetic fields[13,15,17]. For these applications, the splittings of the eigenstates of the *z*-component of the electron spin operator $S_z$ are of particular importance, where *z* refers to the axis connecting the nitrogen atom and vacancy (i.e., the line connecting the red and light gray dots in Figure 1b). Such splittings in the electronic ground state can be understood by considering the relevant (simplified) spin Hamiltonian:

$$\mathcal{H} = DS_z^2 - \gamma_e \mathbf{B} \cdot \mathbf{S}, \qquad (1)$$

where $D = 2.87$ GHz, $\mathbf{S} = [S_x, S_y, S_z]^T$ is the electronic spin operator, $\gamma_e = 2.8$ MHz/G is the electron's gyromagnetic ratio[14], and $\mathbf{B}$ is an external magnetic field that in this work consists of a strong applied field along *z* ($B_0$) plus a weak 'signal field' to be sensed ($B_z^{(\mathrm{sense})}$). The first term on the right hand side of equation 1 describes a zero-field splitting due to spin-spin interactions[17]. The second term describes the Zeeman interaction with the magnetic field. For a sufficiently weak signal field, the NV is largely sensitive to the *z* component such that:

$$\mathcal{H} = DS_z^2 - \gamma_e \left(B_0 + B_z^{(\mathrm{sense})}\right) S_z, \qquad (2)$$

Thus, the signal field has the effect of shifting the spin state states $m_s = \pm 1$ by $\pm\gamma_e B_z^{(\mathrm{sense})}$ (Figure 1a, inset). Transitions between ground electronic states of different $m_s$ are driven by application of a resonant microwave (MW) field. A sensing sequence consists of a specified set of MW pulses followed by readout of the final spin state. Spin-state readout is enabled by the fact that NVs emit fewer PL photons on average when optically excited from $m_s = \pm 1$ than they do from $m_s = 0$, owing to a significant probability of decaying via an alternative pathway mediated by singlet electronic states in the former case (Figure 1a). Thus, an NV ensemble with more population in either $m_s = \pm 1$ fluoresces less brightly than one with more population in $m_s = 0$.

The other important characteristic that enables optically detected magnetic sensing with NVs is the ability to initialize the spin state. At thermal equilibrium, each of the three $m_s$ sublevels in the electronic ground state is roughly equally populated. However, a laser pulse of sufficient duration results in nearly complete polarization into the $m_s = 0$ state, a consequence of the unequal nonradiative decay rates[18]. Thus, an arbitrary (nearly) pure state can be obtained at the onset of a measurement by application of an initializing laser pulse followed by the appropriate MW pulse. This initialization step is a common motif in the nuclear and electronic spin sensing protocols presented herein.

NV-mediated sensing can be implemented with either a single NV or an NV ensemble. On the one hand, single-NV sensing allows for atomic-scale spatial resolution. On the other hand, ensemble averaging leads to a sensitivity improvement roughly proportional to the reciprocal square root of the number of sensor NVs.

**Nuclear spin sensing.** When exposed to a static magnetic field of a few hundred gauss, nuclear spins generate NMR signals, i.e., alternating (AC) magnetic fields ($B_{AC}$), with frequencies between hundreds of kHz and a few MHz. In the protocol presented here, these NMR fields are sensed with a near-surface NV-ensemble-layer. Importantly, each NV is primarily sensitive to NMR fields generated by nuclear spins within a hemisphere volume above the diamond, with the radius set by the depth of the NV beneath the surface[6,19–21]. For instance, NVs that have been implanted 10 nm beneath the diamond surface sense NMR signals from approximately a 10-nm surface layer. In such nanoscale volumes, the number of spins is relatively small, and so the statistical polarization (~ $1/\sqrt{number\ of\ spins}$) exceeds the thermal spin polarization (~ $10^{-7}$-$10^{-5}$ for 0.1-10 T fields at room temperature)[22]. An appropriate AC magnetometry pulse sequence can be used to measure the fluctuations of the statistical polarization[23,24]. As mentioned above, all pulse sequences start with a short laser pulse to initialize the NV into the $m_s = 0$ ground sub-state. Each sequence ends with a laser pulse to read out the quantum state via detection of the PL (Figure 2a). Between these two laser pulses, MW pulse sequences are applied to manipulate the NV quantum state in accordance with a chosen sensing protocol. The AC magnetometry pulse sequence begins with a π/2 pulse, which generates a quantum coherence between the $m_s = 0$ and $m_s = -1$ state by rotating the Bloch vector corresponding to the NV quantum state to the equator of the Bloch sphere (see Figure 2b). This coherent superposition is then allowed to evolve for a specified free precession time, during which it accumulates phase in a manner that depends on the magnetic field being sensed. A final π/2 pulse maps the accumulated phase into a population difference between $m_s = 0$ and $m_s = -1$, which is translated into a change in the NV fluorescence rate during the laser readout pulse. During free precession, a train of π pulses with defined phases, termed a dynamic decoupling sequence (for instance an XY8-N sequence described here[25,26]), is applied. The purpose of this sequence is two-fold: it extends the NV coherence time and creates a narrowband detector for magnetic signals with frequency near $f = 1/(2\tau)$, where $\tau$ is the spacing between pulses. In the case of a XY8-N sequence, the block of eight π pulses is repeated $N$ times, where choice of $N$ depends on $\tau$ and the decoherence properties of the NVs. In subsequent measurements $\tau$ is typically swept. When the condition

$$\tau = \frac{1}{2f_0}, \qquad (3)$$

is satisfied, where $f_0$ is the Larmor frequency of the sample spins, the NV center accumulates maximum phase, leading to a measurable reduction in the NV fluorescence rate during the readout laser pulse. The XY8-N pulse sequence has been usefully applied in sensing surface nuclear spins[6,8,20].

However, experiments have shown that this pulse sequence can also pick up signals from higher harmonics of $f_0$, which can lead to ambiguous results[27]. This issue can be overcome by correlating two consecutive sensing pulse sequences[24,28–30]. This so called "correlation spectroscopy" consists of two XY8-N sequences, separated in time by $t_{corr}$, which is swept during the experiment (Figure 2c). The NV-phase accumulation in a dynamic decoupling sequence depends on the relative phase $\phi$ of the sensed magnetic AC field. Intuitively, if $t_{corr}$ is an integer multiple of $1/f_0$, both XY8-N sequences accumulate the same NV-phase (since $\phi$ is identical) and the correlation signal is at its maximum. If $t_{corr}$ is a half-integer multiple of $1/f_0$ the readout signal is at its minimum, since the magnetic ac field phase $\phi$ is out of phase. As $t_{corr}$ is swept, the resulting PL readout signal oscillates at the nuclear spin Larmor frequency in a manner similar to the free induction decay in conventional NMR.
The NV-center has been successfully used to detect NMR signals from nanoscale sample volumes[6,20,30], single proteins[8], single protons[10] and 2D materials[11]. Many of these experiments have been performed with a single NV sensor, but one can also take advantage of an ensemble of NV sensors for both wide field imaging[20] and enhanced sensitivity[31].

**Limitations of nuclear spin sensing with NV centers.** High-frequency resolution (few Hz) is important for resolving molecular structures via chemical shifts and scalar (i.e., "J") couplings. The nanoscale NV-experiments described above are limited in that they provide only modest frequency resolution of typically 1-100 kHz. This limitation is due to two reasons: (i) Measured linewidths are limited by NV-T2 relaxation when sensing is performed using dynamical decoupling sequences, and by NV-T1 relaxation when correlation spectroscopy is used. Since NV-T2 < NV-T1, correlation spectroscopy gives superior frequency resolution, as good as ~ 100 Hz. (ii) Unfortunately, this frequency resolution is usually not achieved in nanoscale NV-NMR experiments. Sample diffusion limits the interaction time between the NV sensor and the nuclear spin, resulting in short correlations times $\tau_c$ and broadened lines. The linewidth, LW, depends on the diffusion coefficient, $D$, and NV depth, $d$, as described in the following equation:

$$\mathrm{LW} \approx \frac{1}{\pi\tau_c} = \frac{6D}{\pi d^2} \qquad (4)$$

For instance, for a 10-nm deep NV, signals produced by protons in water give rise to linewidths of ~ 40 MHz and viscous oil signals produce linewidths of ~ 10 kHz.

Recent experiments have overcome some of these limitations through two different approaches. The first approach uses nuclear spins in the diamond as a quantum memory to extend the readout time, enabling resolution of chemical shifts at high magnetic fields (~3 T)[32]. In this approach, statistical polarization is detected from nanoscale sample volumes, and linewidths remain limited by sample diffusion. Our group recently demonstrated an alternative “synchronized readout” approach based on XY8-N sequences, achieving linewidths on the order of 1 Hz and thereby enabling chemical shift and J-coupling resolution[33,34]. In these experiments, we overcome the diffusional line broadening by detecting thermal spin polarization in a ~ (10 μm)³ volume. These new developments are not described in further detail in this protocol, since they need advanced technical expertise and equipment.The more basic NV-NMR methods described here are useful for detecting NMR signals from nanoscale surface layers. Practical applications are structural analysis of quadrupolar nuclei in 2D materials[11], NMR microscopy[20] or molecular dynamics at interfaces[30]. Compared to conventional liquid state NMR with an inductive detection scheme, the NV-diamond sensor has two main advantages: i) the detection of very small sample volumes down to a single molecule or to nanometer surface layers (described in this protocol), and ii) the potential to perform NMR microscopy and nanoscopy, enabled by the optical nature of NV readout. On the other hand, for some applications the reduction of the sample volume associated with NV-NMR relative to conventional NMR can pose a disadvantage, e.g. due to the diffusive broadening mentioned above.

**Electronic spin sensing.** Electronic spins resonate at much higher frequencies (~600x) than nuclear spins at the same magnetic field. Such high frequencies (~1 GHz for a few hundred gauss magnetic field) cannot be detected with NVs using dynamic decoupling sequences due to the finite duration of the MW pulses (few tens of nanoseconds). Here, an alternative detection protocol, NV-relaxometry, is used[13,35,36]. Magnetic noise generated by electronic spins is caused by spin flips, with a time scale set by their T1 relaxation time. The noise can be described by its spectral density $S(f) \sim \delta/((f-f_L)^2+\delta^2)$, where $f_L$ is the electronic spin Larmor frequency. The spectral density is broadened by the spectral width $\delta$, which is the inverse of the T1 relaxation time of the electronic spin (Figure 2d). $\delta$ is typically on the order of a few hundreds of MHz for $Cu^{2+}$ (Simpson et al.[37]) to 10 GHz for $Gd^{3+}$ (Sushkov et al.[36]). Thus, $S(f)$ usually features significant noise components around the NV transition frequency (~ GHz), which lead to a reduction of the intrinsic NV-T1 relaxation rate $\Gamma_{total} = \Gamma_{intr} + \Gamma_{induced}$. In a typical experiment, the NV-T1 relaxation time is measured by optically initializing the NV into the $m_s$=0 ground state and allowing its quantum state to thermalize for a time $t$ according to $\Gamma_{total}$. The NV spin state is read out via an optical pulse as a function of $t$ and the decay of the polarization can be fitted to an exponential function. Possible applications for nanoscale electronic spin sensing include the detection of biological important ions[35,37] or metalloproteins[38] in cells.

**Limitations of electronic spin sensing with NVs.** NVs are initialized and read out with green laser pulses. Although the light intensity reaching the sample is reduced by the total internal reflection geometry used in this protocol, the evanescent wave at the sample’s surface may be sufficient to excite the sample. This may cause problems when samples that absorb green light are measured (e.g., various colored transition metal complexes), possibly inducing unwanted photochemistry and sample degradation[39].

## Overview of the Procedure

First, we discuss the experimental design and the required hardware. Second, we describe the technical details of the specific pulse sequences used for sensing. Finally, we describe the experimental procedure itself, which in turn is organized into four parts (Figure 3): (1) the fabrication of NV-diamond chips to be used as sensors (Steps 1-17), (2) the construction of the quantum diamond spectrometer (Steps 18-50), (3) NV diamond characterization experiments (steps 51-59) and (4) quantum sensing procedure for sensing nuclear and electronic spins. The last two parts describe the procedure to run all basic pulse sequences and the procedure to detect NMR and EPR signals on the nanoscale.

## Experimental design

In the following we discuss the technical design choices made and equipment used in the development and implementation of this protocol. Figures 4 and 5 give overviews of the optics and electronics used in this protocol, respectively.

**Choice of diamond substrate and nitrogen implantation.** Ultra pure diamonds with low concentration of defects (nitrogen concentration $<10$ ppb) are needed to maximize probability of creation of NV centers with long coherence times and optimized sensing properties. Traditionally “electronic grade single crystal” diamonds from Element Six (https://e6cvd.com/us/application/all.html) have been used as substrates. Another potential source of diamond substrates are “ultra pure diamond plates” from LakeDiamond (https://lakediamond.ch/products.) Depending on the user’s application, a $^{12}$C-enriched substrate might be necessary to decrease the NMR signal from natural-abundance $^{13}$C. The choice of diamond size and shape depends on the requirements of the experiment and may be limited by the availability of large substrates. Our diamond is cut such that the top face is perpendicular to the <100> crystal axis, and the lateral faces are perpendicular to <110>. Ideally, the edges of the diamond are polished so that the NV layer can be excited through the edge in a total internal reflection geometry. Nanostructured diamond surfaces have been shown to increase the magnetic resonance sensitivities[31]. Shallow NV centers are needed for surface sensing and are typically created through bombardment of the diamond sample with low-energy nitrogen ions. To produce shallow NVs at depths of a few nanometers, the implant energy is typically between 2.5 and 6 keV[21,40]. Such shallow NVs exhibit degraded spin properties[41–43], although NMR signals from samples on the diamond surface are larger due to proximity of the NVs to the sample. Monte Carlo simulations (Stopping and Range of Ions in Matter: SRIM)[44] can be performed to calculate the approximate depth of the created defect as a function of nitrogen ion implantation energy and angle of incidence. An implant angle of ~5-7 degrees from the normal is usually chosen to minimize channeling (i.e., to keep the implanted ions shallow). If one implants directly normal to the surface, then the implanted ions can channel better through the crystal lattice and go further than one would naively expect from calculations[45]. An estimate of the NV depth can be obtained from NV-NMR experiments as described for single NV centers in Pham et al.[21] .
For AC magnetometry, nitrogen ion implantation is usually done with $^{14}$N ($I$ = 1), which has hyperfine structure that does not greatly interfere with NV sensing. In contrast, $^{15}$N ($I$ = 1/2) tends to give background that is sensitive to misalignment of the bias field $B_0$.

Following implantation, the diamonds are annealed under vacuum ($<10^{-6}$ mbar typically, $<2\times10^{-8}$ mbar in this instance) and high temperature (800-1200 °C) to allow migration of vacancies and formation of NV centers. High-temperature processing also allows one to minimize interactions with neighboring spins by annealing out other spin impurities. There are limitations to how much annealing can improve the spin bath properties and formation of NVs; this is still an active area of research, especially for high density shallow implantation diamonds[40,46,47]. We anneal diamonds in a home-built furnace. However, due to the complexity of setting up an annealing system, it is suggested to either send out the diamond sample to other research groups with a working furnace or use a commercial option (e.g., http://www.laserage.com/heat-treating). Following annealing of the sample one can observe and measure several different characteristics of the sample to characterize success or failure of the anneal. Upon annealing a grey tint to the color of the diamond indicates the presence of surface graphitization. Acid cleaning is needed to remove this graphitized layer. For some applications AFM may be needed to check the roughness of the diamond if surface damage is important to the specific application. One can characterize the fluorescence intensity (counts) and the coherence time (NV-T2) of the NVs before and after the anneal to characterize the creation of NVs or the presence of unwanted defects.

The efficiency of NV creation depends on both the ion energy and the ion fluence[48], which should be chosen such that for high-density, shallow layers the nitrogen atom areal density is in the range $10^{12}$-$10^{14}$ cm$^{-2}$, ultimately producing an NV areal density in the range of $10^{11}$-$10^{12}$ cm$^{-2}$. Given that the expected NV density is below the measurement threshold for traditional bulk measurements like UV-Vis, FTIR, or EPR, one needs to perform confocal measurements to characterize the counts. For ultra-low density samples where single NVs are spatially resolvable, one can perform a spatial survey of NVs and count individual centers to calculate the density. For the types of samples used in this paper where single NVs are not resolvable one needs to use the average count rate in a confocal volume to determine the density. Using a rough reference value of a typical single NV count of ~50-100 kcounts/s (around optical saturation), one can obtain an approximate value for the number of NVs in a volume by normalizing to the single NV reference counts.
Moreover, work to understand the surface and mitigate surface noise is an active area of research and there exist other protocols to improve coherence times and fidelity of near-surface NV centers[8,40,41,43,49–52]. Presented in this manuscript is one method of fabrication and preparation which is relatively accessible and is robust in its results. More sensitive applications than the ones demonstrated in this work will require better control of the surface and the implementation of various other methods cited above.

The diamond chip described in this work is an E6 CVD 99.6% $^{12}C$ layer (50-100 µm) on a natural-abundance substrate. Nitrogen ($^{14}N$) implantation was carried out by sending the diamond to a commercial e-beam facility (INNOViON), where it was implanted with a nitrogen ion beam of energy of 6 keV and fluence of 2 x $10^{13}$ $cm^{-2}$. We estimate the final NV density for this diamond to be ≈ 3 x $10^{11}$ $cm^{-2}$.
Surface contamination of the diamond substrate can be removed through the use of a 1:1:1 refluxing mixture of sulfuric, nitric, and perchloric acid. This solution should be used to clean the diamond before implantation to ensure surface contaminants are not present. Cleaning should be repeated before and after annealing to remove any graphitization buildup on the diamond surface. This cleaning procedure is also applied to remove and mitigate any undesirable effects observed during sensing protocols, which can occur due to surface contamination[53,54].

**Choice of magnets.** Permanent magnets are generally preferred to electromagnets for reasons of simplicity and cost, and due to stringent requirements on the stability of the current sources used to power the latter. In principle, any commercial permanent magnets can be used, as long as they generate a field $B_0$ of at least a few 100 G at a distance of a few centimeters from their surface. The field strength $B_0$ is important for nuclear spin sensing because it defines the Larmor frequency $f_L$ according to

$$f_L = \gamma_n B_0, \qquad (5)$$

where $\gamma_n$ is the gyromagnetic ratio of the spin of interest and B0 is the field strength at the sample position. Gyromagnetic ratios for different nuclei are provided e. g. by the Committee on Data of the International Council for Science (www.codata.org). For efficient NV detection with a dynamic decoupling sequence, the Larmor frequencies of the target nuclear spins should be between a few hundred kHz and a few MHz. Magnets can be stacked to increase magnetic field strength.
In all cases, magnetic field gradients should be kept as small as possible in order to suppress inhomogeneous broadening. The use of two opposed identical magnets, each with diameter at least on the order of a few centimeters (i.e., much larger than the laser spot size of ~ 20 µm) helps to minimize field gradients. For a detailed analysis of the magnetic field distribution of a given permanent magnet geometry, the software package Radia (O. Chubar, P. Elleaume, J. C. Radia software package (2017). URL http://www.esrf.eu/Accelerators/ Groups/InsertionDevices/Software/Radia) in Mathematica can be used. A common issue with static magnets is their temperature-dependent magnetization. To mitigate this problem, we employ samarium cobalt magnets (Electronic Energy Cooperation, 2:17 TC-15), which have a low temperature dependence (0.001 %/°C).

**Choice of laser source and acousto-optic modulator.** To excite the NV ensemble, a 532-nm laser with ~1-W output power is employed. We recommend using a high-quality optically pumped semiconductor laser (OPSL) or diode-pumped solid-state (DPSS) laser (for example Coherent Verdi G series or Lighthouse Photonics Sprout series of lasers). However, lower-priced 1-W laser diodes can also be used, at the expense of inferior noise properties.
For the sensing protocol described here, laser pulses on the microsecond time scale are necessary for initializing and reading out the NVs. We recommend using an acousto-optic modulator (AOM) with a drive frequency of 80 MHz or higher (e.g., IntraAction Corp., Model ATM-801A2) to achieve high extinction ratios. The AOM can be driven either by a commercially-available AOM driver (which will include a signal source and amplifier) or by a radio frequency (RF) signal source operating at the specified AOM drive frequency and amplified to reach the required RF power level. We use a commercial AOM driver (IntraAction Corp., Model ME-802N), which is usually modulated by a voltage input. However, we achieved better performance (i.e. a larger extinction ratio) by inserting a switch between its internal signal source and amplifier.

**Choice of excitation and collection geometry.** The quantum diamond spectrometer is optimized for electronic and nuclear spin sensing on the diamond surface. The use of a total internal reflection geometry minimizes back reflection of the laser into the detector and reduces unwanted exposure of the sample to excitation light. Note that energy can still flow from the laser beam into the sample via the evanescent wave produced at the diamond-sample interface. Total internal reflection can be achieved either by sending the laser through an edge of the diamond, or through the light guide into the bottom of the diamond chip (as depicted in Figure 4 and employed here). More glancing angles of incidence can be accessed with through-edge illumination, but this requires polishing of the diamond edge. Through-light-guide excitation has the additional disadvantage that high laser power might degrade the glue used to attach the diamond to the light guide over time. The presence of the glue may also increase background fluorescence.
NV fluorescence is collected with an optical light guide (e.g., Edmund optics) glued with optical epoxy to the bottom of the diamond, which in turn guides the collected light onto a large-area avalanche photodiode (APD) for detection. The light guide diameter should fit the size of the diamond. Moreover, the lightguide scheme makes the experiment fairly insensitive to any optical misalignments. Alternatively, a microscope objective can be used instead of the light guide. Compared to the light-guide geometry described here, the use of an objective offers enhanced contrast and spatial resolution at the expense of increased experimental complexity.

**Choice of photodetector and interference filter.** The photodetector is chosen according to the expected collected photon count rate in a given experiment. Usually, for shallow high-density implanted NV-diamonds, an avalanche photodiode (APD) provides the necessary sensitivity and should have a noise equivalent power of < 0.1 pW/$\sqrt{Hz}$. The bandwidth is usually limited by the data acquisition unit and is typically > 1 MHz. For efficient light collection, the distance between the light guide and the APD surface should be kept as small as possible and the active area of the APD should be larger than the diameter of the light guide. Possible choices of large-area APDs include Luna Optoelectronics SD197-70-72-661 (5 mm active detector diameter, our choice), SD394-70-72-661 (10 mm active detector diameter)], or Laser Components (3 mm active detector diameter A-CUBE-S3000-03). For efficient detection of the red fluorescent light and rejection of the green excitation light, a long-pass interference filter (e.g., Semrock, BLP01-647R) or appropriate band-pass interference filter (e.g., Semrock FF01-736/128) should be used. An additional 532-nm notch filter can also be used to further attenuate stray laser light, if desired.

**Choice of source to generate pulse sequences.** The pulse sequences are generated either by a PulseBlaster card or an arbitrary waveform generator (AWG). For the applications described in this protocol, a PulseBlaster card with high temporal resolution is sufficient (e.g., Spincore PulseBlaster ESR PRO 500 MHz). The card should be compatible with the computer used to control the experiment. The PulseBlaster card generates TTL pulses, which are used both to control the timing of the data acquisition in the experiment, and to switch on and off the laser and MW sources. The latter is accomplished through the use of TTL-controlled RF switches placed in the MW path and in the RF feed path to the AOM. The switches must be rated to handle signals within the relevant frequency ranges (i.e., ~ 80 MHz for the AOM RF feed and ~1-3 GHz for the MW drive) and must operate with rise times of at most a few nanoseconds (e.g., the Minicircuits ZASWA-2-50DR+ switches used in this protocol operate at frequencies ranging from DC to 5 GHz and offer rise times which are typically 5 ns and at most 15 ns).

**Choice of data acquisition unit.** The data acquisition (DAQ) unit is used to read out the APD voltage. Readout is triggered by TTL signals generated by the PulseBlaster card and sent to the DAQ. The DAQ should have a bandwidth which corresponds roughly to the NV repolarization time (in our case ~ 1 μs). For that reason, we use a DAQ with a 700 kHz bandwidth (see specification sheet of our model), a slower bandwidth will lead to a reduced contrast. Our DAQ has a sampling rate of 250 kSa/s, which sets the maximum repetition time of the experiment (i.e. to 4 μs). The quantum diamond spectrometer described here requires a DAQ with at least one analog input channel and two digital input trigger channels. We use a National Instruments USB-6229 DAQ.

**Choice of microwave source, amplifier, and delivery.** Although many options are available, we use a Stanford Research Systems signal source (SG 384) with an internal IQ-mixer for phase control. Any other low-noise and stable signal sources in the frequency range 1-4 GHz can also be used in conjunction with external IQ-mixers. The MWs are amplified by a 16-W Minicircuits amplifier (ZHL-16W-43+). For MW delivery we use a loop described in greater detail below. More sophisticated MW delivery antennas like coplanar waveguides or resonators can also be used[24].

## Pulse sequence basics

The NV-based quantum sensing schemes applied in this protocol have previously been described in the literature[13,23,24]. Here, we outline the technical requirements for the implementation of the pulse sequences used in this protocol. In all pulse sequences, the laser and MWs are pulsed on and off on the nanosecond to microsecond time scale. The timing of this pulsing is controlled by a PulseBlaster card with a 500-MHz clock, which is the heart of the experiment (see Figure 5). The PulseBlaster card outputs TTL signals to the switches, which control the MW drive (orange) and the laser (blue), via an AOM. Data acquisition with the DAQ is also triggered and gated by the PulseBlaster. The DAQ requires one TTL start trigger (red) that defines the start of the pulse sequence, and a gate trigger (green) that instructs it to acquire data each time a data point is to be collected. For the nuclear spin sensing pulse sequence, the phase of the MW pulses must also be controlled. This is done here via the IQ option of an SRS SG384 signal generator. The IQ input is controlled by two switches, which are also controlled by TTL signals (brown and violet) generated by the PulseBlaster. An overview of the pulse sequences used in this protocol is shown in Figure 6.

**NV-electron spin resonance (NV-ESR) pulse sequence.** The most basic sequence is the NV-electron spin resonance (NV-ESR) pulse sequence. We distinguish the term NV-ESR from ESR in order to specify obtaining a spectrum of the NV spin itself rather than that of a target sample using the NV as a sensor. NV-ESR is used to detect the NV resonance frequency and to measure the strength of the applied magnetic field by sweeping the MW frequency $f_M$ and reading out the fluorescence. If the applied MW is on resonance, some population is transferred from $m_s = 0$ to one of the dimmer $m_s = 1$ or $m_s = -1$ states, which causes a dip in the fluorescence signal. In the absence of a strong applied field usually two resonances can be observed around the zero-field splitting D of 2.87 GHz (Figure 6a). Upon application of an external field, the lines shift due to the Zeeman effect. If the magnetic field is not aligned along one of the diamond lattice <111> directions corresponding to a particular NV-axis, up to eight resonances can be observed in the spectrum of an NV ensemble, which might be further split by hyperfine interactions with a nearby nuclear spin (e.g., nitrogen). This is because each of four possible NV orientations experiences a different projection of the applied field and thus a different Zeeman shift. If the field is well-aligned along one of the four NV orientations (as it is in this protocol), the two resonances associated with that orientation are maximally shifted from 2.87 GHz. In this case, the applied field has equal projection on the other three NV orientations, causing their resonances to become degenerate. Thus, only four independent resonance frequencies are observed in the spectrum. The NV-ESR sequence is used to determine the NV resonance frequency and the applied magnetic field, which can be calculated approximately according to the following equation:

$$B_0 = \frac{2870\,MHz - f_{NV}(MHz)}{2.8\,MHz/G} \qquad (6)$$

where $f_{NV}$ is the center of the resonance frequency of the transition at lowest frequency. Keep in mind that $D$ (2.87 GHz) is not an exact constant and might vary depending on temperature and strain in the diamond. This resonance corresponds to the $m_s = 0$ to $m_s = -1$ transition of the NV orientation with the largest magnetic field projection. Usually, the laser light polarization is adjusted by a λ/2 waveplate to maximize the excitation of this NV-orientation. The NV-ESR pulse sequence is shown in Figure 6a. It requires four PulseBlaster (PB) channels: PB_AOM and PB_MW to control the AOM and microwaves, and two channels which act as start (PB-StartTrig) and readout triggers (PB_DAQ) for the DAQ. All the pulse sequences used here have two readouts per sequence for low-frequency (< ~50 kHz) noise cancellation. We refer to the first readout, during which the MW is on, as the “signal”. The second readout, during which the MW is off is referred to as the “reference”. The beginning of the sequence is marked by the DAQ start trigger, which prepares the DAQ for data acquisition. The laser is on during the entire pulse sequence. For each MW frequency $f$, the pulse sequence is repeated $N_{samples}$ times (number of samples), and so the DAQ acquires $2N_{samples}$ data points (signal and reference). The average of the signal data points is then divided by the average of the reference data points to give one value of contrast at $f_M$. The amplitude of the resonance signal depends on MW and laser power as well as the full duration of the pulse sequence and should be optimized by the user. Hyperfine interaction with the $^{14}$N nuclear spin splits each resonance into three lines. At high MW power, the lines are broadened and these hyperfine features are obscured.

**Rabi pulse sequence.** In a Rabi experiment, the MW frequency is tuned to match the NV spin resonance (e.g., to the $m_s = 0$ to $m_s = -1$ transition) and the NV fluorescence is measured as a function of the MW pulse duration. As the NV quantum state undergoes nutation, the expected number of detected fluorescence photons oscillates (Figure 6b). This Rabi oscillation is measured to determine the duration of $\pi/2$ and $\pi$ pulses, which are needed for the sensing sequences later. The Rabi frequency can be tuned by changing the MW power. Typical $\pi$–pulse durations achieved with the suggested amplifier and loop are ~20-60 ns. The Rabi contrast (fractional difference in signal measured at zero applied MW pulse duration and at $\pi$-pulse duration) can be as high as 30% for a single NV. For an ensemble of NVs, the contrast is reduced due to the presence of fluorescence background from off-axis NV centers as well as heterogeneity in NV properties, resulting in typical values in the range 1-6%. The contrast depends on the degree of NV repolarization (related to laser intensity) and on the presence of other sources of background light (light scattering, non-aligned NV centers etc.). Obtaining a high contrast is important since sensitivity improves linearly with contrast.

The Rabi pulse sequence requires four PB channels: PB_AOM, PB_MW, PB_DAQ, and PB_StartTrig. The AOM polarization and readout pulse are usually combined in one pulse. The first part (~ 1 μs) of the AOM pulse is used to read out the NV-state, whereas the subsequent 4 μs repolarize the NV. These durations depend on the laser intensity and must be optimized in each case. We use a laser intensity of ~ 10 kW/cm$^2$. The timing of the DAQ readout pulse relative to the AOM pulse must be chosen carefully. Early readout leads to a reduction of signal-to-background since the population transfer may be probed before completion of the MW pulse sequence. Late readout results in a signal-to-background reduction since most of the information about the NV population is lost once the laser has been on long enough to produce significant repolarization. Although the PulseBlaster outputs the AOM and DAQ trigger pulses at the same time, the arrival times of these pulses are usually delayed with respect to each other because of the non-negligible AOM response and/or differences in cable lengths. For that reason, the DAQ readout pulse output of the PulseBlaster is manually delayed in order to properly synchronize the AOM and DAQ readout pulses. As with the NV-ESR sequence, the second half of the pulse sequence is the reference readout (without MW manipulation) for noise cancellation.

**Spin-echo (Hahn-echo) NV-T2 relaxation pulse sequence.** The most basic AC magnetometry sequence is the spin-echo (Hahn-echo) sequence. It consists of a sequence ($\pi_x/2-\tau-\pi_y-\tau-\pi_{x(-x)}/2$) of MW pulses separated by free precession time $\tau$. As $\tau$ is swept, the signal decays according to the transverse relaxation of the NV with time constant NV-$T2_{Hahn\text{-}echo}$. $\tau$ should be much longer than the $\pi$ pulse duration (we typically use $\tau > 4\pi$ pulse durations). In addition to the four standard PB channels (PB_AOM, PB_MW, PB_DAQ, and PB_StartTrig), two additional channels are needed to control the phase of the MW pulses. Four possible on/off combinations of these two pulses determine the phase output of an IQ mixer (see Table 1). The I and Q pulses should be longer than the MW pulses in order to compensate for the finite response time of the IQ mixer.

| I | Q | Rotational Axis | Phase Shift |
|---|---|---|---|
| off | off | X | 0° |
| on | on | -X | 180° |
| on | off | Y | 90° |
| off | on | -Y | 270° |

**Table 1. IQ modulation**. Input channels, rotational axis and phase output.

For common noise cancellation the pulse sequence is applied twice. In the second sequence, the last $\pi/2$ pulse of the spin echo is phase-shifted by 180° relative to the last $\pi/2$ pulse of the first sequence. With these two readouts ($R_1$ and $R_2$), the contrast ($C$) can be calculated according to:

$$C = \frac{R_1 - R_2}{R_1 + R_2} \quad (7)$$

Keep in mind that this contrast definition is different from that in the ESR and Rabi experiments. We plot contrast versus free precession time for a spin-echo measurement in Figure 6c. In addition to the described decay, two shallow dips can be observed. These are caused by $^{13}C$ spins located within the diamond lattice, which precess at $f$ = 330 kHz at 311 G. The first dip occurs at $\tau = 1.5$ μs = 1/(2$f$), while the second dip occurs at a harmonic: $\tau = 4.5$ μs = 3/(2$f$).

For high-density, shallow NV ensembles, the NV-$T2_{Hahn\text{-}echo}$ time is typically around a few microseconds. NV-$T2_{Hahn\text{-}echo}$ is an important parameter, as it sets the lowest frequency that can be sensed. Usually the NV-$T2_{Hahn\text{-}echo}$ time is extended for higher sensitivity through the application of trains of π−pulses in dynamic decoupling sequences, e.g., an XY8-$N$ as described in the following paragraph.

**XY8-N dynamical decoupling pulse sequence.** The XY8-N sequence consists of trains of pulses of the following form: $\pi_x/2-\tau/2-(\pi_x-\tau-\pi_y-\tau-\pi_x-\tau-\pi_y-\tau-\pi_y-\tau-\pi_x-\tau-\pi_y-\tau-\pi_x)^N-\tau/2-\pi_{x(-x)}/2$. For $N$ = 1, the sequence consists of a train of 8 π pulses where the rotation axis is alternated between x and y (see Figure 2) in order to partially compensate for pulse errors[25,26]. Note that the $\pi/2$ and π pulses are separated by time $\tau/2$, and that the spacing between consecutive π pulses is $\tau$. Sweeping $\tau$ and monitoring the fluorescence reveals a decaying contrast. The decay timescale is extended compared to that obtained when a simple Hahn-echo pulse sequence is applied[55] (Figure 6d). To plot the data on a real time axis, remember to scale the scanned parameter axis $\tau$ by 8· $N$. The dip at $\tau$ = 1.5 μs observed in the data is caused by $^{13}C$ precession at the Larmor frequency $f_L$, which fulfills the condition 1/(2$f_L$) = $\tau$. The dip is more pronounced in the XY8-N than in the Hahn-echo experiment since more phase is accumulated and the sharper filter function narrows the line. Using more π pulses (higher $N$) intensifies these effects. However, $N$ is ultimately limited by the NV-T2 (i.e., there is a reduction in contrast as the pulse sequence duration increases). In addition, pulse errors accumulate over long dynamic decoupling sequence which reduce the contrast. For that reason, the optimum $N$ has to be found experimentally. The pulse sequence requires the same number of PulseBlaster channels (PB_AOM, PB_MW, PB_DAQ, PB_StartTrig, PB_I, and PB_Q) and implements the same noise-cancellation scheme (equation 7) as the Hahn-echo sequence.

**Correlation spectroscopy pulse sequence.** The correlation spectroscopy pulse sequence consists of two XY8-N sequences separated by $t_{corr}$: $\pi_x/2-\tau/2-(\pi_x-\tau-\pi_y-\tau-\pi_x-\tau-\pi_y-\tau-\pi_y-\tau-\pi_x-\tau-\pi_y-\tau-\pi_x)^N-\tau/2-\pi_y/2$-$t_{corr}$- $\pi_x/2-\tau/2-(\pi_x-\tau-\pi_y-\tau-\pi_x-\tau-\pi_y-\tau-\pi_y-\tau-\pi_x-\tau-\pi_y-\tau-\pi_x)^N-\tau/2-\pi_{y(-y)}/2$. Note that the phase of the last $\pi/2$ pulse of each XY8-N sequence is shifted 90°/270° relative to the first $\pi/2$. The time $t_{corr}$ between these two sequences is swept and the fluorescence recorded. The π-pulse spacing $\tau$ in the XY8-N sequences is set to satisfy the condition $\tau$ = 1/(2$f_0$), where $f_0$ is the frequency of the signal we want to sense (and $\tau$ is the time at which the dip in the XY8-N decay occurs). The recorded data oscillates at the signal frequency $f_0$. In the example of Figure 6e, we sense the $^{13}C$ NMR signal in diamond by setting $\tau$ to 1.5 μs, which results in an oscillation at 330 kHz ($f_L$). In order to resolve the Larmor frequency, the sampling of $t_{corr}$ should be high enough to provide at least two points per signal-field period (1/$f_0$), and more may be preferable for straightforward analysis of the data. Of course, the full oscillation can be undersampled in order to speed up acquisition if the experimenter has sufficient prior information about the target frequencies to resolve potential ambiguities. As in the previously-described experiments, low-frequency noise is cancelled by imposing a 180° phase shift between the last $\pi/2$ pulses of two successive correlation spectroscopy sequences (see equation 7). As with the XY8-N sequence, $N$ should be optimized to find the highest signal-to-noise ratio (SNR).

**NV-T1 relaxation pulse sequence.** The NV-T1 relaxation pulse sequence is very simple and, in principle, needs no MW pulses. To measure NV-T1 (longitudinal) relaxation, the time $t$ between AOM laser pulses is swept. As in all previous pulse sequences, the readout AOM pulse and repolarization AOM pulse are combined. By sweeping the total sequence duration $t$, the fluorescence decays exponentially according to the longitudinal relaxation of the NVs from the polarized $m_S$ = 0 state into the thermal equilibrium mixed state (see Figure 6f). For low-frequency noise cancellation, the pulse sequence is repeated, but the relaxation from $m_S$ = 1 to the thermal state is measured by applying a π pulse on the NV after optical polarization. The contrast is calculated according to equation 7. The pulse sequence requires four PB channels: PB_AOM, PB_MW, PB_DAQ, and PB_StartTrig. Typical NV-T1 relaxation times for NVs at room temperature are a few milliseconds.

## Materials:

**Critical:** We list the equipment used for our experimental setup. Unless otherwise specified, items can be replaced by equivalent components from other vendors with similar performance. **!Caution:** Ensure that all chemicals, substances, equipment and apparatus in this protocol are handled and operated safely by obtaining, reading, and following their respective manufacturers' safety instructions.

**REAGENTS:**

**Samples**

- Polydimethylsiloxane (PDMS) (Sigma Aldrich SYLGARD® 184)
- Fomblin oil for $^{19}$F sensing (Sigma Aldrich 317993-100G)
- Copper (II) sulfate pentahydrate (Sigma Aldrich 31293)

**Acid clean reagents**

- Deionized water (e.g., Sigma Aldrich)
- Isopropyl alcohol (e.g., Sigma Aldrich)
- Sulfuric acid (Merck SX1244-6). !Caution: sulfuric acid is strongly corrosive. Protective equipment should be worn (laboratory coat, gloves, safety goggles), and all handling should be performed in a ventilated laboratory fume hood).
- Nitric acid (EMD Millipore NX0409-4). !Caution: nitric acid is strongly corrosive and oxidizing. Protective equipment should be worn (laboratory coat, gloves, safety goggles), and all handling should be performed in a ventilated laboratory fume hood).
- Perchloric acid (VWR BDH4550-500ML). !Caution: perchloric acid is strongly corrosive and oxidizing. Protective equipment should be worn (laboratory coat, gloves, safety goggles), and all handling should be performed in a ventilated laboratory fume hood).

**EQUIPMENT**

**General equipment**

- Oscilloscope (e.g., Keysight DSO7104A)
- Signal source for coil signals (e.g., Rigol Technologies DS1022)
- Voltmeter (e.g., Fluke 77 IV)
- Optical power meter for measuring laser intensity (e.g., Thorlabs PM100D and S121C)
- Laser goggles (e.g., Thorlabs, LG3)
- RF power meter (e.g., Keysight V 3500).
- Soldering iron and solder
- Epoxy glue (e.g., Epoxy technology, Optical epoxy 301-2/8Z KIT and instant adhesive (crazy glue).
- Heat conduction paste !Caution: Consult the safety data sheet for handling.

**Cables**

- BNC cables (e.g., Digikey)
- SMA cables (e.g., Digikey)
- BNC/SMA adaptors (i.e., a selection of BNC-SMA, SMA-SMA and BNC-BNC adapters for all gender combinations, e.g., Digikey)

**Acid clean equipment**

- 1x 25 mL round bottom flask (e.g., Sigma Aldrich)
- Reflux condenser that fits with round bottom flasks, stand, and clamps (e.g., Sigma Aldrich)
- Gas bubbler that fits to the reflux condenser (e.g., Sigma Aldrich)
- Heating mantle for the round bottom flasks (e.g., Sigma Aldrich)
- 3x beakers (e.g., Sigma Aldrich)

- Cleanroom cups (e.g., Sigma Aldrich)
- Ceramic tweezers

**Recommended Protective Equipment**

- Nitrile gloves
- Lab coat
- Acid gloves
- Laboratory coat
- Face shield

**Diamond Substrate**

- Electronic grade diamond substrate (e.g. Element Six)

**Diamond Annealing Equipment**

- Oven (e.g., Applied Test Systems Series 3210)
- Thermocouple type K (e.g., Applied Test Systems Series)
- Quartz tube (e.g., MTI Corporation, 1.5 inch OD, 36 inch length)
- Conflat (CF)-to-quick connect coupling (e.g., Lesker Part No F0275XVC150)
- Gate valve (e.g., Huntington GVA-150-C)
- Turbo pump (e.g.,Pfeiffer TMU071 P)
- Roughing pump (e.g.,Pfeiffer MVP-015 T)
- Vacuum gauge (e.g.,Pfeiffer Vacuum, D - 35614 Asslar)
- ConFlat (CF) tees, elbows, crosses (e.g.,Kurt Lesker)
- Quartz boat (e.g.,MTI Corporation)
- Copper gasket (e.g.,Kurt Lesker, OFHC copper gaskets for CF flanges flange OD: 2-3/4")

**Laser**

- 532-nm continuous-wave laser with ~ 1 W or higher output (e.g.,Coherent Verdi G series). !Caution: exposure of the eyes or skin to the laser can be harmful. Use appropriate laser goggles and follow the general laser safety guidelines.

**Optics and Optomechanics for the Quantum Diamond Spectrometer**

- 1x optical table (e.g.,Thorlabs)
- Opto-mechanics for mounting freestanding optics
- 1x 650-nm long-pass interference filter (e.g.,Semrock BLP01-647R) or suitable bandpass filter (Semrock FF01-736/128), optional 532 notch filter.
- 1x XY-translation stage with a rotating platform (Thorlabs XYR1)
- 1x 25 mm translation stage (Thorlabs PT1)
- 1x manual rotation stage, Ø2" (Thorlabs RP01)
- 2x 1/4" travel single-axis translation stage (Thorlabs MS1S)
- 4x 1.5" pedestal pillar post (Thorlabs RS1.5P)
- 2x 3" pedestal pillar holders (Thorlabs RS3P)
- 1x 1" pedestal post holders (Thorlabs PH082E)
- 2x 0.75" aluminum post (Thorlabs TRA075)
- 4x 1" aluminum post (Thorlabs TRA1)
- 2x 2" mini series mounting posts (Thorlabs MS2R)
- 2x 1" mini series mounting posts (Thorlabs MS05R)
- 1x Base plate (Thorlabs PT101)
- 1x Right angle bracket (Thorlabs PT102)
- 2x dovetail rail carriers (Thorlabs RC1)
- 1x dovetail optical rail (Thorlabs RLA1200)
- 2x compact kinematic mirror mounts (Thorlabs KMS)
- 4x fixed 90° brackets (Thorlabs ER90B)

- 1x cage adaptor plate (Thorlabs SP05)
- 1x removable cage plate (Thorlabs CP90F)
- 1x mounting base (Thorlabs BA2)
- 1x swivel post clamp (Thorlabs MSWC)
- 1x angle clamps (Thorlabs RA90)
- 4x angle clamps (Thorlabs RA180)
- 1x Ø1/2" stackable lens tubes (Thorlabs SM05L03)
- 1x adapter with external SM1 threads and internal SM05 threads (Thorlabs SM1A6T)
- 1x SM1-threated adapters with smooth internal bore, Ø16 mm (Thorlabs AD16F)
- 1x mount for a 4 mm x 25 mm light pipe (Edmund Optics 64-907)
- 1x 4 mm hexagonal light pipe 50 mm (Edmund Optics 49-402)
- 1 x dovetail translation stage with baseplate (Thorlabs DT12XYZ)
- 1 x 50-mm focal length lens (Thorlabs LA1289-A)

**Photodetector**

- Large-area avalanche photodiode (APD) (e.g.,Luna Optoelectronics, SD197-70-72-661); it requires an additional power supply with +12,-12 V, and GND outputs in addition to +5 V and GND for the onboard cooling element. For details consult the APD manual. We employ an additional heat sink and small fan to further facilitate heat dissipation.

**Microwave Parts**

- 1x SRS SG384 signal generator (this is a MW signal generator with IQ option). Note: the qdSpectro software package (Aude Craik, 2019) used in this protocol is designed to work with an SRS SG384 signal generator and has only been tested with this and the SG 386 models. It should be compatible with other SRS SG3800 or SG3900 models, but has not been tested with these. Other pulse generators can be used but will require modification of the qdSpectro code by the user.
- 1x USB/GPIB converter (National Instruments GPIB-USB-HS)
- 1x high-power amplifier (Minicircuits ZHL-16W- 43+)
- Circulator (DiTom D3C2040)
- MW loop: Amphenol 901-9867-RFX and semi rigid coaxial cable (Micro-Coax UT-047C-TP)
- Loop mounting: opto-mechanics for mounting freestanding optics, table clamp (Thorlabs CL5) and right-angle clamp (Thorlabs RA90)
- BNC 50Ω attenuators (e.g.,Pasternack)
- Type N male to SMA female adapter (to connect SMA cable to RF output of SRS SG384 signal generator)

**Pulsing Source**

- 1x 500 MHz PulseBlaster card (Spincore PulseBlasterESR PRO 500 MHz). Note: the qdSpectro software package used in this protocol has only been tested with Spincore PulseBlasterESR PRO 500 MHz card, but should be compatible with other SpinCore PulseBlaster cards. Other signal generators can be used but will require modification of the qdSpectro code by the user.
- 4x MW switches (Minicircuits ZAWSA-2-50DR+)
- 4x SMA 50Ω loads (Amphenol 132360)

**Acousto-Optic Modulator and Optics for the Laser Path**

- 80-MHz AOM (IntraAction Corp., Model ATM-801A2) with 80-MHz AOM driver (IntraAction Corp., Model ME-802N)
- Opto-mechanics for mounting freestanding optics
- 1x 400-mm focal length lens (Thorlabs LA1172-A)
- 1x 50-mm focal length lens (Thorlabs LA1131-A)
- 1x 100-mm focal length lens (Thorlabs LA1509-A)
- 1x 200-mm focal length lens (Thorlabs LA1708-A)

- 200-µm pinhole (Thorlabs P200H) and translating lens mount (Thorlabs LM1XY). Critical: Choice of exact pinhole size depends on required extinction ratio
- 1x five-axis aligner (Newport 9081-M)
- 1x 1" pedestal post holders (Thorlabs PH082E)
- 1x 1" optical post (Thorlabs TR1V)
- 1x iris (Thorlabs IDA12-P5)
- 1x λ/2 waveplate (Thorlabs WPH05M-532) in a continuous rotation mount (Thorlabs RSP1)
- 5 x Ø1" protected silver mirror (Thorlabs PF10-03-P01) mounted in precision kinematic mirror mounts. Note: Number of mirrors depends on the experiment and space restrictions.

**Magnets**

- 2x static magnets with a diameter of > 1cm. We stack two Electronic Energy Cooperation, 2:17 TC-15 magnets in order to increase magnetic field strength and homogeneity.

**Power Supplies**

- +/-5 V power supply for powering MW switches
- 28 V / 4.3 A power supply for powering MW amplifier (Minicircuits ZHL 16W-43+)
- +/-12 V power supply for powering APD (Optoelectronics SD197-70-72-661)
- 5V power supply for powering APD cooling element (you can also use the same power supply as for the switches)
- +/-0.35 V for IQ modulation of the SRS SG384

**Data Acquisition Unit**

- Data acquisition card (DAQ), with a sampling rate of at least 250 kSa/s (e.g., National Instruments NI USB-6229 or NI USB-6211). Note: the qdSpectro software package used in this protocol was tested with NI USB-6229 and NI USB-6211. The code is designed to work with National Instruments DAQs and will require modification by the user if other data acquisition systems are used.

**Computer**

- Personal computer (PC) – the software installation is described in the protocol for a Windows PC and the qdSpectro package has been tested with Windows 10. It should be portable to Linux or Mac operating systems but has not been tested in these platforms and may require some user modification.

**Software Packages**

Critical: Recommended installation procedures for the software used in this protocol are described in the Procedure in the section entitled 'Setup Experimental-Control Software, DAQ, and PulseBlaster Card' (Steps 18-32). Below, the necessary packages, libraries and drivers are listed for reference.

- qdSpectro[56] – Python package developed to run the experiments described in this protocol. It can be downloaded from https://gitlab.com/dplaudecraik/qdSpectro or http://doi.org/10.5281/zenodo.1478113. The current version at the time of writing is v1.0, but the user is encouraged to download the latest version. The package includes a readme file, which users should refer to for any patches or updates, as well as a list of software dependencies (including version numbers with which the package has been tested).
- Python 3 - version 3.6.3 or later and of a bitness which matches the computer's bitness (i.e., install 64-bit Python if running it on a 64-bit computer). https://www.python.org/
- Notepad++ - or any other text editor of your choice for viewing and editing Python scripts. https://notepad-plus-plus.org/

**Drivers**

- National Instruments NI-DAQmx driver compatible with your chosen data acquisition card and operating system

https://www.ni.com/dataacquisition/nidaqmx.htm

- National Instrument drivers for the USB/GPIB converter used for GPIB communication between the PC and the SRS signal generator. qdSpectro has been tested with the National Instruments GPIB-USB-HS converter, which requires the NI-VISA and NI-488.2 drivers to be installed.
- SpinAPI - SpinCore API and Driver Suite for the PulseBlaster card http://www.spincore.com/support/spinapi/SpinAPI_Main.shtml

**Libraries for peripheral instrument control**

- SpinAPI Python3 wrapper – SpinCore's Python wrapper for C functions in SpinAPI, which can be used to communicate with and control the PulseBlaster Card. http://www.spincore.com/support/SpinAPI_Python_Wrapper/Python_Wrapper_Main.shtml If the aforementioned link is no longer active, the required version of spinapi.py can still be retrieved at https://web.archive.org/web/20190208140542/http://www.spincore.com/support/SpinAPI_Python_Wrapper/spinapi.py
- NI-VISA library – this library will likely be included with the drivers for the NI GPIB/USB converter but, if not, can also be separately obtained from https://www.ni.com/visa. It is an implementation of the virtual instrument software architecture (VISA) application programming interface (API). The VISA API facilitates communication with peripheral instruments and must be installed to enable qdSpectro to communicate with the SRS signal generator via GPIB. The bitness of this library must match the Python bitness.
- PyVISA version 1.8 or later - a Python wrapper for the NI-VISA library, which allows the library to be called from Python scripts https://pypi.python.org/pypi/PyVISA

**Python libraries for data manipulation and graphical display**

- Matplotlib – a python library for plotting https://matplotlib.org/index.html
- NumPy – a python library for scientific computing http://www.numpy.org/

# Procedure

## Fabrication of NV-diamond chips (TIMING 2 days)

CRITICAL: Acid cleaning (Steps 1-11) is recommended before sending the diamond out for implantation, before and after annealing, and generally to remove residue from the surface or before changing to a new sample

!Caution: Ensure that proper protective equipment (acid gloves, face shield, and lab coat) is worn during the cleaning procedure. Institutional protocols should be followed regarding waste and chemical usage. The cleaning procedure must be performed in a fume hood. Chemical-resistant ceramic tweezers should be utilized to avoid damaging the diamond surface or chipping edges.

1. ***Acid cleaning (Steps 1-11):*** Set-up round bottom flask, reflux condenser, and bubbler on the heating mantle as shown in Figure 7. Connect cooling water to the reflux condenser. Fill the bubbler halfway with water and connect to a weak air flow. !Caution: This is important to prevent leakage of acid fumes, as perchloric acid fumes are explosive.
2. Transfer diamond to round bottom flask.
3. Pour 5 mL of sulfuric acid into a beaker. Add 5 mL perchloric acid into the beaker. Add 5 mL nitric acid. The order in which you pour acids is related to fuming. Nitric acid fumes the most and sulfuric the least. Pour tri-acid solution into the round flask with the diamonds.
4. Insert condenser into round bottom flask and turn heating mantle on so that the acids are boiling. Keep the fume hood closed. Keep acid solution boiling for one hour. After one hour turn the heater off and let the solution cool for 30 minutes.

5. Prepare a beaker with deionized (DI) water for diluting acid residue.
6. Lift condenser out from the flask. Pour majority of acid out of flask into proper waste container. Critical Step. Be careful not to pour diamonds out of flask during this process.
7. Begin the dilution process. Pour DI water into round bottom flask and swirl around to dilute acid residue. Pour waste water from flask into a waste container (again making sure not to pour diamonds out). Repeat this Step at least twice more.
8. Fill flask with DI water and pour all of the contents of the flask (including diamonds) into a cleanroom cup, or a similarly clean container. Repeat this Step until all diamonds are removed from the flask.
9. Using ceramic tweezers, transfer the diamonds from DI water to a cup of isopropyl alcohol solution.
10. Dry the diamonds with nitrogen gas blower and put them in a clean container for storage.
11. Properly rinse all glassware and dispose of all chemical waste in correct containers.

**PAUSE POINT: Diamonds can be stored in a clean container.**

CRITICAL: The annealing procedure (Steps 12-17) should be done after the diamond has been implanted with ions and acid cleaned. The procedure described here is for use with a home built vacuum furnace system (Figure 8). Due to the complexity of setting up an annealing system, either sending out the diamond sample to other research groups with a working furnace or using a commercial option (e.g., http://www.laserage.com/heat-treating) is suggested. Similar procedures and considerations are applicable for analogous systems. Steps 12-17 are for a starting condition where the furnace is not under vacuum and has been opened, and the turbo pump has been spun down.

12. ***Annealing (Steps 12-17):*** Remove the quartz boat from the quartz tube. Place the quartz boat in an enclosed space (to avoid losing the diamond if it is dropped when transferring). Transfer diamond samples (several samples can be annealed at once in this configuration) to the quartz boat using ceramic tweezers. Place the quartz boat back into the quartz tube and push the quartz boat down to the end of the quartz tube (into the area that will be under the furnace).
13. Use a ConFlat (CF) flange to seal the quartz tube to the rest of the vacuum assembly. Place a copper gasket in between the two metal seals making sure that it is in the proper place relative to the knife edge. Ensure the quality of the vacuum seal through proper tightening of the bolts and nuts for the seal.
14. Open the gate valve. The roughing pump begins pumping down the entire chamber. Consult the manual of the turbo pump being used to determine minimum pressure needed to spin the turbo pump up to speed. Once this pressure is reached, turn on turbo pump and wait for chamber pressure to decrease. The wait time depends on turbo pump used and the volume of the vacuum chamber. Once the pressure reaches an acceptable level ($<10^{-7}$ mbar) then one is ready to start the heating.
15. Program the heating profile (Figure 9) into the furnace controller. Ramp from room temperature to 400 °C over 6 hours. Soak at 400 °C for 6 hours. Ramp from 400 °C to 800 °C over 6 hours. Soak at 800 °C for 2 hours.
16. Let furnace cool to room temperature. Close the gate valve and spin down the turbo pump. Unseal quartz tube from the furnace. **PAUSE POINT: Diamonds can be stored in a clean container.**
17. Acid clean the diamond (steps 1-11)

## Construction of the Quantum Diamond Spectrometer (Timing 3 weeks)

Critical: The following steps (18-50) are only necessary for the initial construction of the setup. If you are using an existing setup, proceed to the section “NV Diamond Characterization Experiments Alignment and sensing (steps 51-59).

18. ***Experimental-Control Software Installation (Steps 18-25):*** Download and install Python 3 version 3.6.3 or later from https://www.python.org/. The Python bitness must match both the NI-VISA library’s bitness (see step 20) and the computer’s/operating system’s bitness. This protocol describes how to run Python scripts from a Windows Command Prompt and how to edit the scripts using Notepad++, a text editor. The user may opt to run and edit the scripts from an Integrated Development Environment (IDE) instead, or to use a different editor.

19. To check that the Python installation was successful, run Python by typing `python` into a Windows Command Prompt and pressing Enter. This should return the Python version number and bitness. To exit Python, type `exit()` (or hold down the Ctrl key and press the Z key) followed by Enter.
20. Download and install the required drivers for the NI USB/GPIB converter (NI GPIB-USB-HS converter). These drivers should include the NI-VISA library but, if not, the library can be separately downloaded and installed from the National Instruments website: e.g. http://www.ni.com/download/ni-visa-16.0/6184/en/ (for version 16.0). Critical Step***:*** Ensure that the bitness of the NI-VISA library matches the Python bitness (i.e., install 64-bit NI-VISA if running it on a 64-bit operating system).
    From a Windows Command Prompt, install pyVISA by running
    `python -m pip install -U pyvisa`
    Check that the library was successfully installed by starting Python (by typing `python` into the command prompt and pressing Enter, as in step 19) and then running
    `import visa`
    If no errors appear, the installation was successful.
21. From a Windows Command Prompt, install matplotlib by running
    `python -m pip install -U matplotlib`
    Check that the library was successfully installed by starting Python and then running
    `import matplotlib`
    If no errors appear, the installation was successful.
22. From a Windows Command Prompt, install numpy by running
    `python -m pip install -U numpy`
    Check that the library was successfully installed by starting Python and then running
    `import numpy`
    If no errors appear, the installation was successful.
23. Download and install Notepad++ from https://notepad-plus-plus.org/.
24. Choose a folder in which to install qdSpectro, the package containing the Python scripts needed to run the experiments described in this protocol. This folder is henceforth referred to as the working directory. Download qdSpectro from https://gitlab.com/dplaudecraik/qdSpectro and save it in the working directory (see Box 1 for a brief description of files included in the package). The current package version at the time of writing is version 1.0, but users are encouraged to download the latest version. Users should check the readme file of the package for any patches and version-specific changes to the instructions given in this paper.
25. Download the SpinAPI Python3 wrapper from
    http://www.spincore.com/support/SpinAPI_Python_Wrapper/Python_Wrapper_Main.shtml
    (if this link is no longer active, the required version of spinapi.py can still be retrieved here: https://web.archive.org/web/20190208140542/http://www.spincore.com/support/SpinAPI_Python_Wrapper/spinapi.py ) Critical Step: Save the file as spinapi.py in the working directory.

26. ***PulseBlaster and DAQ Setup (Steps 26-32):*** Follow the instructions in the "Installation" section of the PulseBlaster manual (e.g., page 9 of the PulseBlasterESR-PRO manual version from September 4th, 2017). This includes downloading the SpinAPI package, inserting the PulseBlaster card into an available Peripheral Component Interconnect (PCI) slot in the computer and testing the PulseBlaster using one of the test programs SpinCore provides.
27. Follow the installation instructions for the National Instruments DAQ (e.g., Chapter 1 of the NI USB-621x manual version from April 2009 [https://www.ni.com/pdf/manuals/371931f.pdf]. This includes downloading the NI-DAQmx driver and connecting the DAQ card to the computer via USB.
28. Refer to the analog input section of the DAQ manual (e.g., chapter 4 in the NI USB-621x manual version from April 2009) for a description of the available connection modes for analog input (AI) signals. The DAQ configuration used in this protocol and the APD signal connection instructions below assume that the APD input to the DAQ is a Referenced Single-Ended (RSE) connection; if the user prefers to use a differential connection, the configureDAQ() function in the DAQcontrol.py script of qdSpectro must be edited accordingly.

29. Choose an analog input (AI) terminal of the DAQ to which you will later connect the APD output voltage signal [note that the signal should be connected across the chosen AI terminal and the Analog Input Ground (AI GND) terminal of the DAQ]. Set up the terminal to be connected via a BNC cable to the APD: depending on the choice of DAQ, this may require soldering a BNC connector to a short section of twisted-pair wires which can be fed into the DAQ's screw terminals.
30. Connect two PulseBlaster channels to two Peripheral Function Interface (PFI) terminals of the DAQ, with ground terminals connected to the DAQ's Digital Ground (D GND). As in the previous step, depending on the choice of DAQ, this may require soldering a BNC connector to a short section of twisted-pair wires that can be fed into the DAQ's screw terminals. These two PulseBlaster channels serve as the sources for the Sample Clock and the Start Trigger signals used by the DAQ to perform hardware-timed acquisition of the APD voltage signal input data (refer to the section of the DAQ manual on Analog Input Timing signals, e.g., pages 4-11 to 4-22 in the NI USB-621x manual version from April 2009).
31. Open connectionConfig.py in Notepad++. Under "DAQ connections", edit the definitions of the variables `DAQ_APDInput`, `DAQ_SampleClk` and `DAQ_StartTrig` to match the names of the DAQ input terminals you have chosen for the APD voltage signal and the PulseBlaster-generated sample clock and start trigger respectively (e.g., if the PulseBlaster channel that will generate the start trigger is connected to DAQ terminal PFI5, the relevant definition in connectionConfig.py should read `DAQ_StartTrig="PFI5"`). When defining the name of the analog input channel connected to the APD, it is useful to run the NIDAQmx Measurement and Automation Explorer (MAX) program to verify what device number has been assigned to the DAQ, since the AI channel name will include this number (e.g., in version 4.0 of the NI MAX program, open the configuration drop-down menu and click to expand "My System", followed by "Devices and Interfaces" and, finally, "NI-DAQmx Devices" to see the device number for the DAQ. If, for instance, the APD is connected to terminal ai1 and the DAQ has device name "Dev2", the AI channel name that should be written into connectionConfig.py is "Dev2/ai1"). Set `DAQ_MaxSamplingRate` to the maximum sampling rate of your National Instruments DAQ in samples per channel per second. Finally, set `minVoltage` and `maxVoltage` (in Volts) to match an AI voltage range which is supported by your DAQ and which accommodates the range of voltages output by your APD (e.g. the 'Analog Input' section of chapter 4 of the NI USB-621x manual version from April 2009 includes a table listing the supported input voltage ranges for the NI DAQ USB-621x series).
32. Also in connectionConfig.py, set the variable `PBclk` equal to the clock frequency of the PulseBlaster card you are using, in MHz. Under "PulseBlaster connections", edit the variable definitions for `PB_DAQ` and `PB_STARTtrig` to match the PulseBlaster output bits that were chosen to output the sample clock and start trigger for the DAQ, respectively. For example, if you are using the SP18A ESR-PRO PulseBlaster board and chose bit 2 (corresponding to the BNC2 connector on the board, as shown in Figure 10 of the September/2017 version of the PulseBlasterESR-PRO manual) to output the start trigger pulses, the relevant definition in connectionConfig.py should read `PB_STARTtrig =2`. The user can check that the PulseBlaster channel definitions are correct by monitoring the channels with an oscilloscope and running `togglePBchan.py` (a Python script that is part of the qdSpecro package), as described in Box 2.

**!Caution** Before proceeding, ensure that the appropriate laser safety training at your institution has been completed. In addition, the laboratory and the laser system itself must comply with the relevant institutional laser safety guidelines. Generally, you should follow the basic laser safety instructions, including wearing laser goggles, refraining from wearing reflective items and avoiding bringing your head to the laser height level.

**!Caution** Before proceeding, ensure that appropriate MW and radio-frequency safety training at your institution has been completed. In particular, ensure that you are familiar with how to handle MW sources, amplifiers and antennas safely.

33. ***Mounting and Alignment of the Acousto-Optic Modulator (AOM) (Steps 33-36):*** Consult the AOM manual for the necessary radio frequency (RF) input power of the AOM. Depending on the choice of AOM driver, there may be a number of options to enable switching the drive to the AOM. In general, the drive consists of an RF oscillator at the frequency required by the AOM (80 MHz in our case), and an RF amplifier. We drive our AOM using a commercially available AOM driver (IntraAction ME-802N). Within the housing of the driver, there is an RF oscillator connected to an amplifier with an ordinary BNC cable. To achieve high extinction switching, we disconnected the oscillator from the amplifier, and inserted a MW switch (Minicircuits, ZASWA-2-50DR+) in between, with the oscillator output connected to the switch input and the switch output (RF 2) connected to the amplifier input. Be careful to avoid unwanted grounding or shorting through these connections by properly isolating the components and cables. Terminate the second switch output (RF1) with a 50-Ω load (see Box 3). Power the switch with +5V and -5V provided by a separate power supply. Choose a PulseBlaster channel to control the switch and connect the channel to the switch's TTL input. Update the definition of the variable `PB_AOM` in connectionConfig.py to match the output bit number of this PulseBlaster channel. Before connecting the amplifier output to the AOM input, be sure you do not exceed the maximum RF input power specified in the AOM manual. First measure the drive power by connecting to an RF power meter and adjust accordingly. Turn off the drive and connect to the AOM after power adjustment. CRITICAL STEP: Depending on the AOM type, the procedure in Steps 33-35 will vary. Consult the manual for proper mounting and alignment. For a general overview use Figures 4 (optics) and 5 (electronics).
34. Install the 532-nm laser on an optical table. One possible arrangement of the optics of the experiment is depicted in Figure 4. The beam waist must be no larger than the clear aperture of the AOM. To meet this condition, we demagnify the beam by a factor of 8 using a telescope consisting of a lens of focal length 400 mm (Thorlabs LA1172-A) and another of focal length 50 mm (Thorlabs LA1131-A), separated by the sum of their focal lengths. The telescope preserves collimation of the beam before passing through the AOM. We choose to modulate the collimated beam in order to decouple divergence from diffraction by the AOM. Place a pedestal post holder (Thorlabs PH082E) with a 1" optical post (Thorlabs TR1V) on a 5-axis aligner (Newport 9081-M). Mount the AOM on the 1" optical post. We recommend attaching a heat sink to the AOM for efficient heat dissipation. Place the 5-axis stage with the AOM in the laser path a few cm in front of the 50-mm focal length lens. Turn on the laser and attenuate to < 1 mW for alignment. Turn on the PulseBlaster channel controlling the AOM switch using the togglePBchan.py script as described in Box 2. Adjust the height and the position of the AOM such that the laser beam passes through the center of the device's aperture. Observe the laser spot on a piece of paper or cardboard placed after the AOM. Laser goggles may help visualization at this point as they block the intense laser light and pass the faint luminescence of the paper or cardboard. Turn on the AOM driver. When the AOM is turned on and is nearly aligned, both a zeroth and first diffraction order spot should be visible. Use an iris (Thorlabs IDA12-P5) to pass the first diffraction order and to block the zeroth diffraction order. Measure the optical power of the first diffraction order after the iris with a power meter. Maximize the power shunted into the first diffraction order by adjusting the degrees of freedom of the 5-axis aligner. Carefully adjusting the RF power of the AOM driver can also improve diffraction efficiency, though if the RF power is set too high several diffraction orders become visible and the AOM might be damaged. Once optimized, measure the ratio of RF power in the first and zeroth diffraction orders. Compare to the diffraction efficiency quoted in the AOM manual. <u>Critical Step</u>: check the extinction ratio, i.e., the ratio of laser power in the first order beam with the AOM switched on and off. A high extinction ratio is necessary to avoid unwanted repolarization of the NVs when the AOM is nominally off. This number should be 8000-10000. If this is not the case, the extinction ratio can be improved in the next step.
35. To improve the extinction ratio, we focus the first diffraction order with a lens of focal length 100 mm (Thorlabs LA1509-A) placed after the iris. At the focal plane of this lens we place a 200-μm pinhole (Thorlabs P200H) mounted on a translating lens mount for lateral fine adjustment (Thorlabs LM1XY). Re-collimate the beam after the pinhole. We use a lens of focal length 200 mm (Thorlabs LA1708-A) placed a focal distance of the pinhole.

36. Place a λ/2 waveplate (Thorlabs WPH05M-532) in a continuous rotation mount (Thorlabs RSP1) between the pinhole and 200-mm focal length lens (see Figure 4). Exact placement of the waveplate in the laser path after the AOM is not important.

37. ***Mounting Magnets on an XY-Translation Stage with a Rotating Platform (Steps 37-39):*** Install an XY translation stage with a rotating platform (Thorlabs XYR1) with four 1.5" pedestal pillar posts (Thorlabs RS1.5P) on the laser table. These define the position of the APD and the center of the NV experiment. Consult Figure 4 for an overview of the entire experiment and the relative positions on the laser table. Place a 1" pedestal post holder (Thorlabs, PH082E) beneath the stage, centered at its rotational axis.
38. Mount two 3" pedestal pillar posts (Thorlabs RS3P) on the same short edge and on the top side of a base plate (Thorlabs PT101). Install a right angle bracket (Thorlabs PT102) with a 25-mm translation stage (Thorlabs PT1) on top of the pedestal pillar holders. The micrometer screw of the translation stage should point up. Mount a manual rotation stage, Ø2" (Thorlabs RP01) on the upper part of the 25-mm translational stage. Mount the assembled piece on the XY translation stage with a rotating platform using the screw holes at the outer edge. The central part of the rotating platform should be kept free for the APD, which will be mounted later.
39. Install two dovetail rail carriers (Thorlabs RC1) on the dovetail optical rail (Thorlabs RLA1200) and place them 3-4" apart. On each rail carrier, mount a 1/4" travel single-axis translation stage (Thorlabs MS1S) that carries a 0.75" aluminum post (Thorlabs TRA075). The translation axis must be parallel to the optical rail. Use instant adhesive to glue two permanent magnets (EEC 2:17 TC-15) to the center of two compact kinematic mirror mounts (KMS). <u>Critical Step</u>: Ensure that one magnet is glued down on its north pole, the other one on its south pole. When the magnets are properly oriented they tend to pull toward each other. Mount each of these mirror mounts to the 0.75" posts via a 1" post (Thorlabs TRA1) and a right-angle end clamp (Thorlabs RA180). The magnets should face each other and should be aligned parallel to the rail carrier. Install the assembled rail on the center of the Ø2" rotation stage (Figure 10).

40. ***Assembling the Excitation and Detection Components (Steps 40-46):*** Mount two fixed 90° brackets (Thorlabs ER90B) with two 2" miniseries posts (Thorlabs MS2R) on the threaded holes at the top of the large area APD (Luna Optoelectronics, SD197 -70-74-591 or SD394 -70-74-591). The posts should be parallel to the edge and to the top side of the APD, as well as to one another. The posts are used to mount the APD and to affix the light guide and fluorescence filter to the APD. Screw the two 1" miniseries mounting posts (Thorlabs MS1R) into the two 90° brackets (Thorlabs ER90B). Mount each of these onto the mini posts on the APD such that the 1" mini posts point perpendicular to the face of the APD. Place them so that they are on different 2" mini posts and on different corners of the APD. Screw in an adapter with external SM1 threads and internal SM05 threads (Thorlabs SM1A6T) into a removable cage plate (Thorlabs CP90F) so that it is flush with the magnetic face. Screw a lens tube (Thorlabs SM05L03) into the adaptor plate on the opposite site of the magnetic face. Use two rings to mount the fluorescence filter (e.g. Semrock BLP01-647R) toward the end of the lens tube. Mount the assembled filter holder on the mini posts of the APD (see Figure 11). The fluorescence filter should sit directly on the APD.
41. Use an 8-32 set screw to combine 1" and 1.5"-long optical posts (Thorlabs TR1V and TR1.5V) into an effective 2.5" post. Mount this elongated post to the center of a mounting base (Thorlabs BA2). The post will be inserted into the pedestal post holder centered beneath the rotating platform (step 37). Use 1/4"-20 set screws to make two additional effective 2.5'' posts by combining the remaining 1" and 1.5" posts (Thorlabs TR1V and TR1.5V). Mount these optical posts on the top of the base, within the same slot. Separate one swivel post clamp (Thorlabs MSWC) into two parts. Remove the thumb screw from the end which has the counterbore hole. Mount both post clamp parts on the 2.5" posts. The through holes should be parallel to the long edge of the base and the thumb screws should point outwards. Adjust the spacing of the 2.5" posts in order to slide the 2" miniseries posts on the APD into the post clamps. Place the assembled APD mount into the pedestal post holder centered beneath the rotating platform (step 37). Keep some distance between the base plate and the rotational stage so that it can be freely moved.
42. Use optical epoxy (Epoxy technology, 301-2/8Z) to glue the diamond centered on the top of the light guide (e.g., 50mm-long hexagonal light pipe with 4mm aperture, (Edmund Optics 49-402), diameter depends on the diamond size). Critical step: Keep the amount of glue as small as possible and ensure that the NV-layer is facing up (see also step 45). Remove one of the retaining rings of the light guide mount (for example 4 mm x 25 mm light pipe, Edmund Optics 64-907) and place it on the light pipe. Place the retaining ring with the light pipe in a SM1-threaded adapter with smooth internal bore (Ø16 mm, Thorlabs AD16F) and use its screws to tighten the light guide. Screw the threaded adapter into the front part of removable cage plate and place it on its magnetic counterpart on the APD. Push down the light guide until it is in contact with the fluorescence filter. Critical step: Do not scratch the filter.

43. Mount a heat sink at the metallic side of the APD. Use thermal paste to improve thermal contact. !Caution: Consult the safety data sheet of the thermal paste for handling. The heat sink should in turn be cooled by a small fan. Consult the APD manual for details about powering: power the APD with +12, -12V, and GND from a power supply. The onboard cooling of the APD has to be connected to +5V and GND.
44. Place two mirrors into the laser path after the pinhole (see Figure 4). They are used to direct and to align the laser beam onto the diamond (see Figure 4 inset). Use a third mirror and flip it 90 degrees with angle clamps [Thorlabs RA90 and a 1" post (Thorlabs TR1V)] to direct the laser beam away from the optical table and toward the diamond (see Figure 4). Ensure that the laser is at low power for alignment (< 1 mW) and turn on the AOM to pass the laser toward the diamond. The laser can be directed into the diamond either directly through a polished side of the diamond, or first through a face of the light guide (Figure 4) and into the bottom face of the diamond. In either case the angle of incidence must be chosen to ensure total internal reflection at the diamond-sample interface. Use a lens (e.g., 30 mm, Thorlabs LA1289-A) to focus the laser onto the diamond. The focal length depends on the laser beam diameter and the desired laser spot size at the NV layer (in our case we design for ~ 30-μm diameter beam waist at focus). To position the laser spot on the diamond, place the lens in a holder (Thorlabs LMR05) attached to posts (2 x 1" post alumina post, Thorlabs TRA1) screwed at the other end into a 1/2" XYZ dovetail translation stage with baseplate (Thorlabs DT12XYZ). Orient the lens in a way such that one translation axis of the stage is perpendicular to the lens axis. This degree of freedom is used the adjust the position of the focal spot on the diamond. Anchor the translation stage to the optical table by mounting with two right-angle end clamps (Thorlabs RA180), a sufficient number of posts (e.g., Thorlabs TR6, 2TR2, and TR1), and a post holder (Thorlabs PH2E) (see Figure 13). The distance of the lens to the diamond should be roughly equal to the focal length.
45. To fine-tune the laser spot alignment, first check that the portion of the laser beam reflected from the top surface of the diamond leaves the apparatus at an angle similar to that of the incoming laser beam. Second, move the laser spot onto the NV region of the diamond. With laser safety goggles on (Thorlabs LG10) turn up the laser power to see a red spot on the diamond caused by the NV fluorescence. We typically use ~ 50-100 mW of continuous-wave laser power after L4. Adjust the focusing lens to minimize the laser spot size at the NV layer position. Usually two weak red spots and one bright spot are observable. The two weak spots are caused by light going through the light guide-glue-diamond interface. If you observe two bright spots on the bottom of the diamond, the diamond is mounted incorrectly with the NV layer pointing down. In this case, use acetone to dissolve the glue and remove the diamond from the light guide. Then, acid clean the diamond (steps 1-11) and glue it with the NV layer up. **Troubleshooting**.
46. Power the APD. Make sure that the APD is cooled by a fan. Check to see that the APD responds to the fluorescent light by monitoring the output voltage on an oscilloscope. You should see a voltage change upon switching the AOM on and off. We typically measure a photovoltage in the range of 100-400 mV (~ few hundred nW) on our DAQ.

47. ***Constructing Loop for MW Delivery (Step 47):*** Cut off a 5-cm piece of the semi rigid coaxial wire (Micro-Coax UT-047C-TP). Use a wire stripper to remove ca. 3 mm of the outer part of the coaxial wire and the isolation so that the inner wire is free. Place the white isolation ring of the SMA adaptor (Amphenol 901-9867-RFX) on that wire (see Figure 12). Plug the golden pin into the white dielectric cylinder. Solder the inner wire to the golden pin. The white isolation ring and cylinder should be in contact. Place the coaxial wire connected to the pin into the SMA connector. The pin should point into the threaded part of the connector. On the opposite side, use solder to fill up the gap between the coaxial wire and the SMA connector for a conductive connection of the ground. Use a wire stripper to remove ca. 3 mm of the outer part at the end of the coaxial wire and the isolation so that the central wire is free. Bend it over to form a loop. Solder the end of the interior wire to the outer part of the coaxial wire. Critical step: the loop size depends on the needs of MW field strength and homogeneity. A smaller loop generates higher MW fields with higher spatial gradients, a bigger loop has the opposite effect.

48. ***Setting up Signal Source (Steps 48-50):*** Connect the GPIB port of the MW signal source (SRS signal generator, e.g.,model SG384) to a USB port on the PC using the GPIB/USB converter. Enable the GPIB interface on the SRS signal generator and select its GPIB address by following the GPIB setup instructions in the SRS manual (e.g.,for models in the SG 380 series, see page 46 of the SG380 series manual Revision 2.04). In the “SRS connections” section of connectionConfig.py, set the variables `GPIBaddr` and `modelName` to be the GPIB address and model name of the SRS (e.g.,`GPIBaddr=27, modelName='SG384'`).
49. **!Caution**: Turn off the RF output of the SRS before continuing. If using the SRS SG384 model, this is done by pressing the Shift button followed by the AMPL button on the SRS. The SRS screen will display “N-type OFF” when this is done, to indicate that the RF output of the SRS is off. Scripts in the qdSpectro package will instruct the PC to send remote commands to turn on the SRS RF output during the experiments, but this output should be kept off while the MW path is being built. Install one MW switch (Minicircuits ZAWSA-2-50DR+) and power it with +/- 5 Volts from a power supply (see Figure 5). Connect the RF output of the SRS signal generator (the N-type output for the SG384 model) to the RF input of the switch using the N-SMA adapter and an SMA cable (Box 3). Choose a PulseBlaster channel to control this switch and connect the channel to the switch’s TTL input. Update the definition of the variable `PB_MW` in connectionConfig.py accordingly. Connect RF Output 2 of the switch to the input of the MW amplifier (Minicircuit ZHL 16W-43+) and terminate RF Output 1 with a 50Ω load. Protect the output of the amplifier with a circulator (e.g., DiTom D3C2040) from potentially damaging MW reflections. **!Caution**: Make sure that you connect the circulator to the amplifier with the correct polarity. Connect the circulator on the amplifier to the MW loop with an SMA cable and SMA female-SMA female adapter (e.g.,Thorlabs T4285). Use the SMA female-female adapter to mount the loop with a clamp (Thorlabs CL5) to two posts connected with a right angle clamp (Thorlabs RA90) to the table. Adjust the height of the mount and bend the coaxial wire so that the loop is in contact with the diamond and the laser spot is centered in the loop (see Figure 13). **!Caution**: Consult the amplifier manual for handling instructions. Do not exceed maximum input power. Ensure that the amplifier is on (i.e.,supplied with its required DC voltages) before feeding RF signals into its input port. General guidelines for turning on and off amplifiers are given in Box 4, but safety guidelines outlined in your amplifier’s manual or by the manufacturer take precedence and must be read and followed.
50. Set up the IQ phase control: for the nuclear spin sensing experiments, the phase of the MW pulses must be controlled, which we achieve by use of the internal IQ mixer of the SRS signal generator SG 384. For a detailed description of the IQ mixer and its calibration, consult the manual. To control the four phases ($x$ = 0°, -$x$ = 180°, $y$ = 90°, -$y$ = 270°, see also Table 1), the SG 384 needs two inputs where each can be switched between +/- $0.5/\sqrt{2}$ ~ 0.35 V. Use a stable power supply with +0.35 V and -0.35 V and split it with a T connector at each output. Install two MW switches (Minicircuits, ZAWSA-2-50DR+). Connect +0.35 V to RF Output 1 and -0.35 V to RF Output 2 of each of the two MW switches. Critical step: Ensure that the absolute value is 0.35 V in both cases. Connect the input of the switches to the IQ inputs of the SG384. Power the switches with +/- 5 V. Connect the TTL inputs of the switches to two PulseBlaster channels and update the definitions of the variables `PB_I` and `PB_Q` in connectionConfig.py accordingly.

## NV-Diamond Characterization Experiments (2 weeks)

51. ***Finding the NV-ESR Transition at Ambient Magnetic Field (Step 51):*** Remove the magnets to work at near-zero applied magnetic field (i.e., at the Earth’s field). This makes it easier to find the NV-ESR resonance for the first time. Refer to Box 1 for a description of the qdSpectro package and how to run experiments. Open ESRconfig.py in Notepad++, read the description of the variables and data-processing options defined in this script and change the frequency sweep values to 2.7 - 3.0GHz with a step size of e.g.,3 MHz (i.e., set `startFreq=2.7e9`, `endFreq=3e9` and `N_scanPts=101`). Set the `microwavePower` variable, which sets the output power level of the SRS signal generator in dBm, to a level which is below your amplifier's maximum input, e.g., we set `microwavePower=0` (units are dBm) when using the Minicircuits ZHL-16W-43+ amplifier. The duration of the pulse sequence (the signal-acquisition half), `t_duration`, should be set to 80μs, with 10000 samples taken per scan point (i.e.,set `t_duration=80*us` and `Nsamples=10000` in `ESRconfig.py`). Turn on the MW amplifier. Run `python mainControl.py ESRconfig`. Check the pulse sequence by monitoring all the PulseBlaster output channels on an oscilloscope and comparing them with the pulse sequence shown in Figure 6. Two dips should be visible in the data (see Figure 6a). The contrast may vary depending on MW power, pulse sequence duration, and properties of the diamond. **Troubleshooting.**

52. ***Running NV-ESR: Alignment and Adjusting Magnetic Field (Step 52):*** Turn off the AOM using `togglePBchan.py` (Box 2). Mount the magnets at the ends of the rail so that each magnet is roughly equidistant from the laser spot on the diamond. Also use x and y of the rotational stage to move the magnets so that the diamond and the laser spot are centered between the two magnets. Rotate the stage so that the magnet faces are parallel/perpendicular to the edges of the diamond [if the diamond is cut along (110) facets] and tilt the rail ~36 degrees out of the horizontal plane. This gives a rough alignment of the $B_0$ field along one of the NV axes. Turn on the AOM and run `python mainControl.py ESRconfig`. Depending on the magnetic field strength, sweep the MW frequency between 1.8 GHz and 4 GHz in 3-MHz steps. If the experiment is not aligned along one of the four possible NV axes (which is very likely, see Figure 14), up to 8 NV-ESR transitions will appear in the NV-ESR spectrum (excluding additional splittings due to hyperfine interactions). **CRITICAL STEP:** Importantly, transitions due to different NV orientations may have unequal amplitudes in their resonance lines and weak transitions can be easily missed. To tune transition rates (and thus resonance line amplitudes), adjust the laser polarization by rotating the half waveplate, adjust the MW polarization by moving the loop (though this is not recommended once high contrast has been achieved), or increase the MW power (without exceeding the maximum input power of your amplifier). Keep in mind that high MW power induces power broadening. Critical step: For the quantum sensing protocol, the $B_0$ field must be aligned along one NV axis. The three inner transitions then overlap (they all see the same projection of $B_0$) and only 4 resonance lines (without hyperfine lines) appear, two on each side of the zero-field splitting (see Figure 14). Try to overlap the resonances of the off-axis NV orientations by using the two rotational degrees of freedom of the magnetic mount. One can also use the mirror holders on which the magnets are glued for fine tuning. If all of these lines overlap within their linewidths the misalignment is typically smaller than a few degrees. After successfully aligning the experiment, tighten the set screws of the rotational stages. Now use the slider on the rail to adjust the magnetic field strength. The magnetic field strength can be calculated from the observed Zeeman shift according to equation 6. For example, if one would like to work at 310 G, then move the lowest NV resonance to ~ 2 GHz. Critical step: Always try to keep the diamond centered between the magnets. If it is not in the center, magnetic field gradients might lead to unwanted line broadening. To check this, reduce the MW power to reduce power broadening and increase the number of samples or averages to improve the signal to noise. Look at the resonance with the lowest (or highest) transition frequency, which corresponds to NVs aligned along $B_0$. If the hyperfine splitting caused by the $^{14}N$ nuclear spin can be observed, then the gradient is less than 1 G over the laser spot, which is sufficient for the following experiments (Figure 14, inset). You can also use the hyperfine amplitudes as an alignment diagnostic if you work at moderate fields in the range 250-1000 G. In the case of poor alignment, the three hyperfine lines exhibit similar amplitudes. In the aligned case, the $^{14}N$ nucleus is polarized[57,58] as can be seen in Figure 14. **Troubleshooting.**

53. ***Running NV-ESR: Determining ESR Resonance Frequency and $B_0$ Field Strength (Step 53):*** Open ESRconfig.py in Notepad++ and change frequency sweep values around the expected resonance frequency after alignment. Typically, we drive the lowest resonance frequency ($m_s = 0$ to $m_s = -1$). Run `python mainControl.py ESRconfig`. Use the center frequency of the dip to calculate $B_0$ according to equation 6.

54. ***Setting the Timing between the Readout Pulses (Step 54):*** Although the PulseBlaster outputs the AOM and DAQ trigger pulse in synchrony, these pulses usually do not coincide in real time because of a finite AOM response time and cable length differences. Open optimReadoutDelay.py in Notepad++. This script sweeps the readout delay between the start of the AOM and the DAQ pulse. Edit the `startDelay` and `endDelay` variables to set the sweep range of the delay, e.g.,we use `startDelay = 10*ns` (note that the `startDelay` must be at least `5*t_min`, where `t_min` is the time resolution of the PulseBlaster which, for a 500MHz-clock board, is 2 ns), `endDelay = 10000*ns`, with 100 points in the sweep (set by `N_scanPts`). Run `python optimReadoutDelay.py`. The recorded data shows temporal overlap of these pulses (see Figure 15). The photovoltage is zero if there is no overlap and increases if both pulses overlap. Critical step: The optimum readout time is sufficiently late in the photovoltage rise (i.e., around 2 µs in Figure 15). Enter this delay as the value of the variable `t_readoutDelay` in the config files of all experiments that follow in this protocol (i.e.,Rabiconfig.py, T2config.py, XY8config.py, correlSpecconfig.py, T1config.py). This setting will remain fixed as long as there are no physical changes to the setup (e.g., changes to AOM alignment or cable lengths).

55. ***Optimizing Rabi Contrast (Steps 55-56):*** Open Rabiconfig.py in Notepad++. Set the `microwaveFrequency` variable to the resonance frequency of the NV (step 53) and set `microwavePower` to approximately 10dB below the amplifier's maximum input power. Set `t_readoutDelay` as determined in step 54. Run `python mainControl.py Rabiconfig`. Rabi oscillations should be observed in the data (Figure 6b and 16). The frequency of the oscillation can be adjusted by changing the MW power. Critical Step: Sensitivity to NMR signals in later steps improves linearly with the contrast (amplitude) of the Rabi oscillation observed here. The contrast should be optimized each time you change diamond, diamond orientation, and/or sample. **Troubleshooting.**
56. Optimize the contrast by changing the laser polarization by rotating the $\lambda/2$ waveplate and by ensuring efficient NV repolarization. During an AOM pulse, the NVs must be reset to their $m_s = 0$ state. If NVs are not repolarized efficiently, the contrast will decrease. In our work, we find efficient repolarization is achieved by using 100-200 mW of laser power in a ~ (30 $\mu$m)$^2$ laser spot with an AOM pulse duration of 5 $\mu$s. One can change any of these three parameters in order to improve the repolarization. Improve SNR by increasing the number of samples and averages as needed to visualize the signal and reference data of the Rabi experiment (see Figure 16). If efficient NV repolarization is achieved, the Rabi oscillation will be observed only in the signal channel. If the NVs are not repolarized efficiently, the oscillation will appear also in the reference channel. The simplest way to optimize NV repolarization is to decrease the laser spot size by translating the focusing lens. If there is insufficient SNR to see the Rabi oscillation in the signal channel, optimize the contrast by adjusting the laser spot size and power, as well as the AOM pulse duration. Importantly, if the laser intensity is too high, the contrast will again be diminished because repolarization occurs faster than the readout time of the DAQ.

57. **Running Rabi: Determining the π and π/2 pulse durations** Run `python mainControl.py Rabiconfig` with the previous parameters. Critical step: Note the MW pulse duration at which the first signal minimum occurs; this is the π-pulse duration (see Figure 6b). To change the Rabi frequency, adjust the power of the MW and/or the position of the laser spot with respect to the MW loop wire. With this experimental setup, we find that π-pulse durations of 20 - 60 ns are achievable. **Troubleshooting.**

58. **Running NV Hahn Echo Experiment: Determine NV-T2 Relaxation Time** Open T2config.py in Notepad++. Set the MW frequency (`microwaveFrequency`) to the resonance frequency of the NV. Set the MW power (`microwavePower`) and the π pulse duration (`t_pi`) to the values obtained in the Rabi (step 57). The minimum sweep time $\tau$ (`startTau`) should be longer than the π-pulse duration (e.g., for `t_pi =24*ns`, set `startTau=100*ns`). Conclude the sweep at `endTau=10*us` (depending on the NV-T2 relaxation time). Set `t_readoutDelay` as determined in step 54. The IQ function of the SRS signal generator must be activated. Set `numberOfPiPulses=1` to produce a Hahn Echo sequence consisting of a single π–pulse sandwiched between two π/2 pulses, with $\tau$ defined as the delay between the π/2 and π pulses, as shown in Figure 6c (if one sets `numberOfPiPulses`>1 to produce a sequence with multiple π pulses, qdSpectro will instead interpret $\tau$ as the delay between the π pulses, as explained in the user-input comments in T2config.py). Run `python mainControl.py T2config`. The acquired data should show a decay, as in Figure 6c. The contrast at short times should be slightly less than half of the contrast in the Rabi experiment. For long $\tau$, the contrast must go to zero. If this is not the case, it indicates that either the π/2 and π–pulse durations are not correct, or the IQ-mixer is not working properly. **Troubleshooting.**

59. **Running NV T1 Experiment: Determine NV-T1 Relaxation Time** Open T1config.py. Change the MW frequency (`microwaveFrequency`) to the resonance frequency of the NV. Set the MW power (`microwavePower`) and the π pulse duration (`t_pi`) to the values obtained in the Rabi experiment (step 57). Set `t_readoutDelay` as determined in step 54. Set the sweep range to 0 to 5 ms and use 1000 samples (i.e.,set `start_t=0*ms`, `end_t=5*ms`, `Nsamples=1000`. Note that, if `start_t` is set to 0 or to any value smaller than `(t_readoutDelay + t_min*round(1*us/t_min))`, qdSpectro will automatically shift it by `(t_readoutDelay + t_min*round(1*us/t_min))` to avoid pulse overlap errors, as detailed in the documentation under the user input section of T1config.py). Keep in mind that NV-T1 experiments might take a long time. Run `python mainControl.py T1config`. The acquired data should show a decay, as in Figure 6f.

## Quantum sensing procedure for the detection of Nuclear and Electronic spins

CRITICAL: We recommend sensing an external AC signal as described in Box 5 before NMR signal detection to check the functionality of the experiment.

60. For running XY8-N Dynamic Decoupling Sequence for surface NMR signal detection, follow Option A. For running correlation spectroscopy for surface NMR signal detection follow Option B. For running NV-T1 experiments for surface electronic spin detection, follow Option C.

**Option A: Running XY8-N Dynamic Decoupling Sequence for Surface NMR Signal Detection (Timing 2 h)**

i. Pipette a droplet of the NMR sample on the diamond's surface at the laser spot. Keep in mind that the refractive index of the sample might change the total internal reflection condition, which in turn may lead to reduced Rabi contrast. When performing this step for the first time, we recommend cleaning the diamond with isopropyl alcohol (IPA). Protons can usually be sensed even on a "clean" diamond surface, likely due to the presence of a nanoscale hydrocarbon film on the surface. Other useful samples for initial experiments are PDMS (for $^1$H) or Fomblin oil (for $^{19}$F), as shown in Figure 18.
ii. Determine the $B_0$ field with the NV-ESR procedure according to step 53 and equation 6. Calculate the nuclear spin resonance frequency $f_L$ by multiplying the $B_0$ field strength in G by the gyromagnetic ratio in MHz/G (equation 5). Calculate $\tau_0 = 1/(2f_L)$, to be set as the time spacing between $\pi$ pulses in the XY8-*N* sequence. Reasonable $\tau_0$ are between 300-2000 ns.
iii. Repeat Rabi (step 57) to determine the π/2 and π-pulse durations. In some cases, the Rabi contrast goes down upon placing a sample on the surface due background fluorescence of the sample or because the TIR condition is no longer met. Note that the sample can also influence MW delivery.
iv. Open XY8config.py in Notepad++. Set the π-pulse duration (`t_pi`) and MW frequency (`microwaveFrequency`) according to the previous steps. Sweep $\tau$ around $\tau_0$ with, for example, 50 points and 4-ns spacing, which is double the smallest step size (`t_min`) allowed by our PulseBlaster card (recall, as shown in Figure 6d, that $\tau$ in the XY8 sequence is defined as the delay between successive π pulses, which is double the delay between the π/2 pulses and π pulses. Since the timing resolution of our 500MHz PulseBlaster card is 2ns, $\tau$ must be set to at least 4ns in this pulse sequence. If using a different PB card, set the spacing to `2*t_min`. Depending on the diamond, one should use $N$ > 8, 10000 samples, and 5-10 averages (by setting the variables `N`, `Nsamples` and `Navg` respectively). Set `t_readoutDelay` as determined in step 54. $N$ should be optimized by comparing the SNR for the same total duration of the experiment.

v. Run `python mainControl.py XY8config`. The acquired data should resemble that in Figure 18. **Troubleshooting**

**Option B: Run Correlation Spectroscopy for Surface NMR Signal Detection (Timing 2 hours)**

i. Pipette a droplet of the sample on the diamond's surface at the laser spot. Keep in mind that the refractive index of the sample might change the total internal reflection condition, which in turn may lead to reduced Rabi contrast. When performing this step for the first time, we recommend cleaning the diamond with isopropyl alcohol (IPA). Protons can usually be sensed on a "clean" diamond surface, likely due to the presence of a nanoscale hydrocarbon film on the surface. Other useful samples for initial experiments are PDMS (for $^{1}H$) or Fomblin oil (for $^{19}F$), as shown in Figure 19.

ii. Determine the $B_0$ field with the ESR according to step 53 and equation 6. Calculate the nuclear spin resonance frequency by multiplying the $B_0$ field strength in gauss by the gyromagnetic ratio in MHz/G (equation 5). Calculate $\tau_0 = 1/(2f_L)$, to be set as the π-pulse spacing in the XY8-N sequence. Reasonable $\tau_0$ values are between 300-2000 ns.

iii. Repeat Rabi (step 57) to determine the $\pi/2$ and π-pulse durations. In some cases, the Rabi contrast goes down upon placing a sample on the surface due background fluorescence of the sample or because the TIR condition is no longer met. Note that the sample can also influence MW delivery.

iv. Open correlSpecconfig.py in Notepad++. Set the π-pulse duration (`t_pi`), the XY8-N π-pulse spacing, $\tau_0$(`tau0`), and the MW frequency (`microwaveFrequency`), as determined in the previous steps. Choose the sampling of the swept delay, `t_corr`, so that at least 2 points per $1/f_L$ are recorded. The full duration of the sweep depends on the requirements of the measurement and can be kept short for initial nuclear spin detection (e.g.,$5/f_L$) or made long (several 10 -100 μs) for recording the nuclear correlation time. Depending on the diamond, one should start by using $N$ = 1 or 2. $N$ should be optimized by comparing the SNR for the same total duration of the experiment. Use 10000 samples and 2-4 averages (by setting the variables `Nsamples` and `Navg` respectively). Set `t_readoutDelay` as determined in step 54.

v. Run `python mainControl.py correlSpecconfig`. The acquired data should show oscillation at $f_L$ and look like those in Figure 19. **Troubleshooting**

**Option C: Run NV-T1 Experiments for Surface Electronic Spin Detection (Timing 4 hours)**

i. Determine the NV-T1 relaxation time of the clean diamond first according to step 59 (it should be around a few milliseconds).

ii. Pipette a droplet of the sample on the diamond's surface at the laser spot. Keep in mind that the refractive index of the sample might change the total internal reflection condition, which in turn may lead to reduced sensitivity. When performing this step for the first time, we recommend using 1-M $Cu^{2+}$, $Gd^{3+}$ or $Mn^{2+}$ solutions.

iii. Repeat Rabi (step 57) to determine the π-pulse duration. In some cases, the Rabi contrast goes down upon placing a sample on the surface due background fluorescence of the sample or because the TIR condition is no longer met. Note that the sample can also influence MW delivery.

iv. Run `python mainControl.py T1config`. The acquired data should show a decrease NV-T1 as shown in Figure 21.

## TIMING

The timing given below are strongly depend on the user's experience. Most steps will be much faster after set up the procedures or can take longer if performed for the first time.

### Steps 1-17, Fabrication of NV-diamond chips. 2 days

Steps 1-11, Acid cleaning. 2 h

Steps 12-17, Annealing. 1 day

**Steps 18-50, Construction of the quantum diamond spectromter (3 weeks)**
Steps 18-25, Experimental Control Software Installation. 1 day
Steps 26-32 PulseBlaster and DAQ Setup. 1 day
Steps 33-36, Mounting and Alignment of the Acousto-Optic Modulator (AOM). 1 week
Steps 37-39, Mounting Magnets on an XY-Translation Stage with a Rotating Platform. 2 days
Steps 40-46, Assembling the Excitation and Detection Components. 2 days
Step 47, Constructing Loop for MW Delivery. 1 h
Steps 48-50, Setting up Signal Source. 1 day

**Steps 51-59, NV Diamond Characterization Experiments. 2 weeks**
Step 51, Finding the NV-ESR Transition at Ambient Magnetic Field. 2 days
Step 52, Running NV-ESR: Alignment and Adjusting Magnetic Field. 2 day
Step 53, Running NV-ESR: Determining ESR Resonance Frequency and B0 Field Strength. 2 h
Step 54, Setting the Timing between the Readout Pulses. 2 h
Steps 55-56, Optimizing Rabi Contrast. 1 day
Step 57, Running Rabi: Determining the π and π/2 pulse durations. 1 h
Step 58, Running NV Hahn Echo Experiment: Determine NV-T2 Relaxation Time. 1h
Step 59, Running NV-T1: Determine NV-T1 Relaxation Time. 2 h

**Step 60, Quantum Sensing Procedure for the Detection of Nuclear and Electronic Spins**
Option A, Running XY8-N Dynamic Decoupling Sequence for Surface NMR Signal Detection. 2h
Option B, Running Correlation Spectroscopy for Surface NMR Signal Detection. 2 h
Option C, Running NV-T1 Experiments for Surface Electronic Spin Detection. 4 h

## Troubleshooting:

Troubleshooting advice can be found in Table 2.

**Table 2.** Troubleshooting table

| Step | Problem | Possible reason | Solution |
|---|---|---|---|
| 45 | Cannot see a red spot due to NV PL on diamond | Laser light power too low | (i) Turn up the laser power; use laser goggles to protect eyes.<br>(ii) Try to focus the laser spot more tightly. |
| 45 | Can only see a red stripe due to NV PL on diamond | Laser focus is not at the NV layer | Move the focusing lens to adjust to adjust the position of the focus. |
| 51 | No ESR signal at zero field | Pulse sequence is not running | Check the pulse sequence on the oscilloscope and check if all channels are active. |
| | | MW power too low | (i) Increase MW power (be careful not to exceed the maximum input of the amplifier).<br>(ii) Move the laser spot closer to the MW wire.<br>(iii) Increase the full pulse sequence duration. |
| 52 | Fewer than 8 lines are observable in the ESR experiment | Not all transitions have the same amplitude and might be weak | (i) Relative resonance intensities can be changed by turning the laser light polarization with the λ/2 wave plate.<br>(ii) Increase MW power (be careful, do not exceed the maximum input of the amplifier). |
| | | Lines are overlapping | Change the magnetic field direction slightly while observing the resonance positions. |
| | | Magnetic field strength is too strong | Check if the field is too strong and the resonance is out of the accessible frequency range. Move the magnets apart to reduce the field strength and record a spectrum at very low field. |
| 52 | Cannot see the hyperfine-splitting in the ESR experiment | MW power too high | Reduce the MW power and increase number of samples or averages. |
| | | $B_0$ field gradient | (i) Center the NVs in the laser spot between the magnets.<br>(ii) Use bigger magnets (> 2 cm diameter) and increase distance of the magnets. |
| | | Intrinsic ESR linewidth is too broad | In very high-density NV-layers, the intrinsic ESR linewidth might be broader than the hyperfine splitting. Change NV implant parameters. Note: Observing the hyperfine splitting is not crucial for the quantum sensing protocol. |
| 55 | Cannot see a Rabi oscillation | Rabi frequency is very low | (i) Extend the MW pulse duration to more than 1 μs and/ or increase MW power (be careful not to exceed the maximum input of the amplifier).<br>(ii) Check if you set the MW frequency to the resonance of the NV measured in an ESR experiment. |
| 56 | Poor Rabi contrast after optimization of laser polarization and optical repolarization | Background light/fluorescence | Shield the experiment from room light. |
| | | Scattered laser light | (i) Clean light guide and diamond surface.<br>(ii) Optimize the laser spot position on the diamond. |
| 57 | Rabi frequency is very low | MW power is low | Increase MW power (be careful not to exceed the maximum input of the amplifier). |
| 58 | Cannot see a T2 relaxation decay even though Rabi experiment works | IQ phase control is off | Turn on IQ phase control. |
| | Cannot see RF loop signal in the XY8-*N* dynamic decoupling sequence | Signal is too weak | (i) Reduce distance between RF loop and diamond<br>(ii) Orient loop so that its B field is aligned along $B_0$. |
| | | | (iii) Use more windings<br>(iv) Use an RF amplifier to increase the signal strength. |
| 60 Option A(v) and Option B (v) | Cannot see a nuclear spin signal in XY8-*N* decay or correlation spectroscopy | SNR is not good enough | (i) Increase number of averages.<br>(ii) Change *N* of the XY8-*N* sequence (higher *N* results in bigger signals but contrast decreases with *N*; must find the optimum for the experiment).<br>(iii) Try to see a synthetic signal from an RF loop (see Box 5). This lends a sense of the sensitivity and functionality of the experiment. |
| | | Sample is not in contact with the diamond | Clean the diamond and replace the sample. Often surface films and contaminants obfuscate the signal. |

## Anticipated Results

To illustrate the capabilities of the quantum diamond spectrometer, we here describe some results of nuclear and electronic spin sensing experiments. In all cases, we worked at a magnetic field of 311 G.

### NV-NMR sensing with XY8-*N* dynamic decoupling sequence

In Figure 18, we show the NV-NMR detection of $^{1}H$ in PDMS and $^{19}F$ in Fomblin oil with an XY8-12 decay. At 311 G, the proton and the fluorine Larmor frequencies are 1.324 MHz and 1.246 MHz, respectively. The magnetic noise at these frequencies causes dips in the XY8-12 decay at 378 ns and 401 ns. Experimental details can be found in the figure caption. Typically, the NV-T2 decay envelop is subtracted and the time axis is converted to a frequency axis with equation 3 (see inset Figure 18a).

### NV-NMR sensing with correlation spectroscopy

In Figure 19, we show the NV-NMR detection of $^{1}H$ in PDMS base and $^{19}F$ in Fomblin oil with correlation spectroscopy. Experimental details can be found in the figure caption. At 311 G, the proton and the fluorine Larmor frequencies are 1.324 MHz and 1.246 MHz, respectively, both of which can be directly seen in the oscillation of the correlation data. Taking the Fourier transform of these data confirms the resonance frequencies (Figure 20). The NMR signal can be identified unambiguously by changing the magnetic field strength, which causes frequency shifts in proportion to the gyromagnetic ratio. The oscillating correlation signal decays over time largely due to spin decoherence or sample physical diffusion[29]. Sample diffusion leads to severe line broadening as compared to experiments that detect thermal nuclear spin polarization from larger volumes. Despite the line broadening, detection of sample diffusion indicates the ability to monitor dynamical molecular processes.

### Electronic spin sensing with NV-T1 relaxometry

In Figure 21, the NV-T1 relaxation of a bare diamond surface and that corresponding to a surface supporting a 1-M $Cu^{2+}$ solution is compared. Experimental details can be found in the figure caption. The electronic noise of the unpaired spins in the $Cu^{2+}$ ion reduces the NV-T1 relaxation time significantly. The relaxation in both cases deviates from a pure mono-exponential decay, indicating an inhomogeneous NV ensemble e.g., due to an NV depth distribution.

## Acknowledgements

This material is based on work supported by, or supported in part by, the US Army Research Laboratory and the US Army Research Office under contract/grant number W911NF1510548. D.B.B. was partially supported by the German Research Foundation (BU 3257/1-1). D.P.L.A.C. was partially supported by the NSF STC "Center for Integrated Quantum Materials" under Cooperative Agreement No. DMR-1231319.

**Author contributions**
D.B.B. led the development of the protocol, build-up of the quantum diamond spectrometer, and acquisition and analysis of data, informed by extensive past work in the Walsworth group and aided closely by D.P.L.A.C. and M.P.B.. D.P.L.A.C. wrote the qdSpectro software package. M.J.T. prepared the NV-diamond chip and provided technical assistance in build-up of the quantum diamond spectrometer. O.B.D. performed a pilot run of the protocol and provided feedback to improve procedures. D.R.G. provided technical guidance to all aspects of the effort. R.L.W. supervised the project. All authors discussed the results and participated in writing the manuscript.

**Competing Financial interest**
The authors declare no competing financial interests.

**Data Availability Statement**
The primary data of this study are available from the corresponding authors upon reasonable request.

**Code Availability Statement** The qdSpectro package is available to download from https://gitlab.com/dplaudecraik/qdSpectro and is licensed under the MIT License. The most recent version at the time of writing is version 1.0, but the user is encouraged to download the latest version and refer to the readme file for any patches and updates. The package is registered under the DOI 10.5281/zenodo.1478113, which points to the latest version.

## References

1. Rule, G. S. and Hitchens, K. T. *Fundamentals of Protein NMR Spectroscopy*. (Springer, 2006).
2. Schweiger, A. and Jeschke, G. *Principles of Pulse Electron Paramagnetic Resonance*. (Oxford University Press, 2001).
3. Zalesskiy, S. S., Danieli, E., Blümich, B. & Ananikov, V. P. Miniaturization of NMR systems: desktop spectrometers, microcoil spectroscopy, and "NMR on a Chip" for chemistry, biochemistry, and industry. *Chem. Rev.* 114, 5641–5694 (2014).
4. Fratila, R. M. & Velders, A. H. Small-volume nuclear magnetic resonance spectroscopy. *Annual Review of Analytical Chemistry* 4, 227–249 (2011).
5. Ardenkjaer-Larsen, J.-H. *et al.* Facing and overcoming sensitivity challenges in biomolecular NMR spectroscopy. *Angew. Chem. Int. Ed.* 54, 9162–9185 (2015).

6. Staudacher, T. *et al.* Nuclear magnetic resonance spectroscopy on a (5-Nanometer)3 sample volume. *Science* 339, 561–563 (2013).

7. Mamin, H. J. *et al.* Nanoscale nuclear magnetic resonance with a nitrogen-vacancy spin sensor. *Science* 339, 557–560 (2013).

8. Lovchinsky, I. *et al.* Nuclear magnetic resonance detection and spectroscopy of single proteins using quantum logic. *Science* 351, 836–841 (2016).

9. Shi, F. *et al.* Single-protein spin resonance spectroscopy under ambient conditions. *Science* 347, 1135–1138 (2015).

10. Sushkov, A. O. *et al.* Magnetic resonance detection of individual proton spins using quantum reporters. *Phys. Rev. Lett.* 113, 197601 (2014).

11. Lovchinsky, I. *et al.* Magnetic resonance spectroscopy of an atomically thin material using a single-spin qubit. *Science* 355, 503–507 (2017).

12. Aharonovich, I. *et al.* Diamond-based single-photon emitters. *Rep. Prog. Phys.* 74, 076501 (2011).

13. Schirhagl, R., Chang, K., Loretz, M. & Degen, C. L. Nitrogen-vacancy centers in diamond: nanoscale sensors for physics and biology. *Annual Review of Physical Chemistry* 65, 83–105 (2014).

14. Doherty, M. W. *et al.* The nitrogen-vacancy colour centre in diamond. *Physics Reports* 528, 1–45 (2013).

15. Rondin, L. *et al.* Magnetometry with nitrogen-vacancy defects in diamond. *Rep. Prog. Phys.* 77, 056503 (2014).

16. Jelezko, F. & Wrachtrup, J. Single defect centres in diamond: a review. *Phys. Stat. Sol. (a)* 203, 3207–3225 (2006).

17. Doherty, M. W. *et al.* Theory of the ground-state spin of the NV- center in diamond. *Phys. Rev. B* 85, 205203 (2012).

18. Hincks, I., Granade, C. & Cory, D. G. Statistical inference with quantum measurements: methodologies for nitrogen vacancy centers in diamond. *New J. Phys.* 20, 013022 (2018).

19. Meriles, C. A. *et al.* Imaging mesoscopic nuclear spin noise with a diamond magnetometer. *J. Chem. Phys.* 133, 124105 (2010).

20. DeVience, S. J. *et al.* Nanoscale NMR spectroscopy and imaging of multiple nuclear species. *Nat. Nano* 10, 129–134 (2015).

21. Pham, L. M. NMR technique for determining the depth of shallow nitrogen-vacancy centers in diamond. *Phys. Rev. B* 93, (2016).

22. Herzog, B. E., Cadeddu, D., Xue, F., Peddibhotla, P. & Poggio, M. Boundary between the thermal and statistical polarization regimes in a nuclear spin ensemble. *Appl. Phys. Lett.* 105, 043112 (2014).

23. Degen, C. L., Reinhard, F. & Cappellaro, P. Quantum sensing. *Rev. Mod. Phys.* 89, 035002 (2017).

24. Abe, E. & Sasaki, K. Tutorial: Magnetic resonance with nitrogen-vacancy centers in diamond—microwave engineering, materials science, and magnetometry. *Journal of Applied Physics* 123, 161101 (2018).

25. Gullion, T., Baker, D. B. & Conradi, M. S. New, compensated Carr-Purcell sequences. *Journal of Magnetic Resonance (1969)* 89, 479–484 (1990).

26. Ryan, C. A., Hodges, J. S. & Cory, D. G. Robust decoupling techniques to extend quantum coherence in diamond. *Phys. Rev. Lett.* 105, 200402 (2010).

27. Loretz, M. *et al.* Spurious harmonic response of multipulse quantum sensing sequences. *Phys. Rev. X* 5, 021009 (2015).

28. Laraoui, A. *et al.* High-resolution correlation spectroscopy of $^{13}C$ spins near a nitrogen-vacancy centre in diamond. *Nature Communications* 4, ncomms2685 (2013).

29. Kong, X., Stark, A., Du, J., McGuinness, L. P. & Jelezko, F. Towards chemical structure resolution with nanoscale nuclear magnetic resonance spectroscopy. *Phys. Rev. Applied* 4, 024004 (2015).

30. Staudacher, T. *et al.* Probing molecular dynamics at the nanoscale via an individual paramagnetic centre. *Nature Communications* 6, ncomms9527 (2015).

31. Kehayias, P. *et al.* Solution nuclear magnetic resonance spectroscopy on a nanostructured diamond chip. *Nature Communications* 8, 188 (2017).

32. Aslam, N. *et al.* Nanoscale nuclear magnetic resonance with chemical resolution. *Science* 357, 67–71 (2017).

33. Glenn, D. R. *et al.* High-resolution magnetic resonance spectroscopy using a solid-state spin sensor. *Nature* 555, 351–354 (2018).

34. Bucher, D. B., Glenn, D. R., Park, H., Lukin, M. D. & Walsworth, R. L. Hyperpolarization-enhanced NMR spectroscopy with femtomole sensitivity using quantum defects in diamond. *arXiv:1810.02408 [physics, physics:quant-ph]* (2018).

35. Steinert, S. *et al.* Magnetic spin imaging under ambient conditions with sub-cellular resolution. *Nature Communications* 4, 1607 (2013).

36. Sushkov, A. O. *et al.* All-optical sensing of a single-molecule electron spin. *Nano Lett.* 14, 6443–6448 (2014).

37. Simpson, D. A. *et al.* Electron paramagnetic resonance microscopy using spins in diamond under ambient conditions. *Nature Communications* 8, 458 (2017).

38. Ermakova, A. *et al.* Detection of a few metallo-protein molecules using color centers in Nanodiamonds. *Nano Lett.* 13, 3305–3309 (2013).

39. Schlipf, L. *et al.* A molecular quantum spin network controlled by a single qubit. *Science Advances* 3, e1701116 (2017).

40. Tetienne, J.-P. *et al.* Spin properties of dense near-surface ensembles of nitrogen-vacancy centers in diamond. *Phys. Rev. B* 97, 085402 (2018).

41. Myers, B. A. *et al.* Probing surface noise with depth-calibrated spins in diamond. *Phys. Rev. Lett.* 113, 027602 (2014).

42. Ofori-Okai, B. K. *et al.* Spin properties of very shallow nitrogen vacancy defects in diamond. *Phys. Rev. B* 86, 081406 (2012).

43. Romach, Y. *et al.* Spectroscopy of surface-induced noise using shallow spins in diamond. *Phys. Rev. Lett.* 114, 017601 (2015).

44. Ziegler, J. F., Ziegler, M. D. & Biersack, J. P. SRIM - the stopping and range of ions in matter (2010). *Nuclear Instruments and Methods in Physics Research B* 268, 1818–1823 (2010).

45. Lehtinen, O. *et al.* Molecular dynamics simulations of shallow nitrogen and silicon implantation into diamond. *Phys. Rev. B* 93, 035202 (2016).

46. Yamamoto, T. *et al.* Extending spin coherence times of diamond qubits by high-temperature annealing. *Phys. Rev. B* 88, 075206 (2013).

47. Haque, A. & Sumaiya, S. An overview on the formation and processing of nitrogen-vacancy photonic centers in diamond by ion implantation. *Journal of Manufacturing and Materials Processing* 1, 6 (2017).

48. Pezzagna, S., Naydenov, B., Jelezko, F., Wrachtrup, J. & Meijer, J. Creation efficiency of nitrogen-vacancy centres in diamond. *New J. Phys.* 12, 065017 (2010).

49. Kim, M. *et al.* Decoherence of near-surface nitrogen-vacancy centers due to electric field noise. *Phys. Rev. Lett.* 115, 087602 (2015).

50. Fávaro de Oliveira, F. *et al.* Tailoring spin defects in diamond by lattice charging. *Nature Communications* 8, 15409 (2017).

51. Rosskopf, T. *et al.* Investigation of surface magnetic noise by shallow spins in diamond. *Phys. Rev. Lett.* 112, 147602 (2014).

52. Sangtawesin, S. *et al.* Origins of diamond surface noise probed by correlating single spin measurements with surface spectroscopy. *arXiv:1811.00144 [cond-mat, physics:quant-ph]* (2018).

53. Atikian, H. A. *et al.* Superconducting nanowire single photon detector on diamond. *Appl. Phys. Lett.* 104, 122602 (2014).

54. Tisler, J., Balasubramania, G., Naydenov, B., Kolesov, R., Grotz, B., Reuter, R., Boudou, J. P., Curmi, P. A., Sennour, M., Thorel, A., Börsch, M., Aulenbacher, K., Erdmann, R., Hemmer, P. R., Jelezko, F., Wrachtrup, J. Fluorescence and spin properties of defects in single digit nanodiamonds. *ACS Nano* 3, 1959–1965 (2009).

55. Pham, L. M. *et al.* Enhanced solid-state multispin metrology using dynamical decoupling. *Phys. Rev. B* 86, 045214 (2012).

56. Diana P L Aude Craik. (2019, April 14). qdSpectro (Version 1.0). Zenodo.

http://doi.org/10.5281/zenodo.1478113.

57. Fischer, R., Jarmola, A., Kehayias, P. and Budker, D. Optical polarization of nuclear ensembles in diamond. *Phys. Rev. B* 87, 125207 (2013).

58. Jacques, V. *et al.* Dynamic polarization of single nuclear spins by optical pumping of nitrogen-vacancy color centers in diamond at room temperature. *Phys. Rev. Lett.* 102, 057403 (2009).

## FIGURE LEGENDS

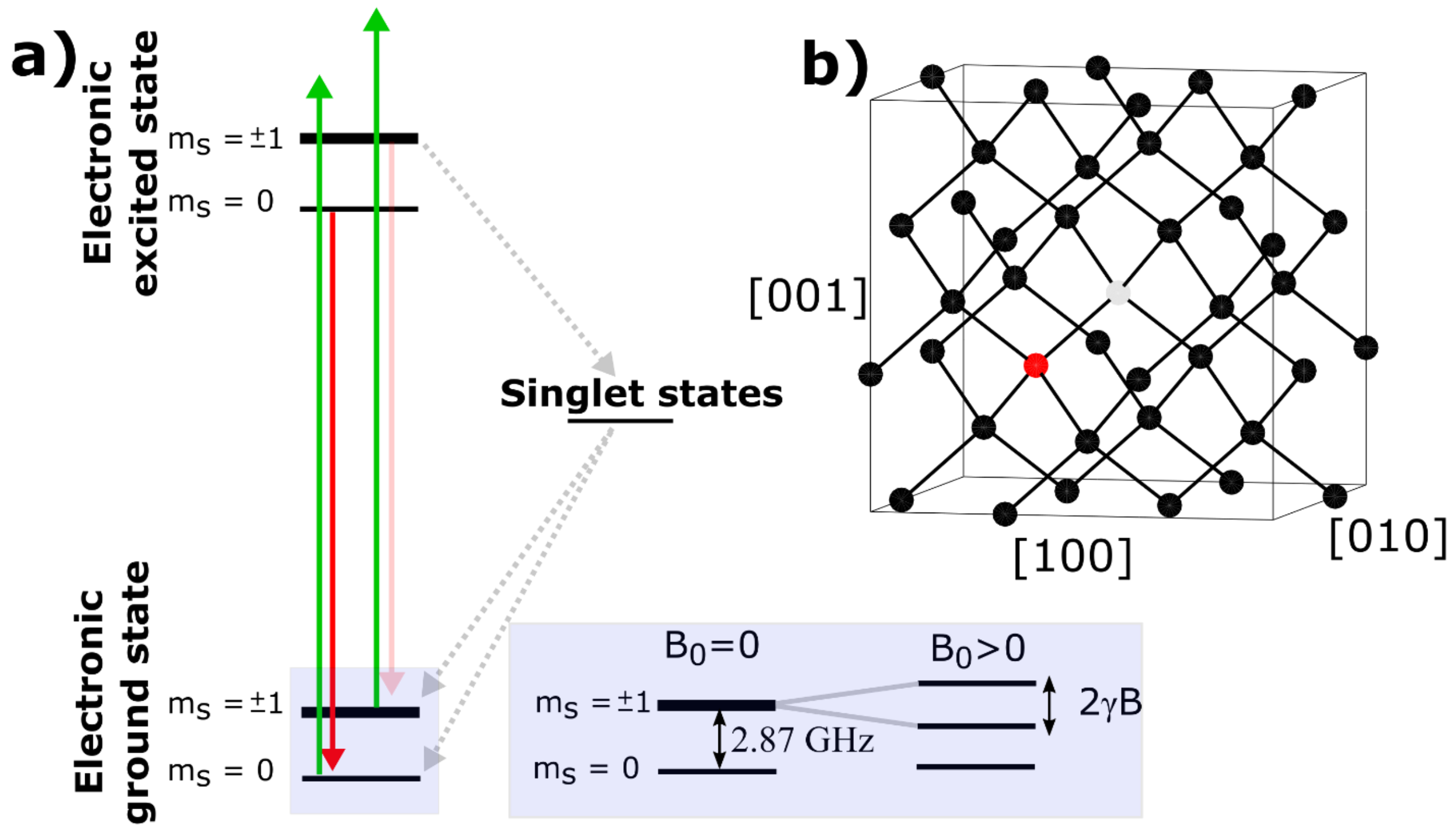

**Figure 1. NV center overview**. **a)** Simplified energy level diagram of NV center. A 532-nm laser (green arrows) can be used to optically excite an NV from the spin-triplet ground state to the spin-triplet excited state. From the excited state manifold, the NV can emit a read shifted photon (red arrows) by photoluminescence (PL). The $m_s$=±1 states are dimmer than $m_s$=0 due to a competing nonradiative decay (grey dotted arrows) to spin-singlet states that lie between the two spin-triplet states in energy. For sensing applications, one typically drives transitions between two spin sublevels (e.g., $m_s$=0 to $m_s$=+1 or $m_s$=-1) of the electronic ground state (inset). **b)** Sketch of NV center and surrounding diamond lattice. An NV is formed by substituting two neighboring carbon atoms (black circles) with a nitrogen atom (red) and a vacancy (light gray).

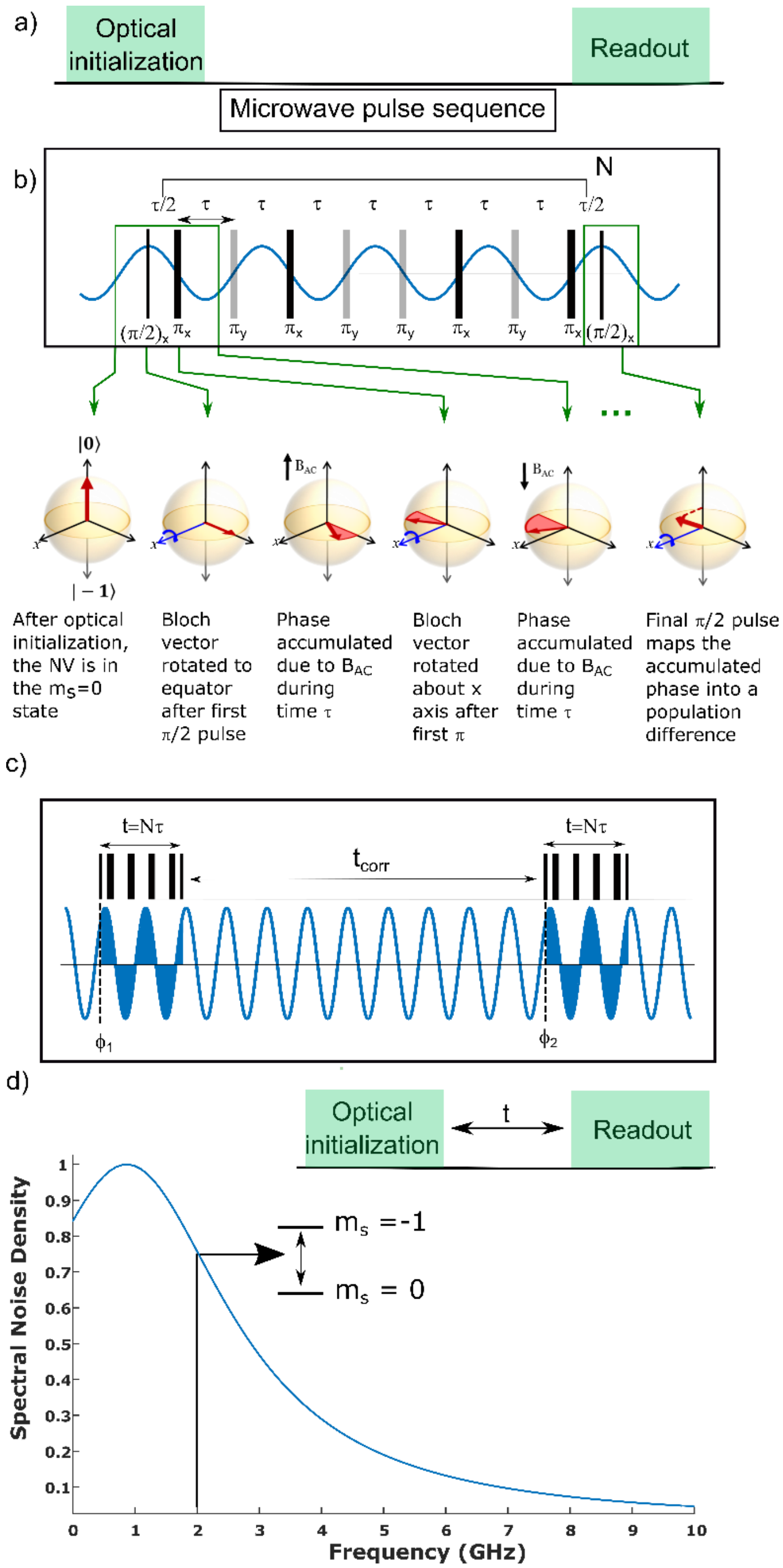

**Figure 2. Nuclear and electronic spin sensing schemes with NV centers. a)** All pulse sequences start with an optical initialization pulse, followed by a microwave pulse sequence for NV-spin control and sensing, and then end with an optical pulse for fluorescence readout. **b)** XY8-N dynamic decoupling pulse sequence is used as a sensing scheme for ac magnetic fields. The NV accumulates phase if half of the ac period coincides with the spacing $\tau$ of the $\pi$ pulses. This accumulated phase can be detected as a dip in the NV-T2 relaxation curve. The rotation axes of the XY8-N pulses are indicated as subscripts, i.e. $\pi_x$ and $\pi_y$. Bottom: the corresponding Bloch sphere picture is shown. **c)** The correlation spectroscopy pulse sequence consists of two XY8-N sequences with fixed $\tau$ at half of the ac field period. The timing $t_{corr}$ is swept, which correlates the phases $\phi$ of the ac magnetic field and generates oscillations in the readout data at the ac frequency. **d)** External nuclear spins (e.g., $Cu^{2+}$ or $Gd^{3+}$) are sensed by the spectral noise around the NV transition frequency (shown here at 2 GHz corresponding to an applied field ~ 300 G), which leads to an increase NV-T1 relaxation. The width of the spectral noise density $\delta$ depends on the rate of electronic spin flips (shown here for $\delta$=2 GHz).

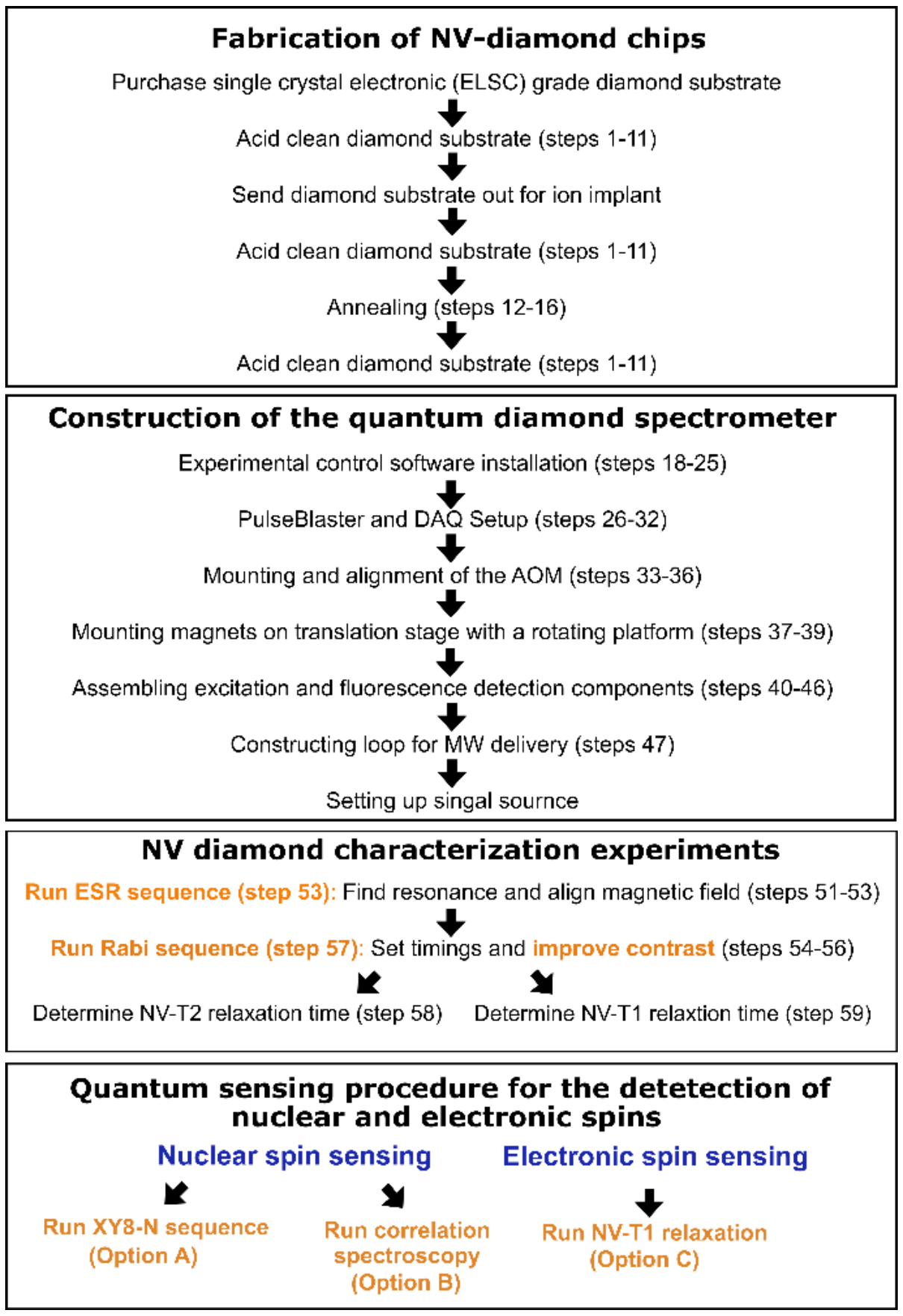

**Figure 3: Overview of the procedure:** The procedure is divided into four parts – (1) fabrication of the NV-diamond chip (steps 1-17), (2) construction of the quantum diamond spectrometer (steps 18-50), (3) NV diamond characterization experiments (steps 51-59) and (4) Quantum sensing procedure for sensing nuclear and electronic spins (step 60 with option A, B and C). Steps for daily use are marked in orange.

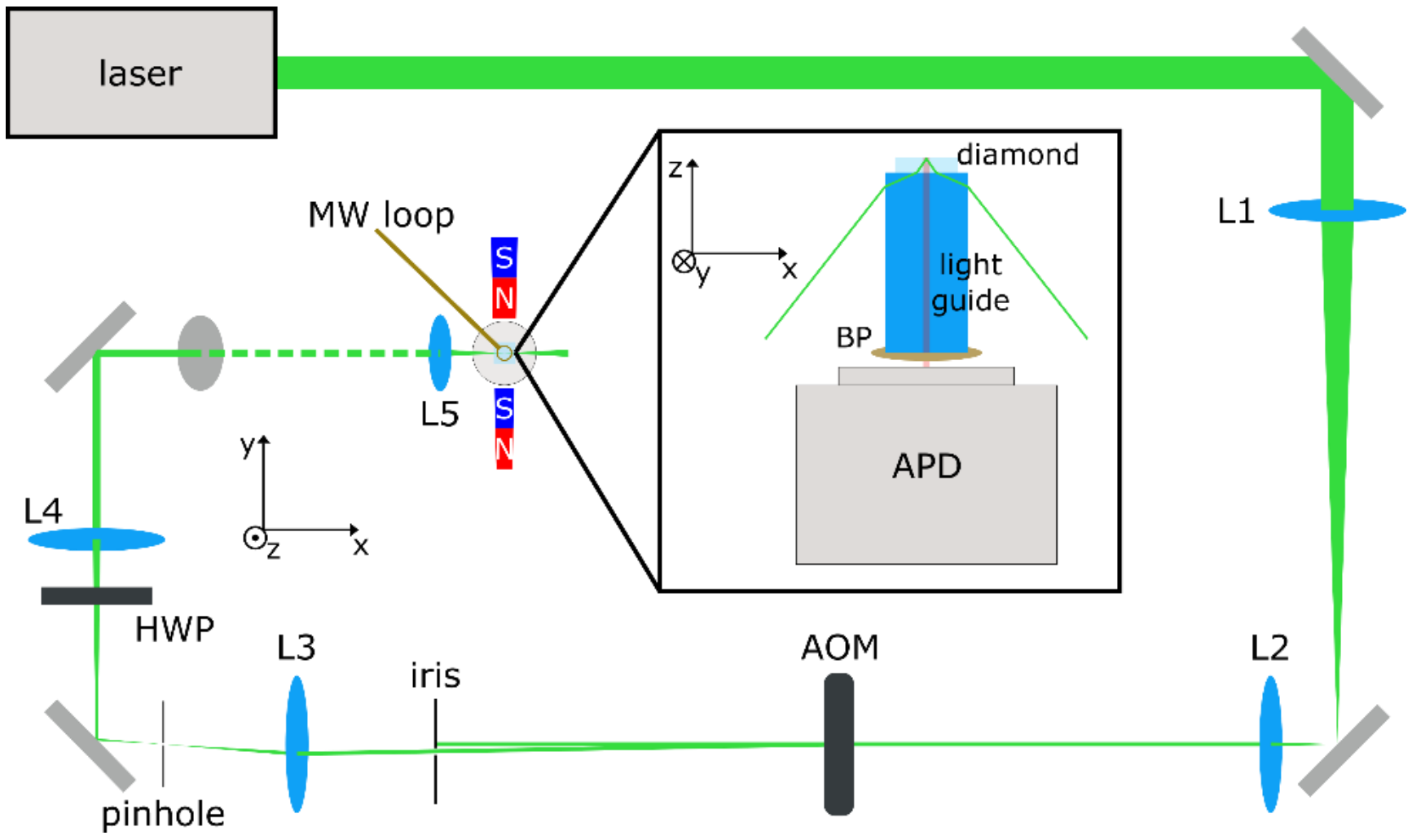

**Figure 4. Schematic of optical components.** The 532 nm laser beam is passed through a telescope consisting of lenses L1 (*f* = 400 mm) and L2 (*f* = 50 mm) in order to demagnify the beam waist by a factor of 8. The small collimated beam is then passed through the acousto-optic modulator (AOM). An iris placed after the AOM blocks the zeroth diffraction order and passes the first order. Additional optical extinction is achieved by focusing the beam through a 200-μm pinhole with lens L3 (*f* = 100 mm). A half wave plate (HWP) is used to rotate the polarization in order to maximize contrast in the NV measurement. The lens L4 (*f* = 200 mm) again collimates the beam. The last mirror in the setup is oriented such that the beam is sent out of the *xy* plane, now with Poynting vector parallel to the *xz* plane (indicated by the dashed green line). A short focal length lens L5 ($f$ = 30 mm) focuses the beam into the NV layer within the diamond under a total internal reflection geometry. Here an example is shown for a diamond chip with unpolished edges such that the light must be coupled in through a light guide. At this point in the setup a microwave (MW) loop and permanent magnet are positioned. The perspective on the bar magnet illustration is meant to convey that it is oriented at an angle of ~ 36 degrees relative to the *xy* plane. Inset: rotated view down the *y* axis of the excitation and detection. The laser is passed through the side of the light guide at an angle such that light is efficiently transmitted from glass into the diamond, then exhibits total internal reflection at the diamond-air interface. PL is collected through the light guide and passed through a band-pass fluorescence filter (BP) before being detected by the APD.

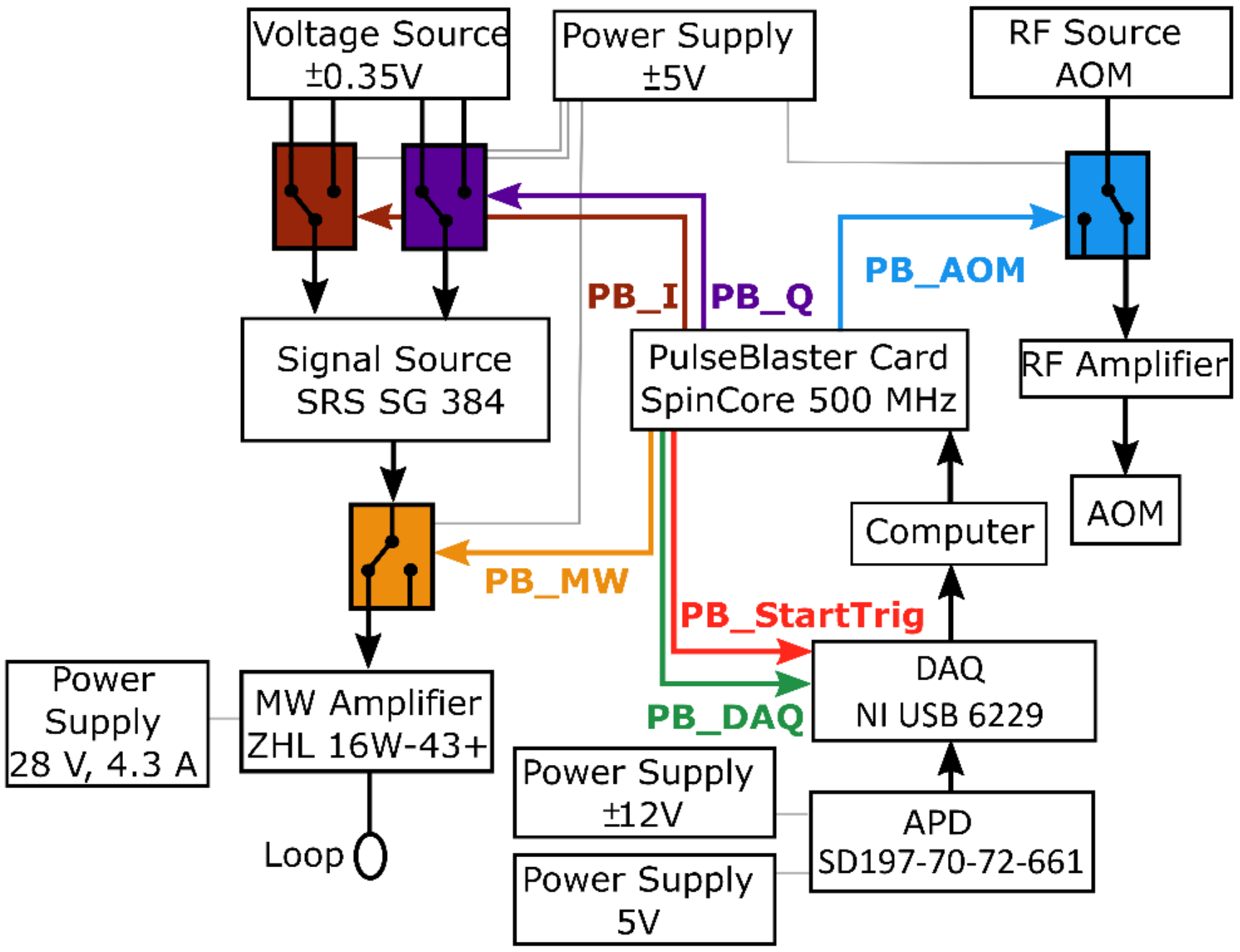

**Figure 5. Schematic of electronic components.** Colors indicate the six PulseBlaster TTL signal outputs used in the pulse sequences outlined in Figure 6. Arrows indicate the direction of information flow.

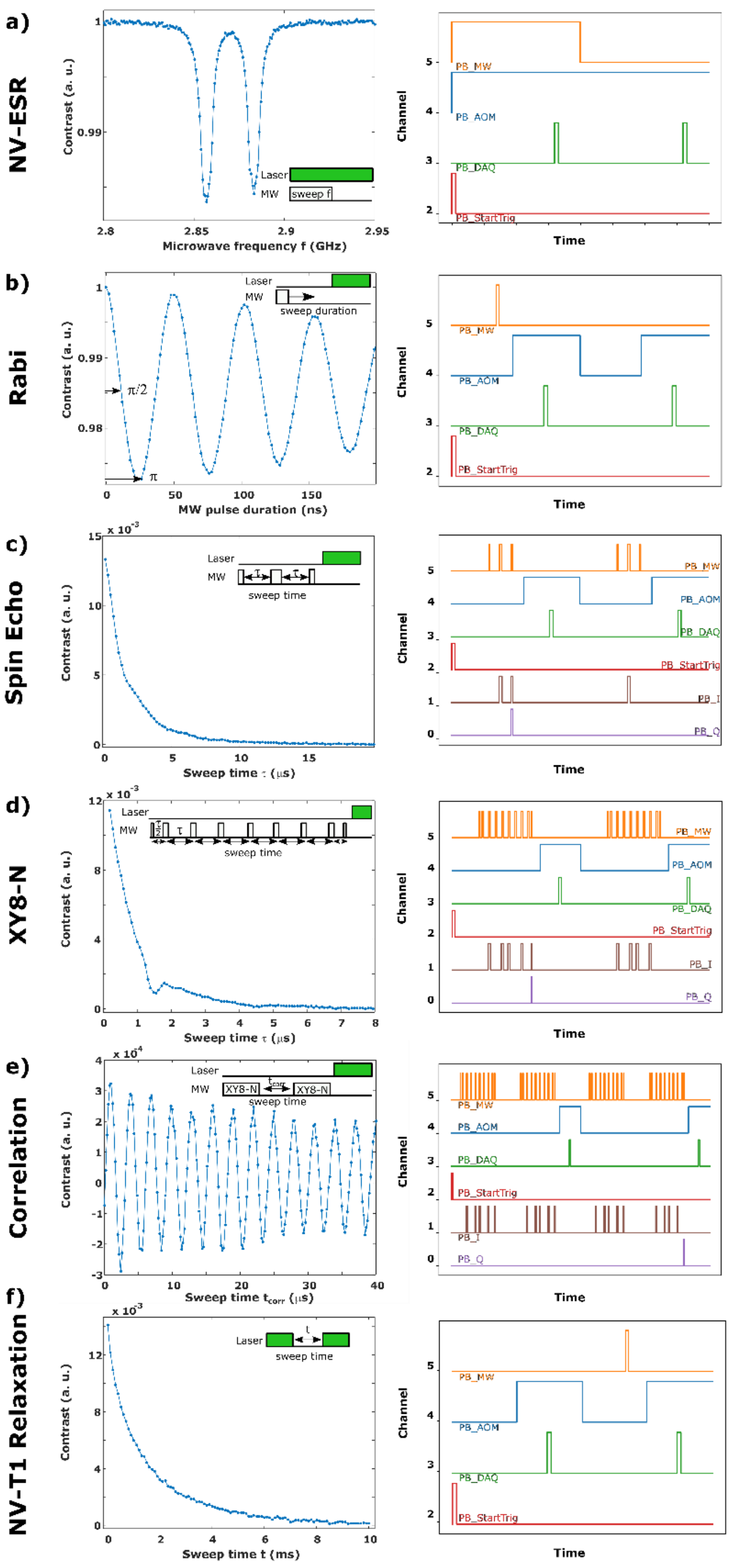

**Figure 6. Overview of the described pulse sequences.** Left side: Experimental data. Inset shows the basic idea of the pulse sequence. Right side: Pulse sequences and channels programmed into the PulseBlaster card. a) ESR experiment at earth magnetic field (1000 samples, 1 average). b) Rabi experiment at 311 G (1000 samples, 1 average). c) Spin echo experiment at 311 G and 24-ns π pulses (10000 samples, 1 average). d) XY8-1 experiment at 311 G and 24-ns π pulses (10000 samples, 1 average). e) Correlation spectroscopy experiment at 311 G, 24-ns π pulses and τ set to 1.5 μs (10000 samples, 1 average). f) NV-T1 experiment at 311 G and 24-ns π pulse (10000 samples, 1 average).

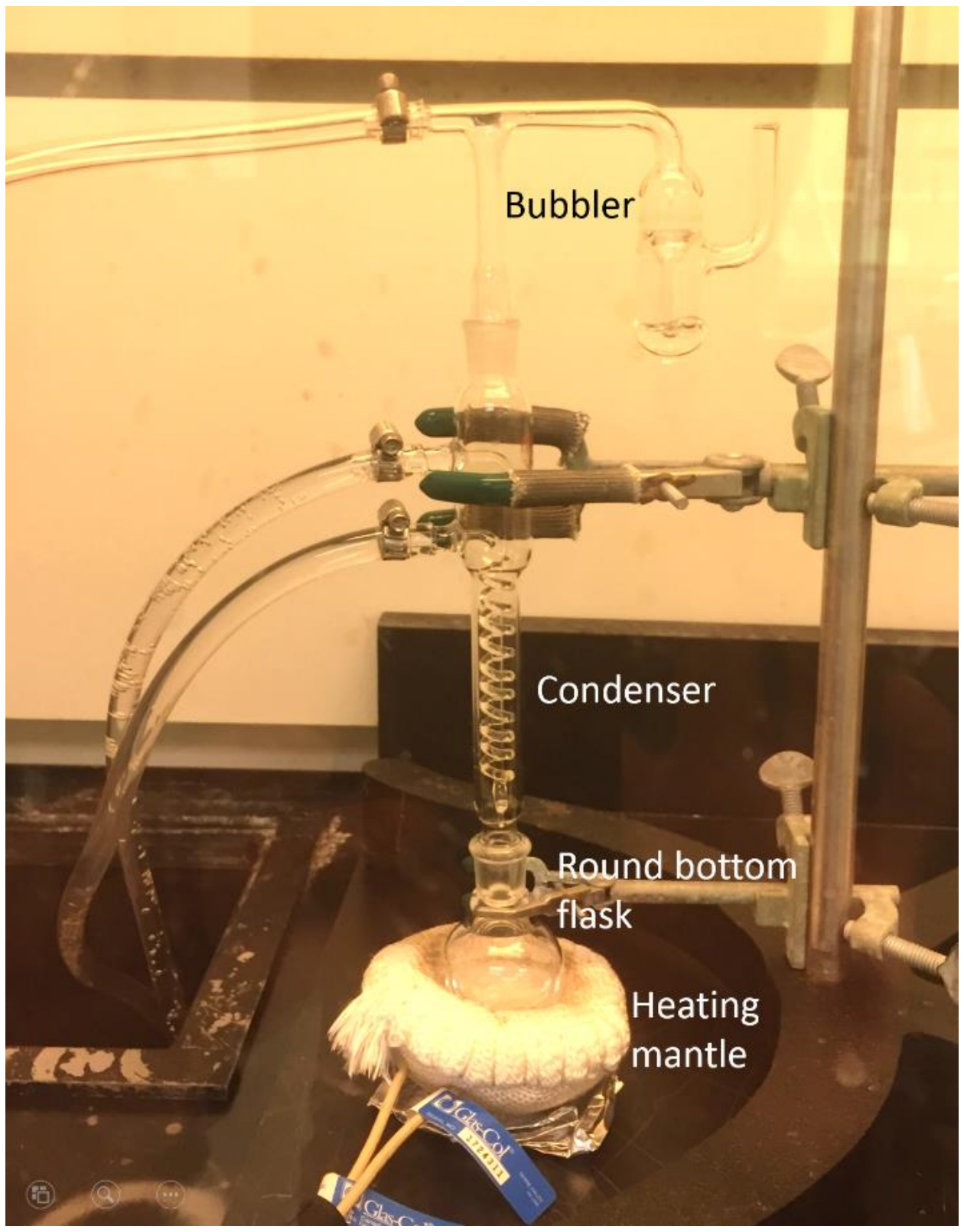

**Figure 7. Setup of the acid clean glassware.**

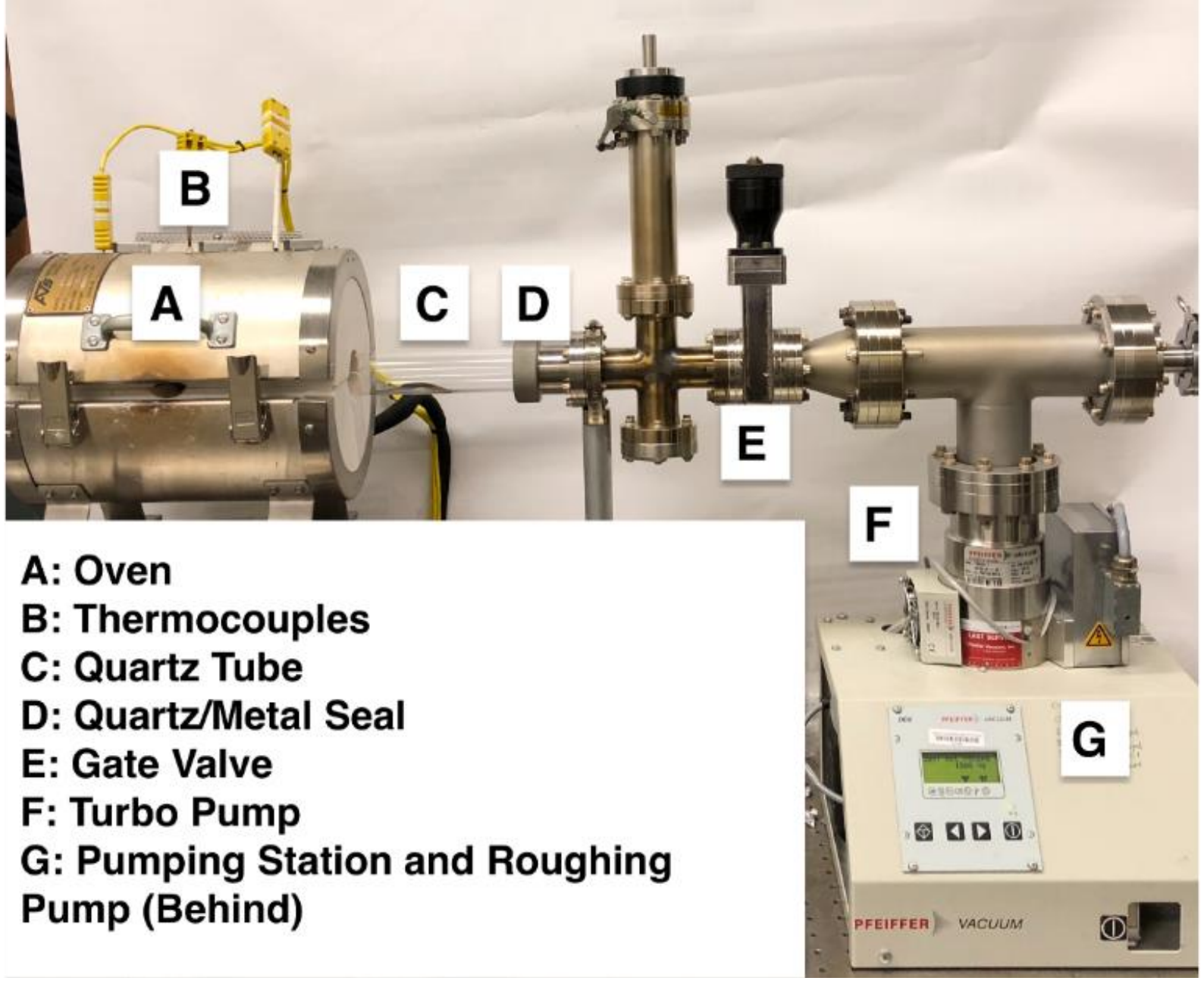

**Figure 8. Vacuum furnace system for diamond annealing.**

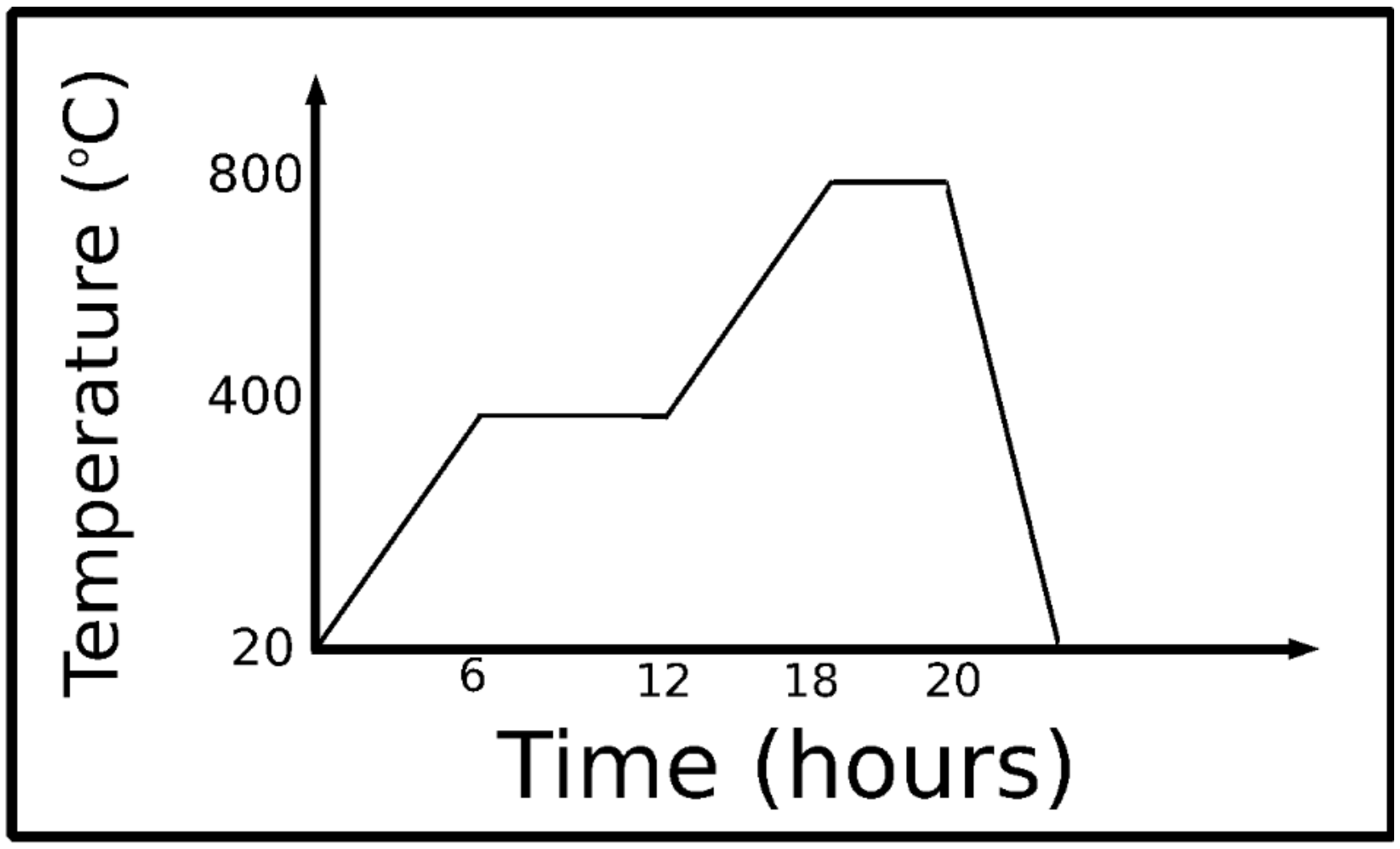

**Figure 9. Example heating profile for diamond annealing.**

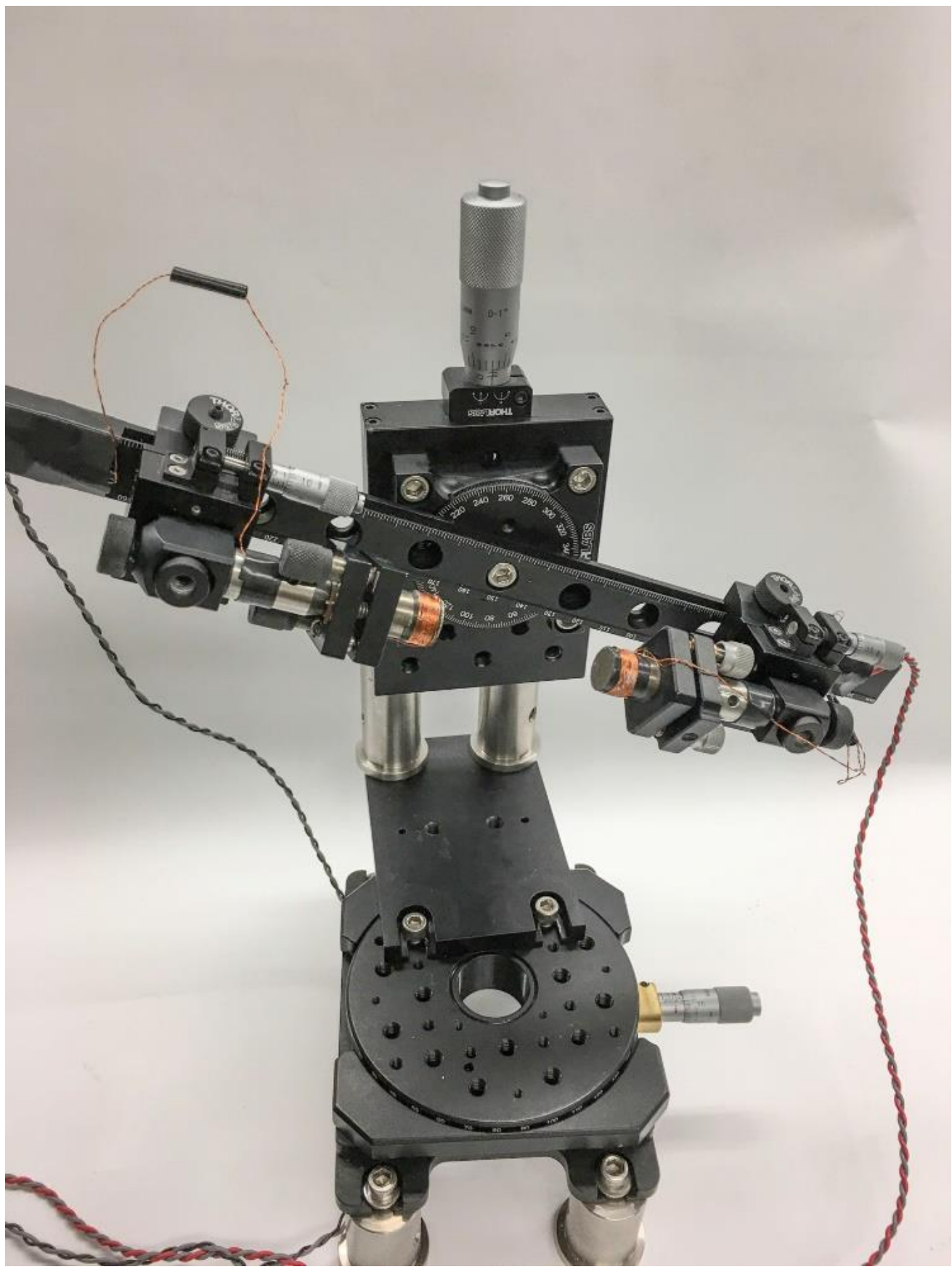

**Figure 10. Picture of the magnet mount with x,z rotation for $B_0$ field alignment.** Coils have been wound around the permanent magnets for small magnetic field adjustments, not described in the protocol.

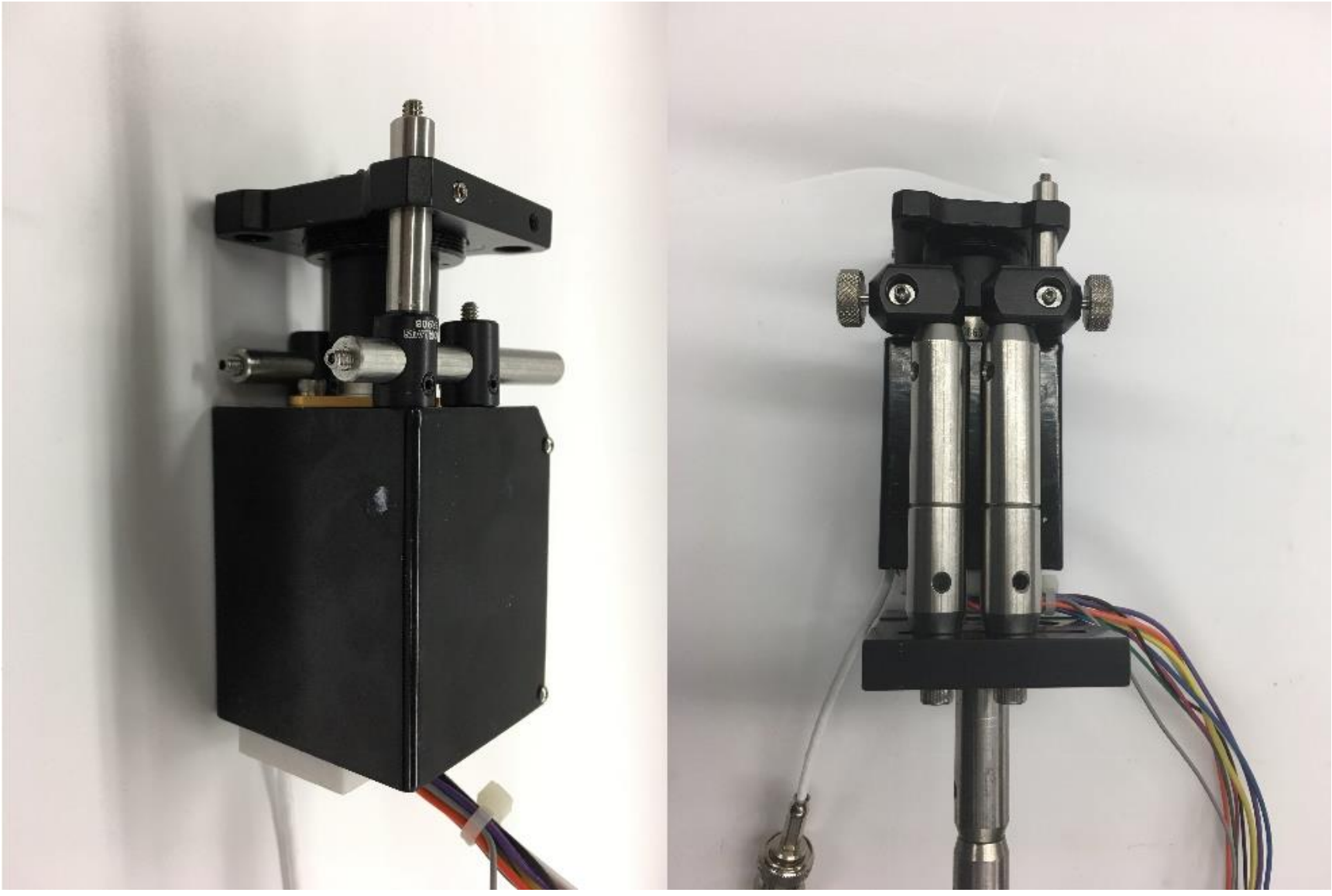

**Figure 11. Photo of the avalanche photodiode (APD) holder.**

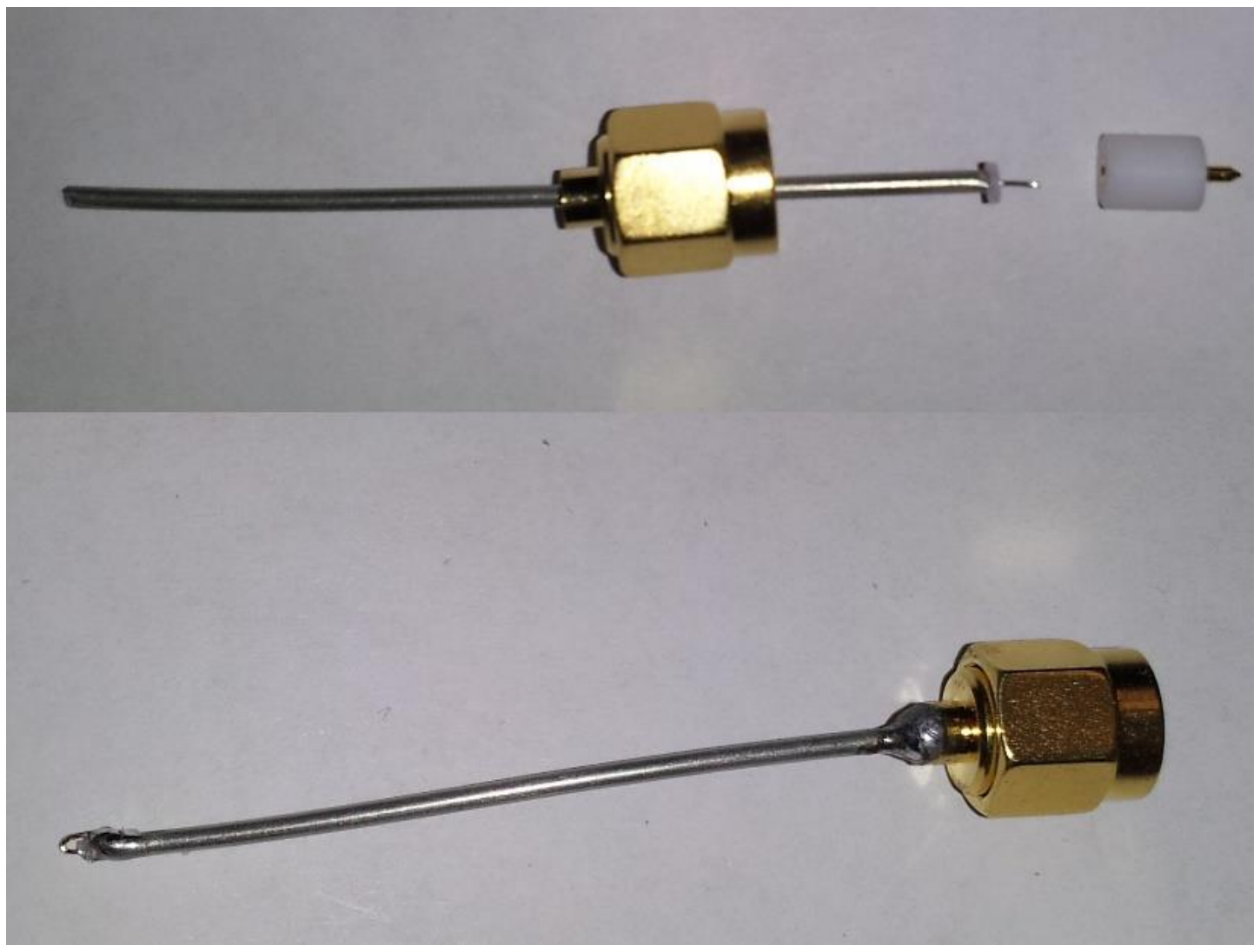

**Figure 12. Photo of the microwave delivery loop.** Top: Individual parts of the MW delivery loop before assembly. Bottom: assembled MW loop.

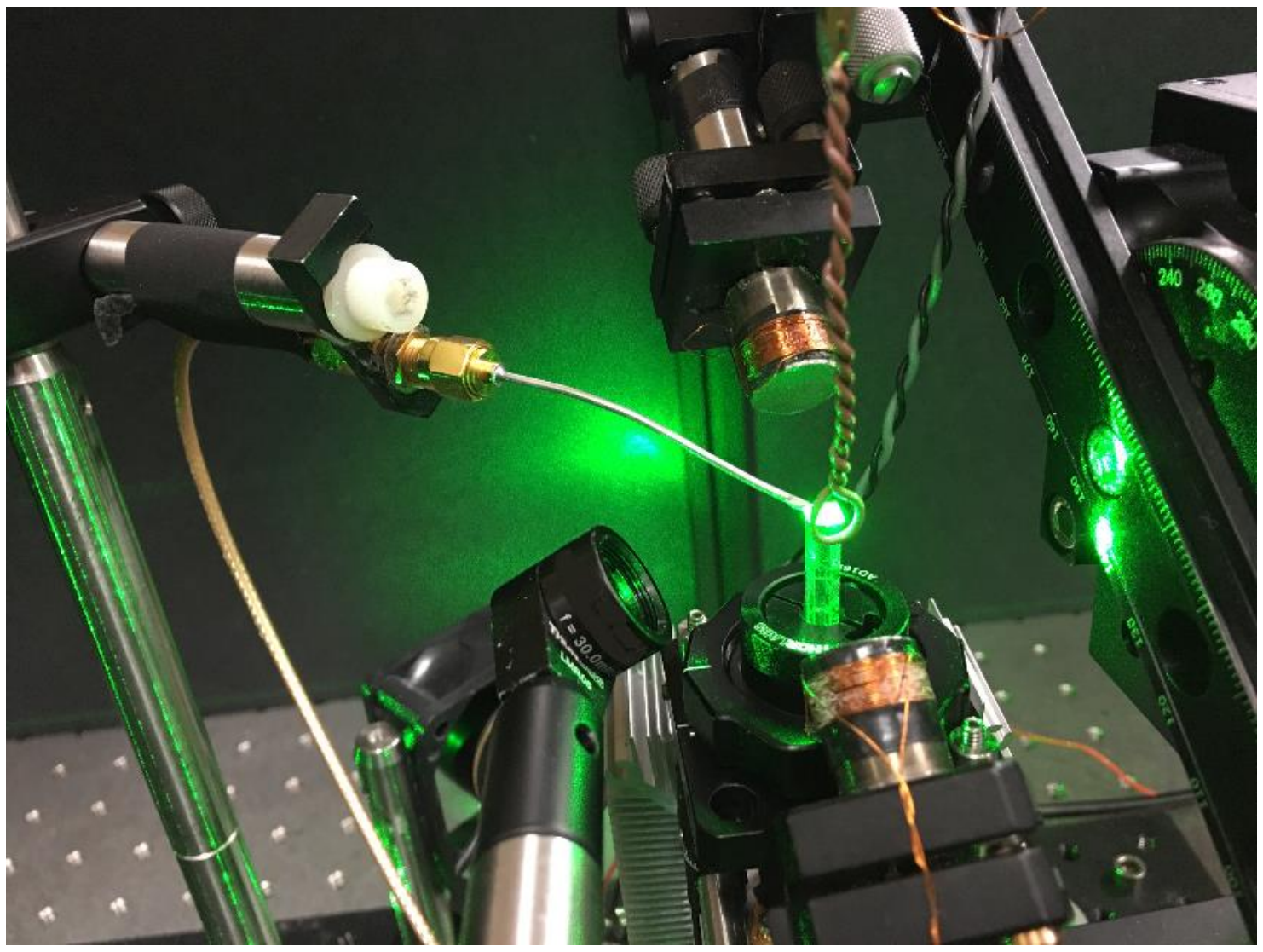

**Figure 13. Photo of the quantum diamond spectrometer.** On the right, the magnet mount is shown (coils have been wound around the static magnets for small magnetic field adjustments, not described in the protocol). In the center of the magnets is the light guide with the diamond (bright green because of illumination). From the top, an RF loop is positioned near the diamond for sensing a generated signal. At the center bottom, the 30-mm lens for focusing the laser beam can be seen. The mount for the microwave delivery loop is located at top left of the picture.

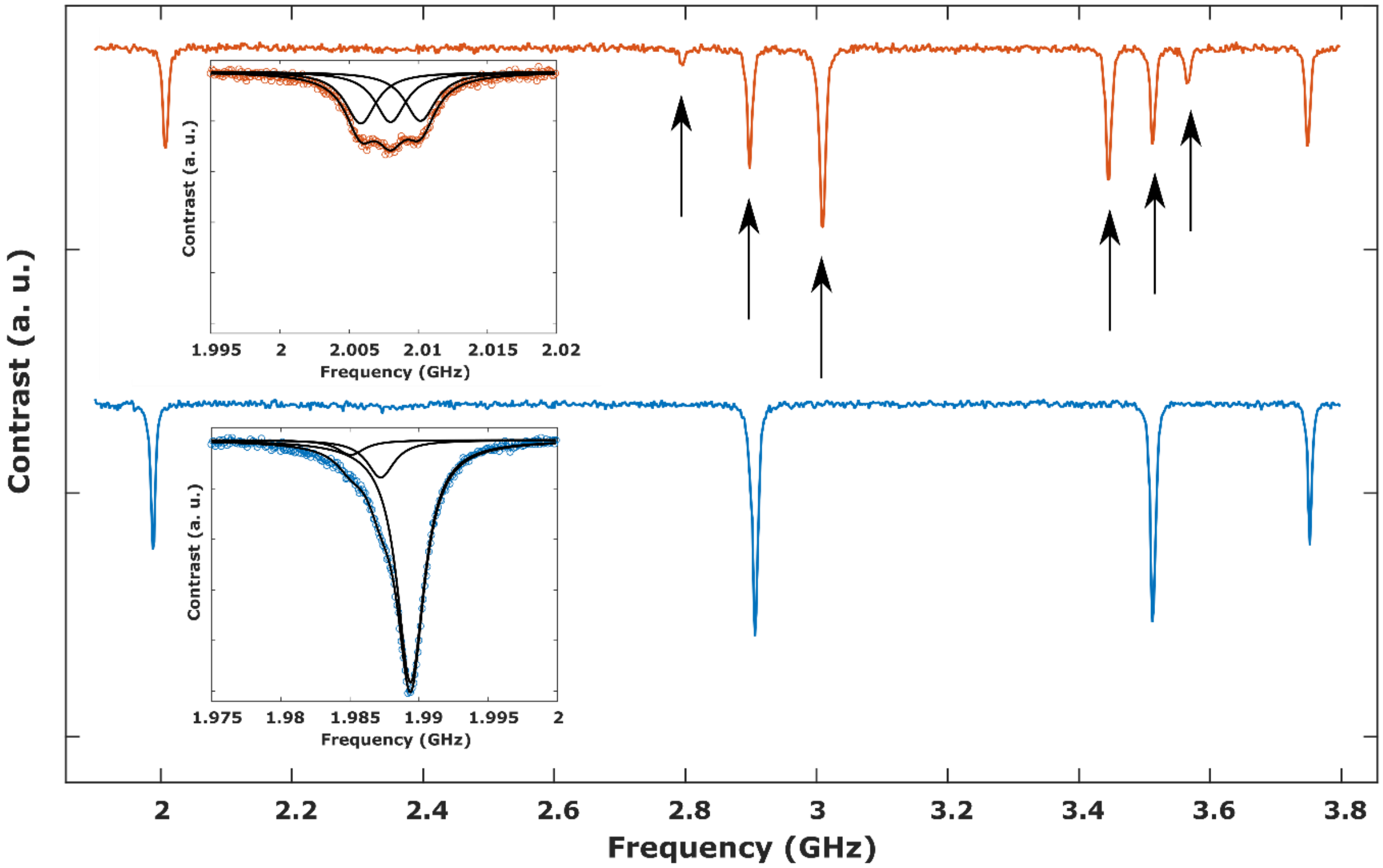

**Figure 14. Magnetic field alignment.** Top: Magnetic field is not aligned along an NV axis. All 2x4 NV-ESR transitions can be observed. The arrow-marked resonances must overlap for alignment. The inset shows that the three hyperfine transitions have equal amplitudes in the misaligned case. Bottom: Magnetic field is aligned along one NV axis. In this case, the magnetic field projection on the other three NV axes is identical and the resonances overlap. Now the hyperfine pattern is polarized as shown in the figure inset.

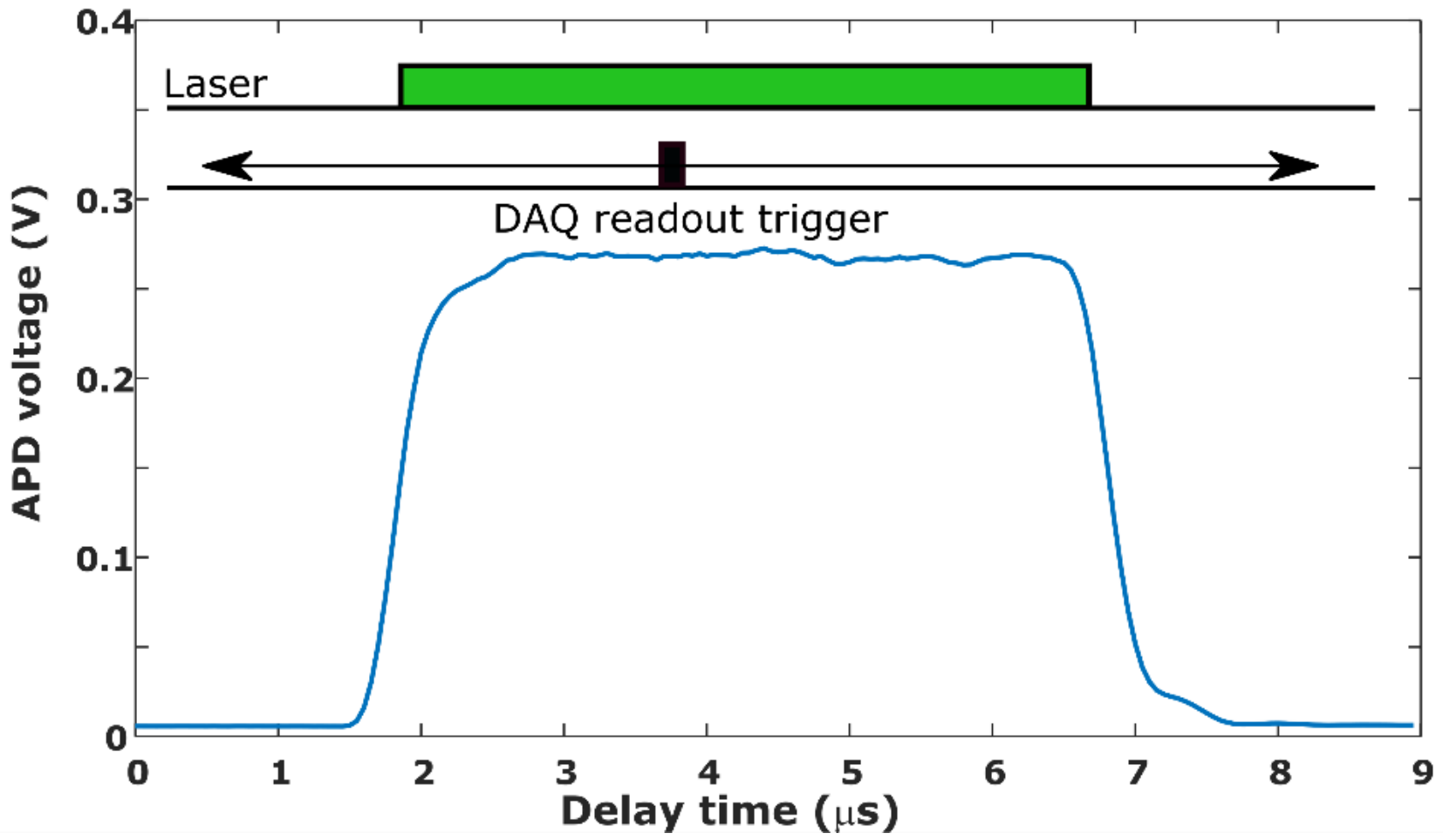

**Figure 15. Set timing between the readout pulses.** DAQ readout pulse (black box) is moved in time relative to the AOM pulse (green box). The data shows the temporal overlap and the optimal timing when the maximum photovoltage is reached (around 2.0 µs in our case).

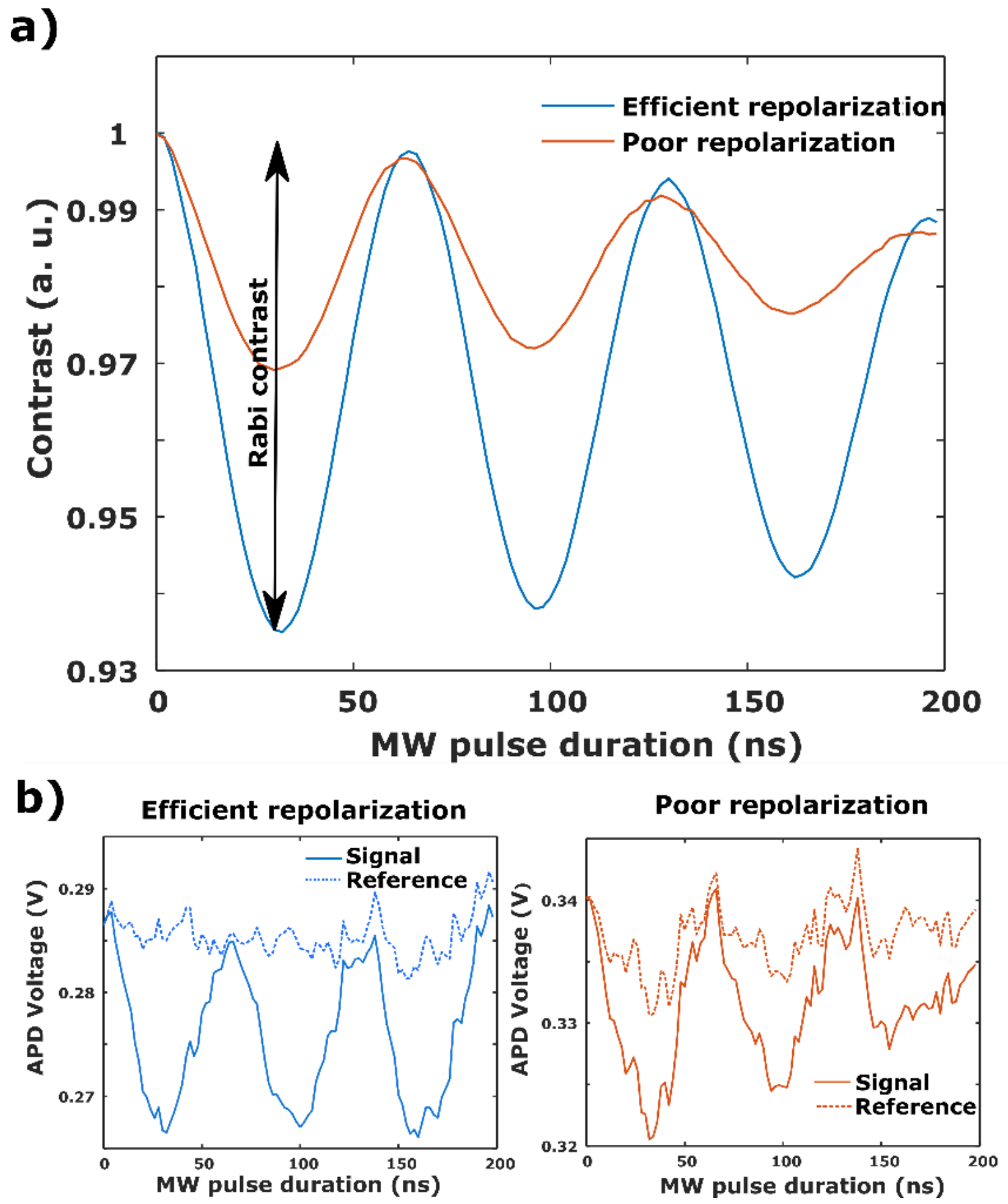

**Figure 16. Rabi contrast. a)** Rabi data shown with efficient and poor repolarization of the NVs. Poor repolarization results in a decrease in contrast. **b)** Raw data shown for these two cases. Left: In the case of efficient repolarization, the reference channel does not show Rabi oscillation. Right: In the case of poor repolarization, the NVs have not been reset and the Rabi oscillation occurs also in the reference. To acquire the poor repolarization data, we increased the laser beam size.

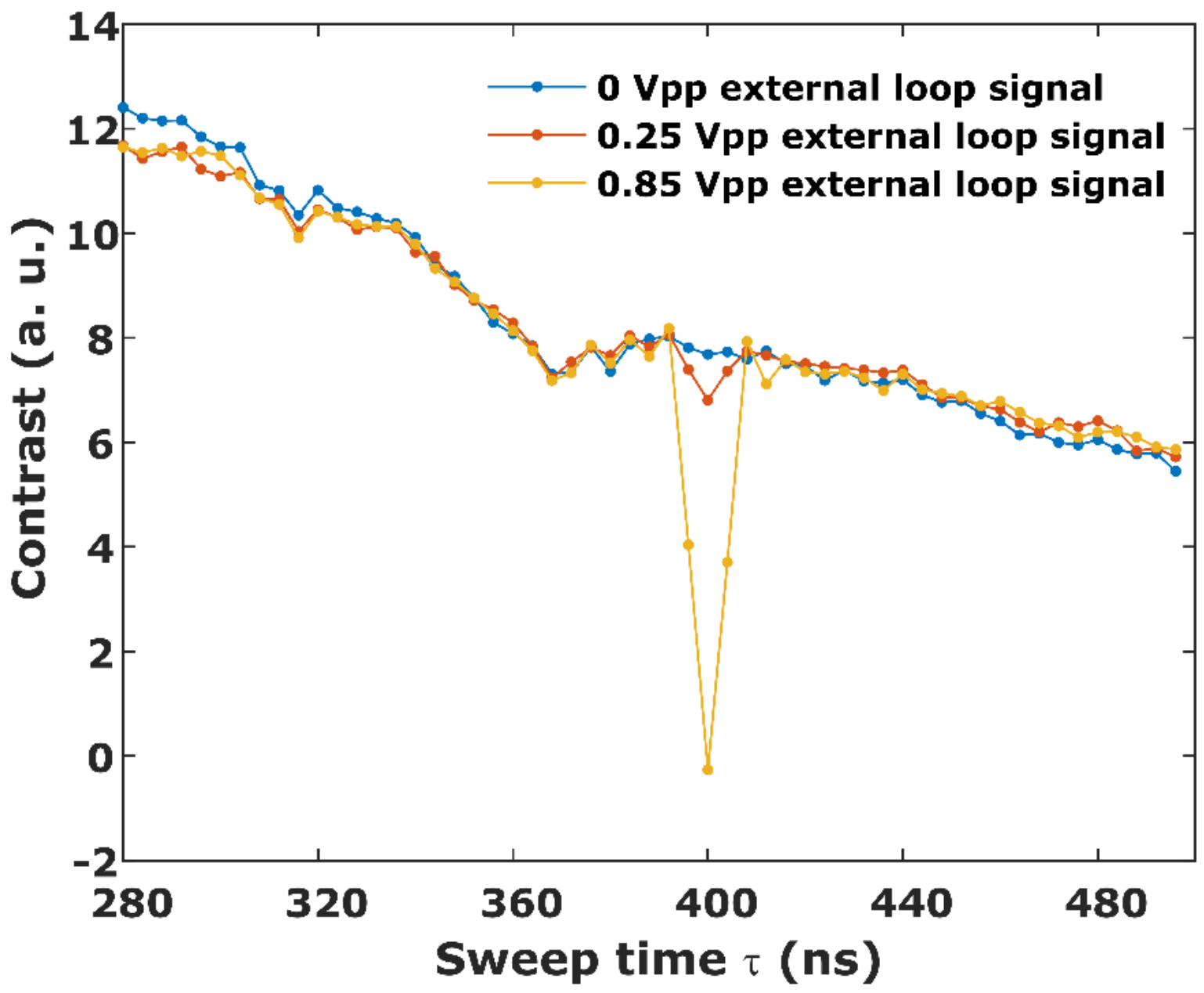

**Figure 17. Detection of an external AC signal from an RF loop at 1.25 MHz with an XY8-12 dynamic decoupling sequence.** Keep in mind that the depicted voltages are relative values and depend on the loop design, orientation, distance to the diamond etc.

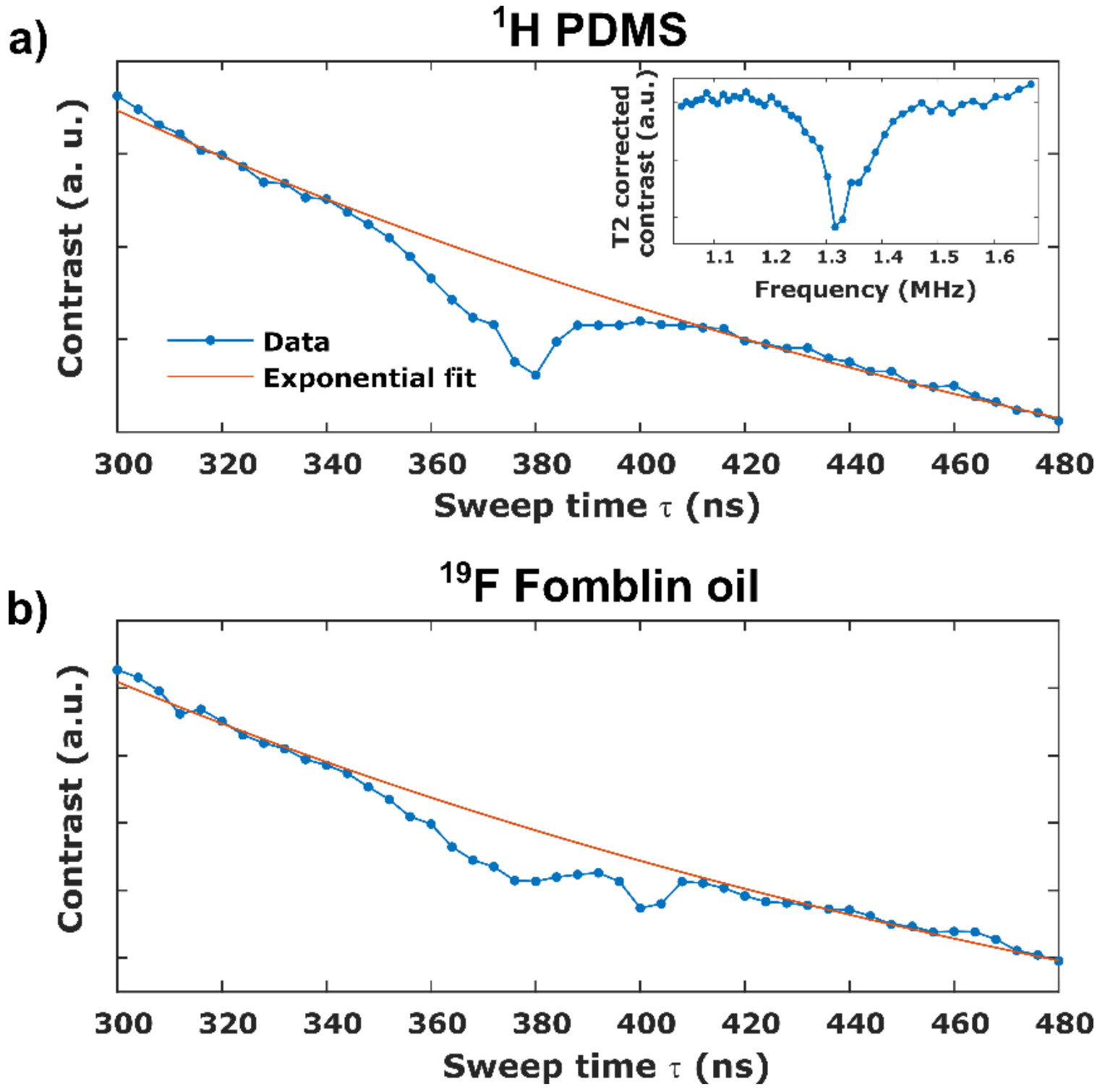

**Figure 18. Nuclear spin sensing with a dynamic decoupling sequence at 311 G. a)** $^1$H NMR detection of PDMS on the diamond surface. Red line is a mono-exponential fit of the NV-T2 decay. Inset: Decay-subtracted data plotted on a frequency axis. **b)** $^{19}$F NMR detection in Fomblin oil. The broad feature around 378 ns is due to background protons. Red line is a mono-exponential fit of the NV-T2 decay. In both cases 40-ns π pulses, 10000 samples, 20 averages, 4 ns sampling intervals, and $N$ = 12 are used.

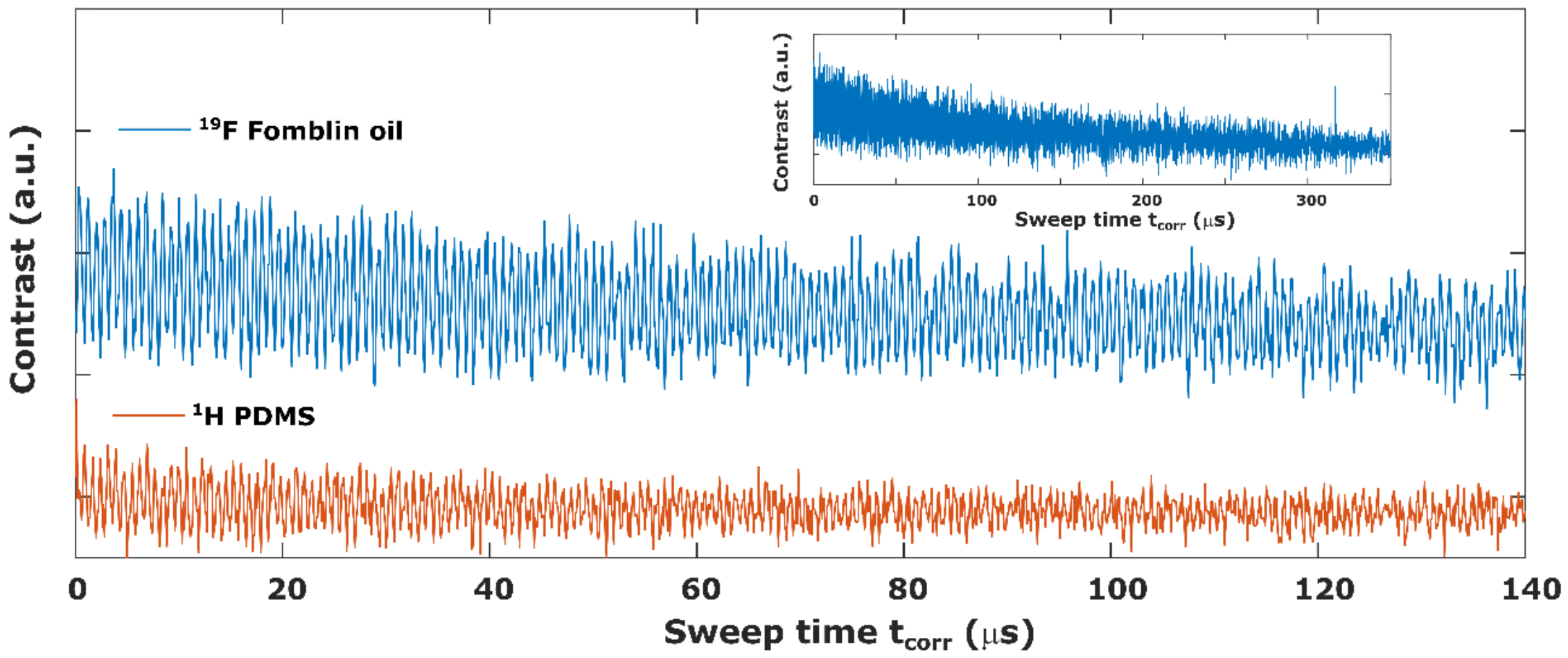

**Figure 19. Correlation spectroscopy for NV-NMR sensing at 311 G.** $^{19}F$ NMR detection in Fomblin oil (blue) and $^{1}H$ NMR detection in PDMS (orange). The oscillations are at the nuclear Larmor frequencies. Inset shows the full recorded data over 350 μs for the Fomblin oil. In both cases 40-ns π pulses, an XY8-4 pulse sequence, 100-ns sampling intervals, and 10000 samples were used. The $^{19}F$ data is averaged 2 times, the $^{1}H$ data is averaged 10 times.

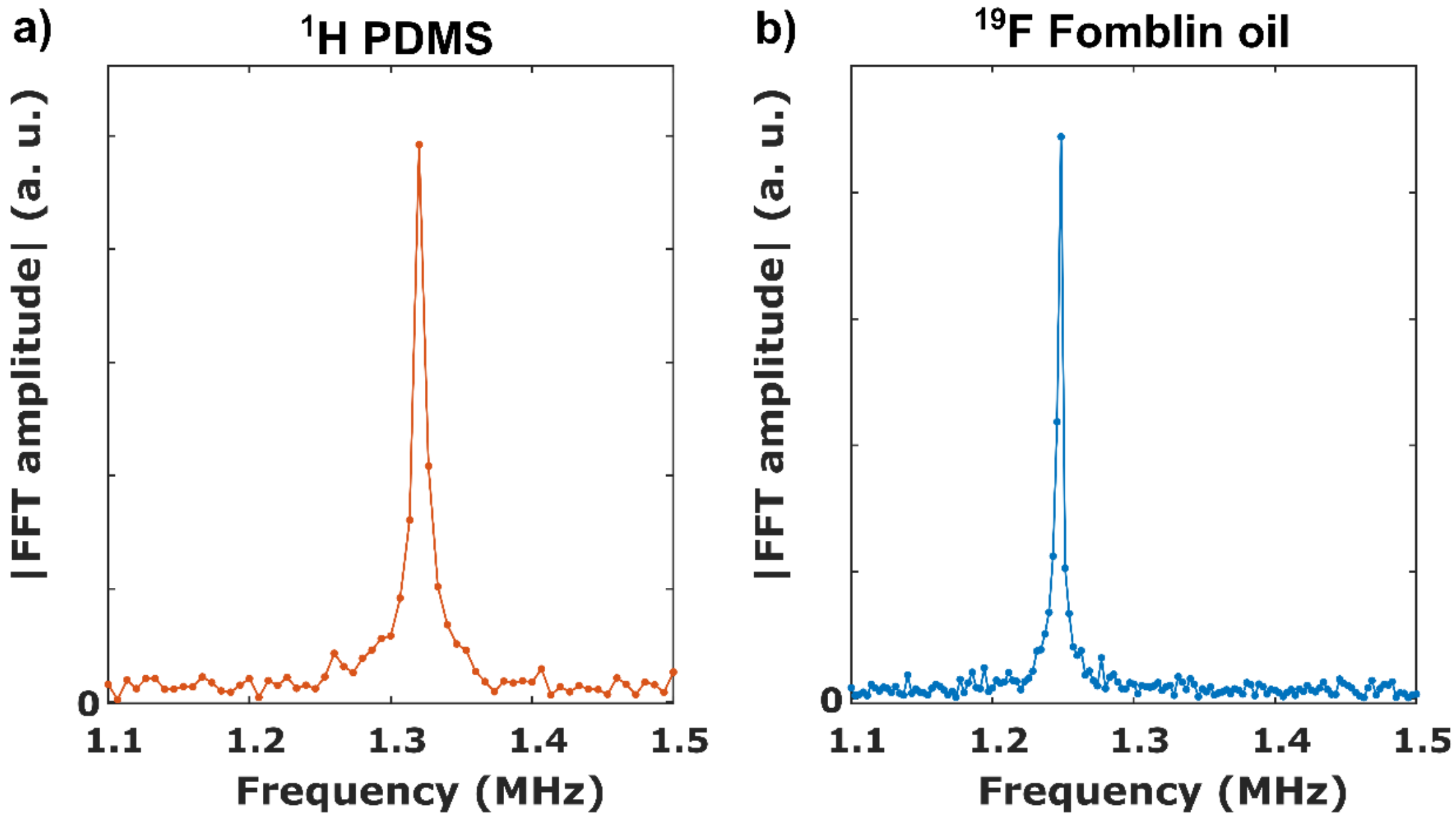

**Figure 20. Fourier transform of the NV-NMR correlation signal at 311 G.** Fourier transform data of time series displayed in Figure 19. **a)** $^{1}H$ in PDMS. **b)** $^{19}F$ in Fomblin oil. Note that the sampling rate is different in both cases.

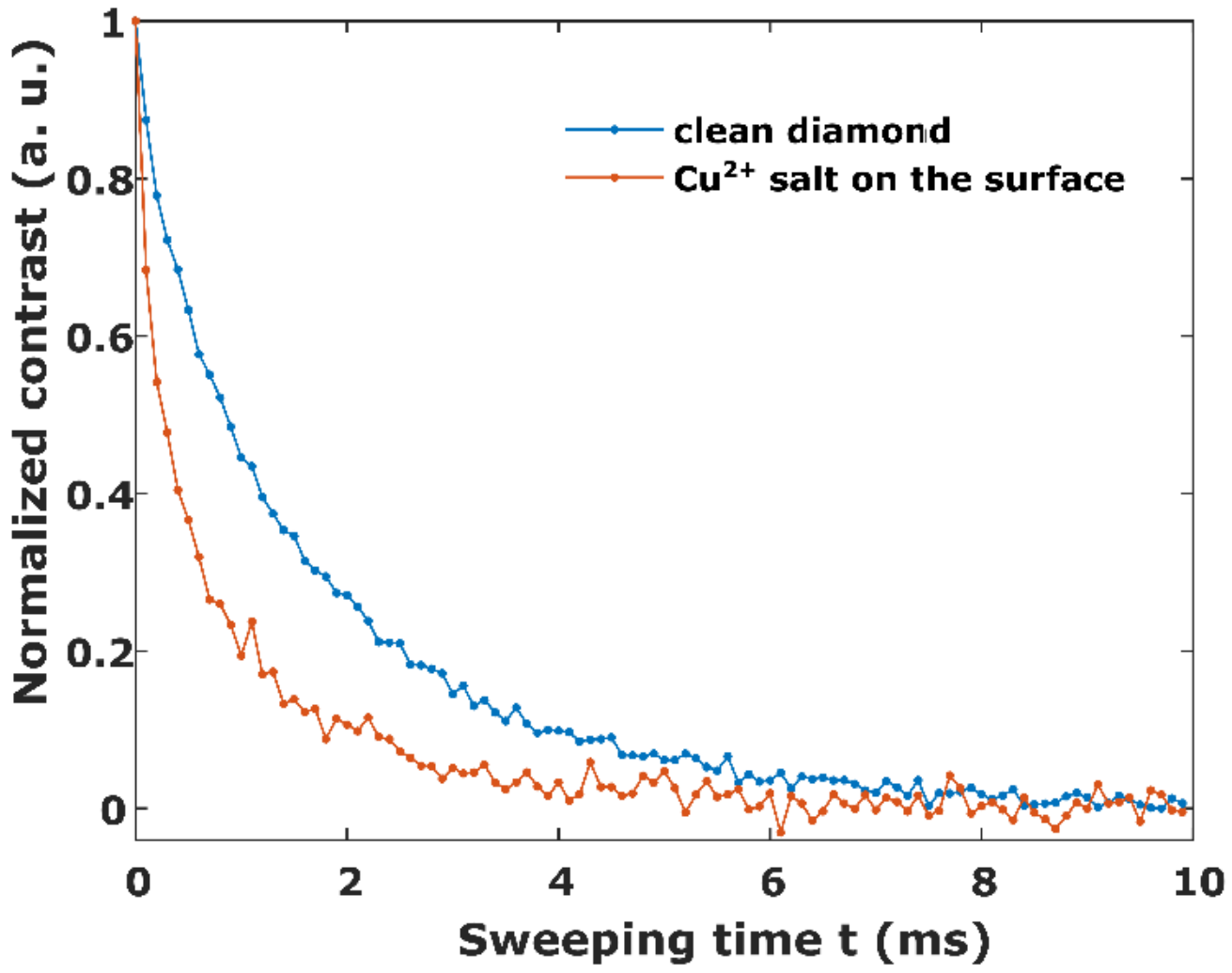

**Figure 21. Electronic spin detection with NV-T1 relaxometry.** NV-T1 decay of clean diamond (blue) compared to that with electronic spins ($Cu^{2+}$) on the diamonds surface (orange). Data are normalized to 1 at zero decay time. In both cases the π pulse duration is 36 ns and the number of samples is 1000. The number of averages is 4 and 40 for the clean and $Cu^{2+}$ case, respectively. The $Cu^{2+}$ data had to be averaged more due reduced contrast.

## BOXES

**---------------------------------- Box 1. Using the qdSpectro package ------------------------------------------------**

Once the qdSpectro package is downloaded, the working directory should contain the files listed below.

**User-input configuration files:**

- connectionConfig.py – configuration file for PulseBlaster, DAQ and SRS connections to the PC. This file is edited by the user, as directed in the protocol, before any of the package scripts are run.
- __config.py – experiment configuration files. Each experiment has its own configuration file (e.g., the configuration file for the ESR experiment is ESRconfig.py), which consists mainly of a “user input” section, where the user can edit experimental parameters and configure options relating to how the data will be processed, plotted and saved.

**Main control and auxiliary libraries:**

- mainControl.py – all experiments described in this protocol are run from the mainControl.py script, which takes an experiment-specific configuration file as an argument. Given the input parameters defined in the configuration file, mainControl.py runs the experiment, generates plots and saves the results.
- DAQcontrol.py – contains functions that configure the DAQ
- SRScontrol.py – contains functions that control the SRS signal generator
- PBcontrol.py – contains functions that configure and program the PulseBlaster card
- sequenceControl.py – contains functions that create the pulse sequences required to run the experiments in this protocol

Before running any experiments with qdSpectro, the user should read the readme file provided with the package, where any upgrades and patches will be described, and edit connectionConfig.py, as directed in this protocol.

To run an experiment with qdSpectro:

i. Open the relevant ___config.py file in notepad++. Read the description of the experimental parameters and data-processing options defined in this script.
ii. Edit the experimental parameters and configure the data-processing options in the “User Inputs” section of this script, as required.
iii. To run the experiment, open a windows command prompt and, from the working directory, run:
iv. `python mainControl.py __config`
v. To quit an experiment before it finishes running, press Ctrl+C.

A note on units: units for user-input parameters (entered in step ii above) are specified in the comments accompanying the user-input section of the ___config.py files. For added clarity, we also note here that the default unit for time variables in version 1.0 of the qdSpectro package (the current version at the time of writing) is nanoseconds. The user may either enter time variables in nanoseconds or use one of the following unit multipliers: `ns = 1`, `us = 1e3`, `ms = 1e6`. For example, if setting the variable `endTau` to 10 microseconds, the user may either enter `endTau = 10000` or `endTau = 10*us` in the user-input section of the relevant ___config.py file. The latter format is used throughout the instructions given in this paper. For completeness, we also note that, in version 1.0 of qdSpectro, microwave frequencies are entered in hertz (e.g. if setting the variable `startFreq` to 2.7GHz, the user should enter `startFreq=2.7e9`) and microwave powers in dBm (e.g. if setting the variable `microwavePower` to 0 dBm, the user should enter `microwavePower=0`). Users running a different version of qdSpectro should refer to that version's readme file for any version-specific user-input instructions.

**-------------------------------- Box 2. Turning PulseBlaster Channels On and Off ----------------------------------**

At several points throughout the setup of the apparatus described in this protocol, PulseBlaster (PB) channels have to be toggled on and off (e.g., to test if the AOM is working and well aligned). Use the script togglePBchan.py to toggle any PulseBlaster channel. In a Windows Command Prompt, start Python from the working directory and run `togglePBchan.py`. A key will be displayed relating a letter to a PulseBlaster channel, as below:

- A = PB channel connected to the switch on the RF source driving the AOM
- M = PB channel connected to the switch on the MW output of the SRS signal generator
- I = PB channel connected to the switch on the I input of the SRS signal generator
- Q = PB channel connected to the switch on the Q input of the SRS signal generator
- D = PB channel connected to the DAQ’s sample clock input
- S = PB channel connected to the DAQ’s start trigger input

To turn on a given PB channel, type in the letter corresponding to it and press Enter. To turn the channel off, type in the same key again. Check the functionality by measuring the PB output voltage on an oscilloscope.

**-------------------- Box 3. Connecting Signals to the Minicircuits ZAWSA-2-50DR+ Switch--------------------**

The RF switch used in this protocol directs power applied to its RF input to one of its two RF outputs, depending on whether the signal at its TTL input is high or low (refer to datasheet for threshold high and low voltages). The version of the Minicircuits ZAWSA-2-50DR+ switch used in this protocol, for example, directs power to RF Output 2 when the TTL control is high and to RF Output 1 when the TTL control is low. We terminate RF Output 1 with a 50Ω load and use the TTL control to switch on and off RF Output 2 (i.e., Output 2 is on when the TTL input is high and off when the TTL input is low). To avoid damaging the switch, the user should ensure that the unused RF output is terminated with a matched (50Ω) load and that the switch is powered with the appropriate DC supply voltages before connecting an RF signal to the input port.

**---------------------------------- Box 4 Amplifier Turn-On and Turn-Off Order ----------------------------------------**

When testing and operating MW and RF amplifiers, the user must read and follow all amplifier safety guidelines given by the manufacturer of your amplifier. Some general good-practice guidelines are given below, but these do not replace manufacturer safety guidelines and should only be followed if they do not contradict manufacturer instructions/guidelines, which always take precedence. In general, amplifier turn-on and turn-off order is as described below:

**Amplifier turn-on order:**

1. Terminate the amplifier – either with a matched load or with the load it will be operated with (e.g., antenna loop). If the load is not matched, as is the case with the loop antenna we use in this protocol, it is good practice to place a circulator or isolator directly after the amplifier to prevent the formation of standing waves. Ensure that the isolator/circulator is connected in the right direction and can handle signals at the power and frequency range output from your amplifier.
2. Once the amplifier is terminated, turn on its voltage supply to power it with the required DC voltages.
3. Lastly, connect the MW source to the input of the amplifier and turn the source on, ensuring that you do not exceed the rated input power of the amplifier.

**Amplifier turn-off order**

1. Turn off the MW source.
2. Turn off the amplifier's DC power supply.
3. You may now disconnect the load from the output of the amplifier.

**-----------------Box 5. Detecting external AC magnetic fields with XY8-N for calibration --------------------**

Often it is useful to sense synthetic signals generated by an RF loop, e.g., to calibrate the experiment and/or for troubleshooting if nuclear spin signals cannot be observed. Follow this procedure to check that the XY8-N pulse sequence is working correctly:

1. Determine the NV resonance frequency according to step 53.
2. Repeat the Rabi experiment (step 57) to determine the $\pi/2$ and π durations.
3. Open XY8config.py in Notepad++. Set the MW frequency to the resonance frequency of the NV. Set the MW power and the π-pulse duration to the values obtained in the Rabi. The starting value of the swept time should be longer than the π-pulse duration (e.g., for `t_pi=24*ns`, we use `startTau=280*ns`) and conclude the sweep around 500 ns. Use maximum allowed sampling rate (i.e. 4 ns). Use `N` = 1 to begin with. Set `t_readoutDelay` as determined in step 54. Run `python mainControl.py XY8config`. One should see a decay as shown in Figures 6d and 17.
4. Place a wire loop next to the diamond (e.g., see RF loop in Figure 13) and connect it to a signal source (e.g., Rigol DS1022). If possible, try to mount the loop so that its magnetic field points along the $B_0$ direction. Turn on the signal source, set the frequency to, e.g., 1.25 MHz and use a safe output power level.
5. Run `python mainControl.py XY8config`. Sweep $\tau$ from 280 ns to 500 ns. Use the MW frequency and power as well as the π-pulse duration determined in the previous NV-ESR and Rabi experiment. It is easier to see the signal for higher *N* e.g., in our case *N* = 12. A 1.25-MHz frequency signal should give a dip at 400 ns [ = 1/(2 x 1.25 MHz)] in the XY8-*N* decay curve (see Figure 17). This indicates that the pulse sequence works as expected. *N* should be optimized by comparing the SNR for the same total duration of the experiment.